\documentclass[twocolumn,twocolappendix]{aastex631}
\usepackage{newtxtext,newtxmath}
\usepackage{amsmath}
\usepackage{footmisc}
\usepackage{soul}  
\usepackage{xcolor}  
\sethlcolor{red!60}
\usepackage{booktabs}
\usepackage{graphicx}	
\graphicspath{ {./images/} }
\usepackage{amsmath}	
\usepackage{cellspace} 
\usepackage[version=4]{mhchem} 
\usepackage{multirow}

\def\tablecommentsnormal#1{\vskip1pt{\small\vskip1sp\indent\vrule height 11pt depth 2pt
width 0pt\currtabletypesize{\sc Note}---{#1}\vskip1pt}} 

\begin{document}

\title{A Detailed Investigation of HD~209458~b HST \& JWST Transmission Spectra with \texttt{SANSAR}}

\correspondingauthor{Jayesh Goyal}
\email{jgoyal@niser.ac.in}

\author[0009-0000-9750-7745]{Avinash Verma}
\affiliation{School of Earth \& Planetary Sciences,  
National Institute of Science Education and Research (NISER), \\
Homi Bhabha National Institute (HBNI),
Jatni, Khordha-752050, 
Odisha, India}

\author[0000-0002-8515-7204]{Jayesh Goyal}
\affiliation{School of Earth \& Planetary Sciences,  
National Institute of Science Education and Research (NISER), \\
Homi Bhabha National Institute (HBNI),
Jatni, Khordha-752050, 
Odisha, India}

\author[0009-0000-7698-7057]{Swaroop Avarsekar}
\affiliation{School of Earth \& Planetary Sciences,  
National Institute of Science Education and Research (NISER), \\
Homi Bhabha National Institute (HBNI),
Jatni, Khordha-752050,
Odisha, India}

\author[0009-0006-6904-234X]{Gaurav Shukla}
\affiliation{School of Earth \& Planetary Sciences,  
National Institute of Science Education and Research (NISER), \\
Homi Bhabha National Institute (HBNI),
Jatni, Khordha-752050,
Odisha, India}

\begin{abstract}
HD~209458~b is the first exoplanet on which an atmosphere was detected. Since then, its atmosphere has been investigated using multiple telescopes and instruments. However, many of its atmospheric constraints remain debatable. While HST observations suggested a highly subsolar metallicity, recent JWST NIRCam observations by \citet{Xue2024} constrained a super-solar metallicity with highly sub-solar C/O. In this work, we show a detailed investigation of HD~209458~b transmission spectra observations from JWST and HST using \texttt{SANSAR}, a newly developed planetary atmosphere modeling framework, with free, equilibrium chemistry and self-consistent grid retrievals. The overall best-fitting model with free retrievals ($\chi^2_{\rm{red}}$=1.21) constrains its metallicity and C/O to be highly sub-solar, while equilibrium chemistry and grid retrievals ($\chi^2_{\rm{red}}$=1.27 and 1.30, respectively) are consistent with solar values using STIS+WFC3+NIRCam observations. 
The retrieved abundances of H$_2$O and CO$_2$ are almost three orders of magnitude lower (highly sub-solar) with STIS+WFC3+NIRCam compared to NIRCam, using free retrievals. NIRCam observations alone also result in misleading constraints on metallicity and C/O, with equilibrium chemistry and grid retrieval. We find that the model choice of varying C/H or O/H to vary the C/O  in equilibrium chemistry retrievals leads to different metallicity constraints with NIRCam, but similar constraints with STIS+WFC3+NIRCam. We conclude that NIRCam observations can lead to overestimation of abundances for exoplanet atmospheres and, therefore, should be used in combination with UV/Optical and near-infrared observations to obtain robust constraints on abundances, C/O, and metallicity. Specifically, even though we can detect the \ce{CO2} feature with just NIRCam, we cannot constrain its abundances robustly without the optical baseline.

\end{abstract}


\section{Introduction} \label{sec:intro}

With the launch of the James Webb Space Telescope (JWST), we are obtaining unprecedented observations of exoplanet atmospheres covering a wide range of wavelengths and spectral resolution. The Transiting Exoplanet Community Early Release Science Program (ERS) has revealed the extensive details of the atmosphere of WASP-39~b \citep[e.g.,][]{ERS_Main2023, Rustamkulov2023, Alderson2023, Ahrer2023}. Similarly, various other JWST programs have given intriguing insights on the physical and chemical properties as well as processes in these planets \citep[e.g.,][]{grant2023jwsttst, Bell2023, Fu2024, Welbanks2024, Sing2024}. These unprecedented JWST observations have also led to many new challenges in their interpretation using atmospheric models, making it necessary to have a detailed investigation of these observations with different model choices \citep[e.g.,][]{Niraula2023, Lueber2024}. 

The interpretation of exoplanet atmosphere observations can be done with various approaches. The traditional approach is fitting to a grid of atmosphere model simulations with varying levels of complexity \citep{Seager2000, Molliere2015, Goyal2021, ERS_Main2023}. The other approach is the inverse modeling approach, where a forward model (with relaxed physical/chemical constraints) is coupled to a Bayesian sampler, so-called atmospheric retrieval \citep[e.g.][]{Madhusudhan2009, Line2013, Waldmann2015, Macdonald2017, Barstow2017, Molliere2019, Lewis2020, Cubillos2021}. Each of these approaches has its own advantages and disadvantages. In this work, we use both approaches to interpret the JWST and the Hubble Space Telescope (HST) observations of HD~209458~b. 

HD~209458~b is the first exoplanet on which an atmosphere was detected \citep{Charbonneau2002}. This was followed by the robust detection of water in its atmosphere \citep{demingwfc3} using the HST Wide Field Camera 3 (WFC3) instrument, after it was tentatively detected by some previous studies \citep{Barman2007, Beaulieu2010}. Even high-dispersion ground-based spectroscopy observations provided intriguing insights by detecting CO and high-altitude winds in the atmosphere of HD~209458~b \citep{Snellen2010}, followed by detection of \ce{H2O} \citep{Lopez2019}. \citet{Sing2016} presented the combined HST Space Telescope Imaging Spectrograph (STIS) plus HST WFC3 observations of HD~209458~b, highlighting the importance of clouds and hazes in muting spectral features. Following this, there have been many studies to interpret the HST observations of HD~209458~b using different modeling approaches and choices, leading to different constraints on its H$_2$O abundances, metallicity, and C/O ratio \citep[for e.g.][]{Line2016, Barstow2017, Macdonald2017, Goyal2018, Pinhas2019, Min2020, Welbanks2021, fairman2024, Novais2025}. The H$_2$O abundance was constrained to be sub-solar \citep{Macdonald2017, Pinhas2019}, and weak evidence of NH$_3$ was reported in the atmosphere of HD~209458~b by \citet{Macdonald2017b}. \citet{Macdonald2017} also reported detection of patchy clouds, further confirmed by \citet{Barstow2020}. \citet{Giacobbe2021} further claimed detection of \ce{NH3}, \ce{CH4} and \ce{C2H2}, CO along with \ce{H2O} using high resolution cross-correlation analysis with ground-based observations.  More recently, \citet{Xue2024} presented the JWST's Near-Infrared Camera (NIRCam) transmission spectrum observations of HD~209458~b for the first time from $\sim$ 2.3 to 5.1 $\mu$m. They reported the detection of H$_2$O and CO$_2$ in its atmosphere and constrained its atmospheric metallicity to $3.5^{+3.4}_{-1.4}$ and $4.9^{+5.8}_{-2.2}$ $\times$ solar metallicity, and C/O ratio to $0.11^{+0.12}_{-0.06}$ and $0.23^{+0.12}_{-0.15}$, using JWST NIRCam observations and HST WFC3+NIRCam observations, respectively. It is also important to note that these constraints on metallicity and C/O ratio by \citet{Xue2024} were obtained using hybrid retrieval assuming chemical equilibrium abundances for all the species, except \ce{CH4}, \ce{NH3}, \ce{C2H2}, and \ce{HCN}, for which the abundances were free parameters, to confirm the detections claimed by previous studies. They did not find any evidence of \ce{CH4}, \ce{HCN}, \ce{NH3} and \ce{C2H2}, which is in disagreement with previous works.

In this work, we present \texttt{SANSAR} (Suite of Adaptable plaNetary atmoSphere model And Retrieval), a flexible planetary atmosphere modeling framework. This Python based modeling framework includes various tools and techniques to model and interpret the observations of a wide range of planetary atmospheres, including exoplanets, planets in our solar system, as well as the Earth. In this work, however, we only focus on the exoplanet transmission spectra part of the framework. A large number of retrieval models currently exist to interpret exoplanet atmosphere transmission spectra observations, for example, POSEIDON \citep{Macdonald2017}, petitRADTRANS \citep{Molliere2019}, PLATON \citep{2020ApJ...899...27Z}, and many more. The exhaustive list can be found in \citet{MacDonald2023}. Each of these models makes different choices, which have their own advantages and limitations. The model choices made while performing atmospheric retrievals can lead to differences in the constraints on the retrieved parameters \citep[for e.g.][]{Rocchetto2016, Barstow2020, Niraula2023, Schleich2024}. Therefore, in this work, we also demonstrate the effects of these different model choices with \texttt{SANSAR}. For example, the choice of opacity treatment method is extremely important for any retrieval model. Opacity sampling and correlated-k are the most widely used methods for opacity treatment in atmospheric retrievals of exoplanet atmospheres. In opacity sampling, the nearest wavelength/wavenumber point in the required resolution grid is sampled from a high resolution ($\sim$R$\ge$1000000) Line-by-Line (LBL) cross-section of a species, while in the correlated-k method, the cross-sections are integrated in the wavelength/wavenumber band by rearranging them. These methods are explained in detail in further sections. Some of the retrieval models use opacity sampling \citep{Macdonald2017} while some use correlated-k \citep{Line2013} as an opacity treatment method. In \texttt{SANSAR}, we have implemented both these methods, and in this work, we demonstrate the differences between these two widely adopted methods and at what spectral resolution they tend to agree, comparing with Line-by-Line (LBL) models within \texttt{SANSAR}. Furthermore, some of the models choose to vary carbon (C/H) while some vary oxygen (O/H) elemental abundances, to change the C/O ratio while computing equilibrium chemical abundances in retrievals. Therefore, we also show the differences in the retrieved constraints when O/H is varied compared to C/H in \texttt{SANSAR} equilibrium chemistry retrievals, as highlighted in \citet{Drummond2019}. In this work, we also demonstrate the differences in retrieved constraints with observations from different HST and JWST instrument combinations using \texttt{SANSAR}. These differences in retrieved constraints can inform us which instrument combinations can give us the most robust constraints and therefore should be used to accurately characterize the exoplanet atmospheres in the JWST era.

\citet{Xue2024} interpreted the JWST spectrum of HD~209458~b with a hybrid inverse modeling approach, assuming equilibrium chemistry for all the species, including the ones that they detect, while keeping the abundances of \ce{CH4}, \ce{NH3}, \ce{C2H2}, and \ce{HCN} free, which they do not detect. They constrained its atmospheric metallicity [$\log (Z)$] and C/O ratio, using this approach. However, they did not perform a complete free retrieval analysis to obtain the detection significance of individual detected species and their abundances. They also performed a joint retrieval analysis that included JWST as well as HST WFC3 observations. However, in this analysis, they did not include HST STIS observations and, therefore, also did not include Na and K opacities, which have prominent features in the part of the spectrum measured by HST STIS. However, neglecting the optical part of the transmission spectrum can have implications on the abundance constraints \citep{Wakeford2018, Pinhas2019, fairman2024}; therefore, in this work, we perform our retrieval analysis on the entire currently available spectrum for HD~209458~b. Moreover, \citet{Xue2024} assumed isothermal Pressure-Temperature ($P$-$T$) profiles in their retrieval, which could lead to biases in measured abundances \citep{Rocchetto2016, Schleich2024} and, therefore, the $\log (Z)$ as well as the C/O ratio. In this work, we perform free retrievals with isothermal as well as two different Pressure-Temperature ($P$-$T$) profile parametrizations. We also perform equilibrium chemistry retrievals on the same dataset as used in \citet{Xue2024}, first comparing the constraints and then further highlighting the differences when a complete dataset with STIS observations is used. Furthermore, we also perform grid-retrievals with self-consistent radiative-convective equilibrium $P$-$T$ profiles, consistent with equilibrium chemistry, as detailed in \citet{Goyal2020}. 

Besides HD~209458~b, we also perform atmospheric transmission spectra retrievals for WASP-96 b using SANSAR. WASP-96 b, an inflated hot-Saturn exoplanet, was observed by JWST in 2022 using NIRISS/SOSS instrument, covering the wavelength range from 0.6 to 2.85 $\mu$m. This observation was released by the JWST Early Release Science (ERS) team \citep{Radica2023} and has been studied extensively with different forward and retrieval models \citep{Taylor2023}. Hence, we choose this exoplanet for benchmarking \texttt{SANSAR} transmission spectra retrievals.

In Section \ref{sec:SANSAR}, we describe the details of the \texttt{SANSAR} transmission spectra model. In Section \ref{sec:benchmark}, we benchmark transmission spectra computation from the \texttt{SANSAR} model with other published results. In Section \ref{subsec:free}, we present free chemistry retrieval results of the interpretation of HD~209458~b with \texttt{SANSAR} using different instrument combinations, followed by equilibrium chemistry results in Section \ref{subsec:eq}. In Section \ref{subsec:grid}, we present the results of the grid of self-consistent model atmospheres of HD~209458~b computed using ATMO, with transmission spectra computation using \texttt{SANSAR}. We discuss and compare the results with different retrieval methods, instrument combinations, and previous works in Section \ref{sec:discussion}, and finally, we conclude in Section \ref{sec:conclusions}.

\section{SANSAR Model}
\label{sec:SANSAR}
In this section, we describe the details of the \texttt{SANSAR} model used in this work for simulating transmission spectra as well as to perform free retrieval, equilibrium chemistry retrieval, and self-consistent grid-retrieval to interpret exoplanet atmosphere observations. 

\subsection{Transmission Spectrum Computation}
\label{subsec:method_trans}

When a planetary body transits between the line of sight of the star and the observer, it causes a decrease in the flux of the star, as it is measured by the observer. This decrease in flux can be different in different wavelengths, depending on the characteristics of the exoplanet atmosphere. This wavelength-dependent decrease in flux can be defined using the transit depth $(\delta_\lambda)$ given by,

\begin{equation}
\delta_\lambda = \frac{F_{\text{out},\lambda} - F_{\text{in},\lambda}}{F_{\text{out},\lambda}},
\label{eq:transit depth}
\end{equation}
where the spectral flux $F_{\lambda}$ represents the spectral intensity integrated over the solid angle subtended by the star as seen by the observer. Subscript `out' and `in' represent the flux before/after and during the transit, respectively.

Spectral Flux \citep{Thomasstammes1999} is defined as
\begin{equation}
F_\lambda=\int_{\Omega} I_\lambda \hat{\boldsymbol{n}} \cdot \hat{\boldsymbol{k}} d \Omega, 
\end{equation}
where $I_{\lambda}$ is the spectral intensity, $\hat{\boldsymbol{n}}$ is the unit vector in the direction of beam propagation, and $\hat{\boldsymbol{k}}$ is the unit vector in the direction of the observer.
Since the distance to the star is much greater than its radius, we can approximate all rays reaching the observer as parallel. Thus $\hat{\boldsymbol{n}} \cdot \hat{\boldsymbol{k}}$ = 1, this also implies that the direction of the ray is constant before and after entering the atmosphere, i.e., refraction is neglected in this model. In this geometry, a solid angle subtended by a star on the observer is given by $d \Omega = \frac{dA}{D^2}$, where $dA$ is the projected source area onto the observer and $D$ is the distance of the planetary system from the observer.

Neglecting planetary emission and considering stellar brightness to be uniform throughout the stellar disk, we get
\begin{equation}
F_{\text{out},\lambda} =  \frac{I_{\lambda}A_{\text{out}}}{D^2} 
;~
F_{\text{in},\lambda} = \frac{ I_{\lambda}A_{\text{in}}}{D^2}, 
\label{eq:flux value}
\end{equation}
where $A_{\text{out}}$ and $A_{\text{in}}$ are the areas projected by the source before and during transit from the observer's perspective.
For a more general derivation, you can refer to \citet{MacDonald_2022}.

Substituting equation \ref{eq:flux value} in equation \ref{eq:transit depth}, we obtain
\begin{align*}
\delta_\lambda = \frac{A_{\text{out}} - A_{\text{in}}}{A_{\text{out}}}.
\end{align*}

We know that 
\begin{align*}
A_{\text{out}} = \pi R_{\star}^2\,\,\,(\text{Area of stellar disk}),  \text{and} \\
A_{\text{in}} = \pi R_{\star}^2 - \pi R^2_p(\lambda)\,\,\,(\text{Planetary disk area}).
\end{align*}

Therefore, we obtain,

\begin{equation}
    \delta_\lambda = \frac{\pi R_\star^2 - (\pi R_\star^2- \pi R_p^2)}{\pi R_\star^2} = \frac{R^2_p(\lambda)}{R_\star^2}
    \label{eq:rprstar}
\end{equation}

where, $R_p(\lambda)$ is the wavelength-dependent radius of the planet

Taking absorption into consideration, equation \ref{eq:rprstar} can be written as,

\begin{equation}
\delta_{\lambda} = \frac{R_{\text{low}}^2 + 2\int_{R_{\text{low}}}^{R_{\text{low}}+H_{\text{atm}}} b (1-e^{-\tau(b)})db}{R_\star^2}
\label{eq:td main eq}
\end{equation}

Here $R_{\text{low}}^2$ is the lowest radius below which no radiation can pass the atmosphere, $H_{\text{atm}}$ is the total height of the atmosphere, and $b$ is the impact parameter of the ray passing through the atmosphere. $R_{\text{low}}$ is a model-dependent radius here and is determined uniquely in each iteration of retrieval (or forward model) by following the hydrostatic equilibrium from the reference pressure and reference radius pair to the point of highest pressure.
A detailed derivation of equation \ref{eq:td main eq} is presented in Appendix \ref{ap:trans-derive}
This equation is a mathematical representation of the addition of the cross-sectional area of each layer, weighted by the amount of radiation they block.
The optical depth, $\tau$ in the equation \ref{eq:td main eq}, is a characteristic of the molecules that might be present in the planet's atmosphere and the atmospheric structure. The computation of $\tau$ is further detailed in Section \ref{sec:model atmosphere}.

\subsection{Model Atmosphere}
\label{sec:model atmosphere}
In the \texttt{SANSAR} model, the atmosphere is assumed to be in hydrostatic equilibrium given by;

\begin{equation}
\frac{dP}{dr} = -\rho(r)g(r),
\label{eq:hydrostatic}
\end{equation}

where, $\frac{dP}{dr}$ is the change in pressure with change in radius,  $\rho(r)$ is the mass density at radius $r$ and 
$g(r)$ is the gravitational acceleration at radius $r$. We have adopted a 1D approximation of the atmosphere where chemical and atmospheric properties are only radius-dependent. Therefore, we discretize the atmosphere in $n$ layers, each defined by its own pressure and temperature and constrained by hydrostatic equilibrium. While each column is treated as 1D, SANSAR also allows for azimuthal asymmetries in aerosol properties, enabling the modeling of patchy or spatially variable cloud and haze distributions across the planetary terminator as discussed in Section \ref{sec:cloud haze}. Variation of gravitational acceleration $g(r)$ at any radius $r$ in the model is given by,

\begin{equation}
g(r) = \frac{g_{\text{obs}}R_p^2}{r^2},
\label{eq:gobs}
\end{equation}

Here, \( g_{\text{obs}} \) represents the observed (measured) gravitational acceleration of the planet, determined from its observed mass and white light radius (R$_p^2$ in equation \ref{eq:gobs}).  In some retrieval studies \citep{Barat2024}, the surface gravity \( g_{\text{obs}} \) is not fixed to a measured value but instead treated as a free parameter either directly or by allowing the planetary mass to vary. However, when the planetary mass is well constrained from observations, as is often the case for Hot \-Jupiters, the measured surface gravity is typically used directly in the model. Conversely, for planets with poorly constrained masses, retrievals commonly allow for flexibility by fitting for \( g_{\text{obs}} \) or the mass as a free parameter. Substituting equation \ref{eq:hydrostatic} in equation \ref{eq:gobs} and using the mass density definition from the ideal gas law, we get,

\begin{equation}
\frac{dP}{dr} = -\frac{\mu P }{k_B T } \frac{g_{\text{obs}} R_p^2}{r^2} , \quad \text{where $\mu$ is the mean molecular mass}.
\end{equation}

Here, $k_B$ is the Boltzmann constant. Rearranging the equation for radius, we obtain

\begin{equation}
dr = - \frac{k_B T}{\mu P g(r)} dP , \quad \text{where } g(r) = \frac{g_{\text{obs}}R_p^2}{r^2}.
\end{equation}

Here $\frac{k_B T}{\mu g(r)}$ is called the scale height of the atmosphere, which gives us the vertical distance over which atmospheric pressure drops by a factor of `e'. Transmission spectroscopy is well-suited for planets with higher scale heights.

To solve this differential equation, an initial radius and pressure pair, referred to as the reference radius and reference pressure, is required. This pair can be chosen arbitrarily; however, since the observed radius is often determined through other methods, it is common to use the observed radius as the reference radius and assign a reference pressure of 1 mbar or 10 mbar at this radius. In theory, any known radius-pressure pair can serve as a reference \citep{Welbanks2019}. During retrieval, either the reference radius or the reference pressure can be fixed, allowing the other to be retrieved.

\begin{figure}
    \includegraphics[width=10cm]{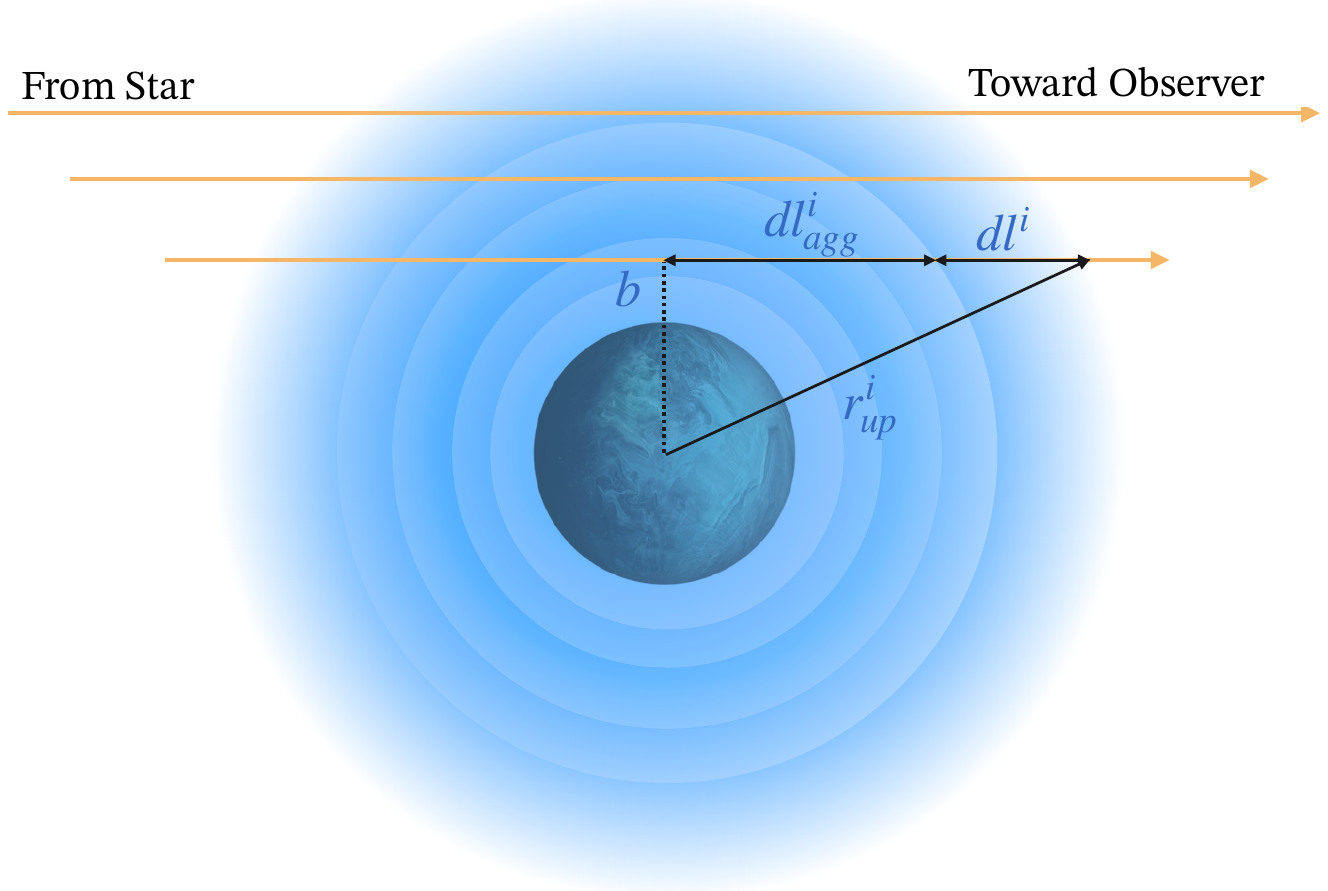}
    \caption{Ray diagram showing the calculation of path length traveled by a ray in an exoplanetary atmosphere at an arbitrary impact parameter $b$. Here $r^{i}_{\rm{up}}$ is the upper boundary radius of a layer, $d l_{\text{agg}}^i$ is the total path length traveled by the ray in the half geometry till the $i-1$ layer, and $d l^i$ is the path length of the ray in the $i^{\text{th}}$ layer.}
    \label{fig:path_length}
\end{figure}

Solving the equation numerically gives us the thickness and radial profile of each layer used in calculating the path length of a ray with impact parameter $b$ in the $i^{\text{th}}$ layer while computing the transmission spectrum using equation \ref{eq:td main eq} as shown in Figure \ref{fig:path_length}.

Therefore, the length $dl^i$ traversed by a ray with impact parameter $b$ in $i^{\text{th}}$ layer is given by,
\begin{equation}
    d l^i = \sqrt{(r_{\text{up}}^i)^2 - b^2} - d l_{\text{agg}}^{i}, 
\end{equation}

where $r_{\text{up}}^i$ is the distance of the top of the layer $i$ from the center, $b$ is the impact parameter of the ray, and $d l_{\text{agg}}^i$ is the aggregated length of ray that it traveled till layer $i-1$. Utilizing the symmetry of the model atmosphere, the total length traveled in layer $i$ will be twice that of $d l^i$. Finally, we can use this path length to calculate the slant optical depth, $\tau_{\lambda}(b)$  by multiplying the path length with the total extinction coefficient $\kappa_{\lambda}(i)$ at the impact parameter $b$ given as,
\begin{equation}
    \tau_{\lambda}(b) = \sum^{n}_{i=1}\kappa_{\lambda}(I) d l(b,i),  
    \label{eq:tau calc}
\end{equation}

where $n$ is the number of layers. Here, the contribution to the total extinction coefficient can come from different sources, given as,

\begin{equation}
    \kappa_{\lambda,\text{tot}} = \kappa_{\lambda,\text{chem}} + \kappa_{\lambda,\text{pair}} + \kappa_{\lambda,\text{Rayleigh}} +
    \kappa_{\lambda,\text{aerosol}}
\end{equation}

Here, $\kappa_{\lambda,\text{chem}}$ is the extinction caused by chemical species (molecules, atoms and ions), $\kappa_{\lambda,\text{pair}}$ refers to the extinction originating from pair species (CIA pairs or free-free pairs) extinction, $\kappa_{\lambda,\text{Rayleigh}}$ is the extinction due to Rayleigh scattering processes and $\kappa_{\lambda,\text{aerosol}}$ is attributed to cloud and hazes.
\subsection{Opacity Database}
\label{sec:opac-database}
For accurate modeling of the transmission spectrum of a planet, accurate computation of $\tau$ described in the previous section is important, which is dependent on the 
absorption cross-section (opacity) of different species in the planetary atmosphere as shown in equation \ref{eq:tau calc}. Therefore, a well-tested opacity database spanning a large range of wavelengths, temperatures, and pressures of various atoms and molecules is needed. The opacity database that is used in this work is the same as adopted in \citet{Goyal2020} for ATMO \citep{Tremblin2015, Amundsen2014, Goyal2018}. This database is well tested and has been utilized to interpret the observations of various exoplanet atmospheres \citep[for e.g.][]{Evans_2017, ERS_Main2023}. This opacity database consists of absorption cross-sections of CO$_2$, CO, H$_2$O, Na, K, TiO, VO, Li, Rb, Cs, HCN, C$_2$H$_2$, PH$_3$, Fe, CH$_4$, H$_2$S, SO$_2$, FeH at the resolution of 0.001 cm$^{-1}$. The database also includes Collision Induced Absorption (CIA) due to H$_2$-H$_2$ and H$_2$-He. The absorption cross-sections in this database are computed for 20 temperature grid points ranging from 70 K to 3000 K and 40 pressure grid points ranging from 10$^{-9}$ bar to 10$^3$ bar, giving a total of 800 $P$-$T$ points for each species. The details of the line-list sources and broadening parameters for each species can be found in \citet{Goyal2020}. Line list sources for a subset of species used in this work are also provided in Table \ref{tab:line-list sources} of Appendix \ref{app: linelist_source}

\subsection{Opacity Treatment}
Atoms and molecules can have billions of quantum mechanical transitions between different states, depending on the temperature and pressure of the planetary atmospheres. Therefore, it would be accurate to include all these transitions while computing the transmission spectra. Ideally, this would mean computing the transmission spectra at the highest spectral resolution possible, in this case, the spectral resolution of our absorption cross-sections, so-called line-by-line (LBL) cross-sections. These LBL cross-sections are at spectral resolution, R $\sim$ $10^7$. However, the spectral resolution that we encounter in HST (R$\sim$100) and/or JWST observations maximum can reach up to R$\sim$ 3500 (G395H). Moreover, computing transmission spectra at LBL resolution can be computationally expensive, especially for retrievals where we have to perform millions of these transmission spectra computations. Therefore, a few techniques have been developed and adopted widely to sample these LBL cross-sections to low resolutions, as needed by the spectral resolution of the observations or for simulating particular processes in the theoretical model (e.g., generating radiative-convective equilibrium $P$-$T$ profiles). In \texttt{SANSAR}, we have implemented two widely used techniques, opacity sampling \citep{Johnson1996} and correlated-k technique \citep{Lacis1991}. We detail their implementation in \texttt{SANSAR}, further in this section.

\subsubsection{Opacity sampling}
\begin{figure}
    \includegraphics[width=8cm]{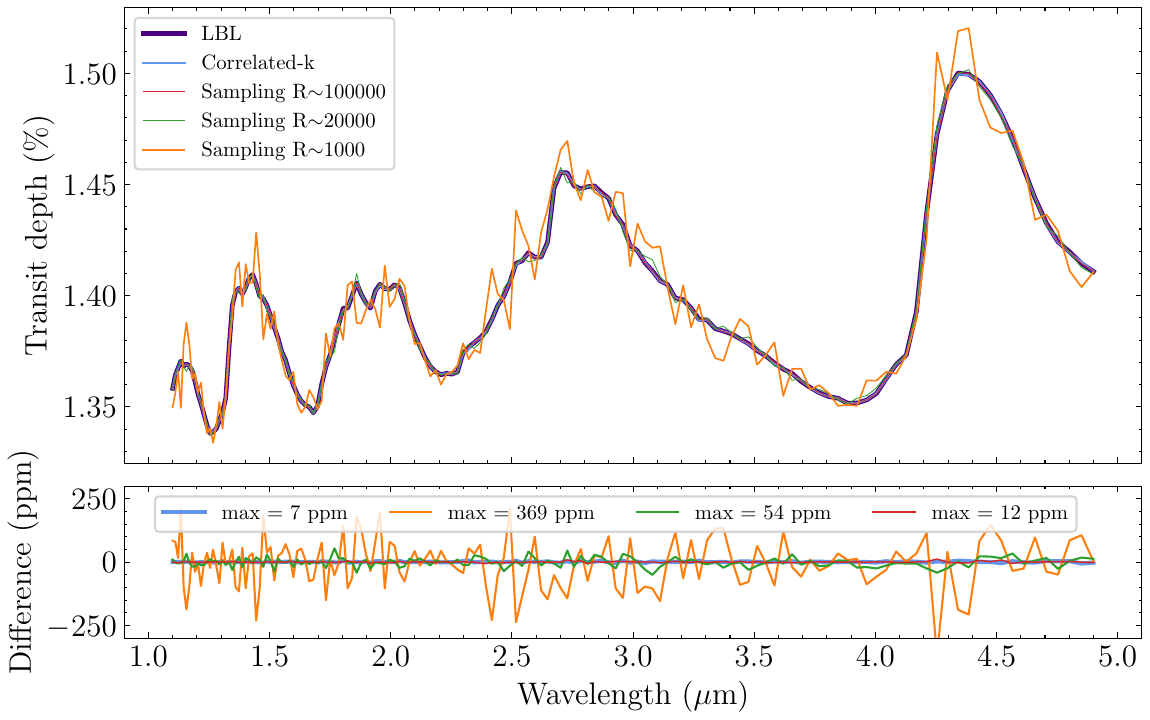}
    \caption{Figure showing line by line (LBL) transmission spectra comparison with opacity sampling at a resolution of 1000, 20000, and 100000  (all resolutions finally binned to R$\sim$100 for comparison) and with correlated-k at R$\sim$1000. Residuals compared with LBL are shown at the bottom, and the maximum error in each case is shown in the legend.}
    \label{fig:sampling}
    
\end{figure}

The Opacity sampling technique was originally developed for stellar atmospheres to reduce the number of radiative opacity calculations for a large number of atomic and molecular lines in an atmosphere \citep{Johnson1996}. Usually, the observations are at very low resolution as compared to the high-resolution LBL cross-section. We can sample the opacities at a moderate resolution (higher than observation but much lower than LBL) by choosing the nearest wavenumber points from the LBL wavenumber grid, this is called opacity sampling. As long as we sample enough wavenumber points in each opacity bin, the error is very low. We demonstrate this by generating example model transmission spectra for a WASP-96 b like planet for three different spectral resolutions at 1000, 20000, and 100000, with opacity sampling and comparing them with LBL transmission spectra as shown in Figure \ref{fig:sampling} (all the spectra has been binned to R$\sim$100 for comparison). As shown in the figure, the transit depth error introduced by opacity sampling at a resolving power of R$\sim$1000, relative to line-by-line (LBL) calculations, is substantial, with an absolute mean difference of 74 ppm and a maximum deviation reaching 369 ppm. Increasing the resolution to R$\sim$20000 significantly reduces the discrepancy, yielding an absolute mean difference of 14 ppm and a maximum error of 54 ppm. At R$\sim$100000, the errors become negligible for most applications, with an absolute mean difference of approximately 2 ppm and maximum deviation around 12 ppm. Therefore, the optimized choice of opacity sampling resolution should be made according to the spectral resolution of the observations. In this work, we use opacity sampling at a resolution of 20,000 to interpret the transmission spectra observations of WASP-96 b with the NIRISS SOSS instrument for benchmarking \texttt{SANSAR} as detailed in Section \ref{sub:inverse-benchmark}.

\subsubsection{Correlated-k}
\label{sec:corrk}
Absorption cross sections can vary substantially from one wavelength to another. The calculation of the transmission function, $\mathcal{T}(u)$ using these absorption cross-sections is given by:

\begin{equation}
\mathcal{T}(u)=\frac{1}{\tilde{\nu}_2-\tilde{\nu}_1} \int_{\tilde{\nu}_1}^{\tilde{\nu}_2} \mathrm{~d} \tilde{\nu} e^{-k_\rho(\tilde{\nu}) u},
\end{equation}

where $\tilde{\nu}_1$ and $\tilde{\nu}_2$ are the limits of the spectral interval, $k_\rho(\tilde{\nu})$ is the mass absorption coefficient at wavenumber $\tilde{\nu}$, and $u$ is the mass column density.
High-resolution radiative transfer calculations are necessary to account for the highly variable nature of absorption coefficients across wavelengths. The core concept of the correlated-k method lies in the fact that radiative transfer calculations depend only on the value of the absorption coefficient rather than the specific wavelength itself. This property allows the absorption coefficients to be reordered into a smooth, monotonic distribution by arranging them in order. Using this rearranged distribution, absorption coefficients with similar values can be grouped together and represented collectively by a single coefficient k. The total transmission is then expressed as
\begin{equation}
\mathcal{T}(u)=\int_{k_{\min }}^{k_{\max }} \mathrm{d} k f(k) e^{-k u}
\label{eq:corrk-fkdk}
\end{equation}

where $f(k)$ is a probability density function such that $\mathrm{d}kf(k)$ represents the probability of the absorption coefficient falling between $k$ and $k$+d$k$. The detailed calculation of these probabilities can be found in \citet{Amundsen2017}. The key insight of this method is that it transforms the highly variable absorption coefficients at different wavenumbers into a smoothly varying probability distribution. This transformation is possible because the wavenumber values themselves do not play any role in radiative transfer computations.

We can approximate the integration in equation \ref{eq:corrk-fkdk} using Gaussian quadrature by introducing cumulative probability $g$, where d$g$ = $f(k)$d$k$ and 

\begin{equation}
g(k)=\int_0^k \mathrm{~d} k^{\prime} f\left(k^{\prime}\right).
\end{equation}

where $g(k)$ is the probability of the absorption coefficient being less than $k$. Therefore, the transmission function can be written as

\begin{equation}
\mathcal{T}(u)=\int_0^1 \mathrm{~d} g e^{-k(g) u} \approx \sum_{l=1}^{n_k} w_l e^{k_l u}. 
\end{equation}
Here, $n_k$ is the number of Gauss points, $w_l$ are the weights, and $k_l$ is the mass absorption coefficient at the Gauss nodes.

We have adopted the method of random overlap to treat the k-coefficients of a mixture of gases as described in \citet{Amundsen2017}. Here, we assume that the k-coefficients of different gases are uncorrelated within a band. As the coefficients are uncorrelated, the total transmittance can be written as the product of the individual gas transmittances. 

\begin{figure}
    \includegraphics[width=8cm]{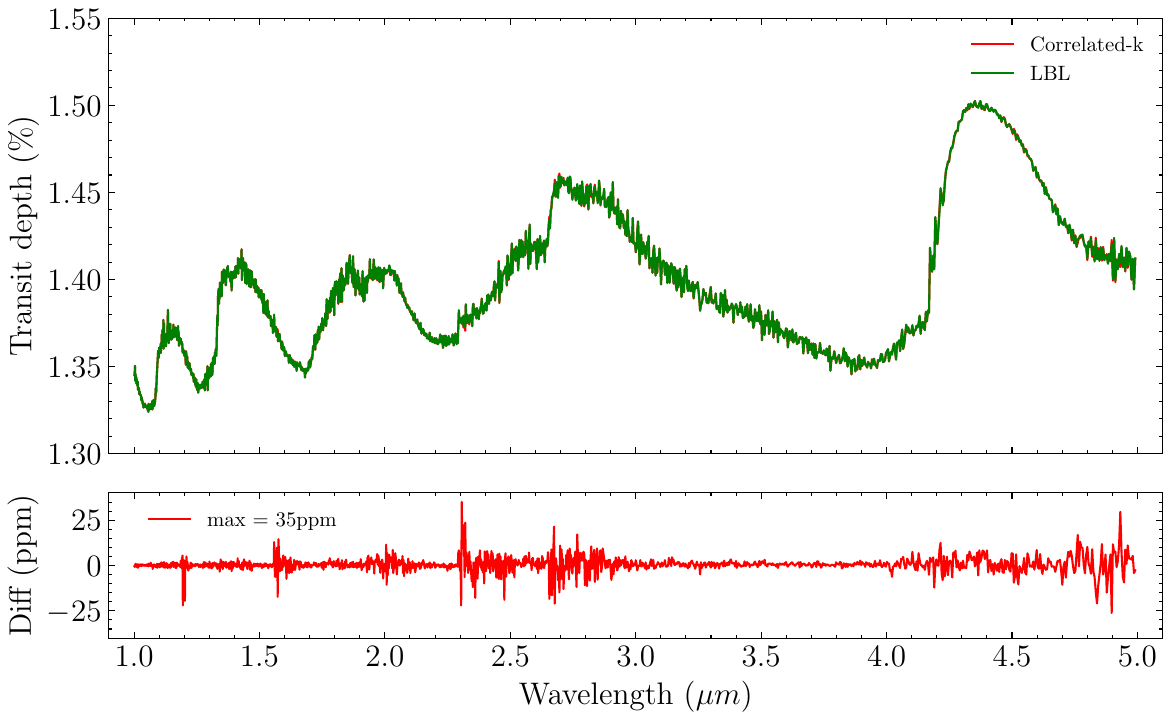}
    \caption{Comparison between transmission spectra computed using the correlated-k opacity method and those from line-by-line (LBL) calculations. The LBL spectra have been binned to correlated-k wavelength grid (R$\sim$1000). Residuals are shown in the bottom panel.}
    \label{fig:corrk_lbl}
\end{figure}

The k-tables in \texttt{SANSAR} are generated using \texttt{Exo\_k} \citep{Leconte_2020}, an open-source tool to compute k-tables using LBL cross-sections at any spectral resolution. \texttt{Exo\_k} provides the control to choose resolution and gauss points while generating k-tables, hence we have used that. We have used the LBL cross-section database as described in Section \ref{sec:opac-database} for k-table generation. These k-tables were generated with 10 Gauss points. The \texttt{Exo\_k} gives the flexibility in \texttt{SANSAR} to generate k-tables at any given resolution, as required. 
To verify our implementation of the correlated-k method with respect to the LBL method, we generated transmission spectra of WASP-96 b with H$_2$ dominated atmosphere where \ce{H2O}, \ce{CO}, and \ce{CO2} are trace gases, using both the methods. The LBL spectrum was generated at a resolution of R $\sim$ $10^7$, while the correlated-k spectrum was generated at R $\sim$1000. We then binned the LBL spectrum on the correlated-k wavelength grid. Figure \ref{fig:corrk_lbl} shows the comparison between LBL and correlated-k transmission spectra. The agreement between them is excellent, with a maximum difference of 35 ppm.  Furthermore, we compare correlated-k spectra with opacity sampling spectra at three different sampling resolutions (all binned to R $\sim$ 100) as described in the previous section and shown in Figure \ref{fig:sampling}. We can see that the accuracy of correlated-k at just R$\sim$1000 with 10 Gauss points can match the opacity sampling at R$\sim$100000 while being $\sim$10 times faster than opacity sampling, as discussed in Section \ref{sec:perf bench}.

\subsection{Rayleigh Scattering}
In \texttt{SANSAR}, the wavelength-dependent Rayleigh scattering cross section, $\sigma_{\lambda}$ is calculated using the following equation from \citet{Liou1980};

\begin{equation}
\sigma_{\lambda} = \frac{8}{3}\frac{\pi^3 (\eta^2 -1)^2}{\lambda^4 N_{\text{ref}}^2}\frac{(6 + 3\text{dp})}{(6-7\text{dp})},
\end{equation}

where $\eta$ is the refractive index of the scattering gas, $N_{\text{ref}}$ is the Loschmidt constant (number density at standard temperature and pressure), and dp is the depolarization factor, which takes into account the nonspherical nature of the particles (also known as King correction). {Most of the depolarization factors have been back calculated from the King coefficients from \citet{SNEEP2005293}. The current version of \texttt{SANSAR} incorporates Rayleigh scattering by H\textsubscript{2} and He, using refractive indices from \citet{Leonard1974} and \citet{Mansfield1969}, respectively, along with depolarization factors from \citet{Penndorf1957}. For CO, N\textsubscript{2}, CH\textsubscript{4}, SF\textsubscript{6}, N\textsubscript{2}O, and CO\textsubscript{2}, both the refractive indices and depolarization factors are adopted from \citet{SNEEP2005293}, while the values for NH\textsubscript{3} and H\textsubscript{2}O are taken from \citet{Allen2000}}.

\subsection{Clouds \& Haze}
\label{sec:cloud haze}

We have adopted a 2-D cloud and haze prescription from \citet{Macdonald2017}

\begin{equation}
{\kappa}_{\text{cloud }}(\lambda, P)= \begin{cases}n_{\mathrm{H}_2} a \sigma_0\left(\lambda / \lambda_0\right)^\gamma, & P<P_{\text {cloud }} \\ \infty, & P \geqslant P_{\text {cloud }}\end{cases}
\end{equation}

Here haze is defined as enhanced \ce{H2} Rayleigh scattering, where $a$ is the Rayleigh-enhancement factor, $\sigma_0$ is the \ce{H2} Rayleigh scattering cross-section at $350$ nm (taken as 5.31 $\times$10$^{-31}$ m$^2$ from \citet{refId0}), $\gamma$ is scattering slope and $\kappa_{\text{cloud}}$ is the cloud extinction coefficient. \\
Here, we assume a cloud deck that blocks all the light irrespective of its wavelength below a certain pressure, $P_{\text{cloud}}$. Also, assuming this cloud spans over a certain fraction ($\phi$) along the terminator region. The way it is achieved is by taking a weighted average of the transmission spectra with clouds and without clouds.

\begin{equation}
\delta_\lambda={\phi} \delta_{\lambda, \text { cloudy }}+(1-{\phi}) \delta_{\lambda, \text { clear }}
\end{equation}
Here $\delta_{\lambda, \text { cloudy }}$ is the transit depth with cloudy atmosphere and $\delta_{\lambda, \text { clear }}$ is the transit depth by assuming a clear atmosphere

\subsection{Interpolation}
 Absorption of radiation traveling from one point to another in an atmosphere is governed by the optical depth (Section \ref{sec:model atmosphere}). Since the absorption cross-section depends on the pressure-temperature ($P$-$T$) profile of the atmosphere, it is crucial to obtain accurate cross-section values corresponding to the defined atmospheric $P$-$T$ profile (see Section \ref{sec:opac-database}). Storing and utilizing cross-section values on a densely sampled $P$-$T$ grid is computationally impractical. Therefore, we interpolate the absorption cross-section values from a sparse absorption cross-section grid to match the actual atmospheric $P$-$T$ conditions. Unlike wavenumber, cross-section values vary smoothly over a $P$-$T$ range, thus allowing the use of a variety of interpolation techniques. In this work, we have adopted bicubic convolution-based interpolation \citep{1981ITASS..29.1153K}, which uses a 4*4 convolution kernel. The bicubic interpolation is an extension of 1-D cubic interpolation in two dimensions. A convolution kernel defines how neighboring $P$-$T$ from the grid in an opacity database contribute to the atmospheric $P$-$T$. A 1-D convolution kernel is defined as:

\begin{equation}
W(x)= \begin{cases}(a+2)|x|^3-(a+3)|x|^2+1, & \text { for }|x| \leq 1 \\ a|x|^3-5 a|x|^2+8 a|x|-4 a, & \text { for } 1<|x|<2 \\ 0, & \text { otherwise }\end{cases}
\end{equation}

Here, $x$ is the positional value of atmospheric pressure (or temperature), and $a$ is the smoothening factor. 
$x$ is determined based on the position of atmospheric pressure (or temperature) relative to the grid values of the pressure (or temperature) in the opacity database. A convolution kernel calculation is followed by interpolation along both axes. This approach facilitates the use of a single kernel for all the molecules, thus providing computational efficiency as the kernel needs to be calculated only once. 

Here, we have used $a$ = -0.5, which is a good balance between smoothness and sharpness. The plot demonstrating the efficiency of bicubic convolution interpolation in \texttt{SANSAR} is shown in Figure \ref{fig:A bicubic comp} of Appendix \ref{app: benchmarking}

\subsection{Parametric $P$-$T$ Profiles}
\label{sec:atm pt}
Unlike forward models, where radiative-convective equilibrium $P$-$T$ profiles are computed numerically, $P$-$T$ profiles are retrieved and constrained using the observations in inverse modeling (retrievals). Therefore, the $P$-$T$ profile needs to be parameterized with certain parameters that would be variables in the retrieval process. Two widely used parameterizations of $P$-$T$ profile for exoplanet atmospheres are the Guillot profile \citep{Guillot2010} and the Madhusudhan profile \citep{Madhusudhan2009}.  We have implemented both of these parameterizations in \texttt{SANSAR}. Each of these profiles has its own advantages and disadvantages. 

Guillot  $P$-$T$ profile assumes the atmosphere to be in thermal equilibrium and is constrained by the balance between total incoming star irradiation and outgoing planetary radiation. It is based on a three-channel (two downwelling visible and one upwelling thermal) approximation and has five free parameters. The five free parameters in the Guillot profile are Planck mean thermal IR opacity ($\kappa_{\text{IR}}$ in cm$^2$ g$^{-1}$), ratios of the Planck mean opacities in the first visible stream to thermal ($\gamma_1$), ratios of the Planck mean opacities in the second visible stream to thermal ($\gamma_2$), albedo/redistribution factor ($\beta$) and factor partitioning the flux between the two visible streams ($\alpha$). Apart from these variables, Guillot parameterization also requires certain constant parameters of the star-planet system for energy balance, which are stellar radius, stellar temperature, the distance between the star and planet, and the internal temperature of the planet. The complete derivation and equation of the Guillot  $P$-$T$ profile parameterization can be found in \citet{Guillot2010} and \citet{Line2013}.

In contrast, the Madhusudhan profile does not take into account the balance between total incoming star irradiation and outgoing planetary radiation. Unlike the Guillot profile, the Madhusudhan profile is not physically motivated but is very flexible and can fit a wide range of $P$-$T$ profiles. In this parameterization, the atmosphere is divided into three regions with the freedom to have inversion in the middle layer. The temperature in the three regions is parameterized as:

\begin{equation}
\begin{array}{r}
P=P_0 e^{\alpha_1\left(T-T_0\right)^{\beta_1}}\,\,\,\,\,\,\text{for}\,\,\,\,\,\,P_0<P<P_1\,\,\,\,\,\,\text{(Layer 1)} \\
P=P_2 e^{\alpha_2\left(T-T_2\right)^{\beta_2}}\,\,\,\,\,\,\text{for}\,\,\,\,\,\,P_1<P<P_3\,\,\,\,\,\,\text {(Layer 2)} \\
T=T_3\,\,\,\,\,\,\text{for}\,\,\,\,\,\,P>P_3\,\,\,\,\,\,\text{(Layer 3)}
\end{array}
\end{equation}

Here $P_0$ is the pressure at top of the atmosphere, $P_1$ is the pressure at the boundary of layer 1 \& 2, $P_2$ is the inversion pressure, $P_3$ is the boundary pressure at layer 2 \& 3, $\beta_1$ = $\beta_2$ = 0.5, $\alpha_1$ and $\alpha_2$ are free parameters, $T_3$ is the isothermal temperature in layer 3.

\subsection{Chemistry}
In \texttt{SANSAR}, currently, we can perform free retrievals as well as equilibrium chemistry retrievals. In free retrievals, the abundances (number density/mixing ratios) of each species (for e.g. \ce{H2O}, \ce{CO2}) are free parameters. However, during free retrieval, we ensure that the mixing ratios of all the chemical species are normalized to one given by, 




\begin{equation}
X_i = \frac{\tilde{X_i}}{\tilde{X}_{\text{total}}}, \quad \text{where }\\ \tilde{X}_{\text{total}} = \sum\limits^{\text{bulk}}\tilde{X}_{i} + \sum\limits^{\text{trace}}\tilde{X}_i.
\end{equation}

Here, $\tilde{X}_{\text{total}}$ is the sum of unnormalized mixing ratios of all species, $\tilde{X}_i$ is the unnormalized mixing ratios of each species, whereas $X_i$ represents the mixing ratio after normalization. In this study, the bulk atmospheric constituents are H$_2$ and He, whose unnormalized mixing ratios are fixed at 0.84 and 0.15, respectively; the rest of the chemical species are considered to be trace. 

In \texttt{SANSAR}, equilibrium chemistry retrievals are currently performed by coupling \textsc{pyfastchem}, which calls \textsc{FastChem 2} \citep{stock2022fastchem} to \texttt{SANSAR}. \textsc{FastChem} solves a nonlinear system of coupled equations obtained by the law of mass action and element conservation equations in a semianalytical method. In equilibrium chemistry retrievals, carbon to oxygen ratio (\ce{C/O}) and metallicity, $\log(Z)$ are used as free parameters instead of mixing ratios. In order to change the metallicity, we scale all the abundances of the elements from \cite{asplund2009}, except \ce{H} and \ce{He}. To vary the \ce{C/O}, we change the elemental abundance of either \ce{C} or \ce{O}. We detail the effect of changing either \ce{C} or \ce{O} on our results in Section \ref{sec:results}. Therefore, the elemental abundances, C/O, $\log(Z)$, and the $P$-$T$ profile are given as input to \textsc{pyfastchem}. We obtain thermochemical equilibrium number densities of the gases along with the mean molecular mass of the atmosphere from \textsc{pyfastchem}, which are used as input to \texttt{SANSAR} to compute the transmission spectra. We note that in our current retrieval setup, for each call, \textsc{pyfastchem} calculates the number densities of gases `on the fly' (otf). \textsc{pyfastchem} provides options to compute equilibrium chemical abundances with gas phase only, gas phase with local condensation, and gas phase with rainout condensation using \textsc{FastChem Cond} \citep{fastchem_cond}. In this work, we perform equilibrium chemistry retrievals with gas phase only, as required for interpretation or comparison with previous works.

\subsection{Retrieval}
As mentioned before, the observations can be interpreted using forward as well as inverse models (retrieval). In the previous sections, we described the implementation of \texttt{SANSAR} forward model transmission spectra. In this section, we detail the coupling of a Bayesian sampler to a forward model in \texttt{SANSAR} to perform exoplanet atmosphere retrievals and determine parameters that represent the observed spectra.  

Given a set of models ($M$) and their corresponding parameters ($\boldsymbol{\theta}$), Bayes' theorem is used to get constraints on the posteriors of the parameters. This is given by 

\begin{equation}
p\left(\boldsymbol{\theta_i} \mid \boldsymbol{y}_{\text {obs }}, M_i\right)=\frac{p\left(\boldsymbol{y}_{\text {obs }} \mid \boldsymbol{\theta_i}, M_i\right) p\left(\boldsymbol{\theta_i} \mid M_i\right)}{p\left(\boldsymbol{y}_{\text {obs }} \mid M_i\right)}, 
\end{equation}

where, $p\left(\boldsymbol{y}_{\text {obs }} \mid \boldsymbol{\theta_i}, M_i\right)$ is defined from the log-likelihood, $p\left(\boldsymbol{\theta_i} \mid M_i\right)$ is the prior probability distribution and $p\left(\boldsymbol{y}_{\text {obs }} \mid M_i\right)$ is the total evidence.\\

In \texttt{SANSAR}, we use \texttt{PyMultiNest} \citep{pymultinest}, a Python wrapper for \texttt{MultiNest} \citep{2009MNRAS.398.1601F} as our Bayesian sampler. \texttt{PyMultiNest} uses nested sampling to explore the prior space of the parameters by maximizing the log-likelihood in each iteration. Log-likelihood in this is defined as 

\begin{equation}
    \begin{aligned}
        \log\mathcal{L} &= \text{norm} - 0.5~\chi^2\\
        &= -0.5\sum\limits^N_{i=1}\ln(2\pi\epsilon^2_i)   -0.5\chi^2\\
        & = -0.5\sum\limits^N_{i=1}\ln(2\pi\epsilon^2_i) - 0.5\sum\limits^N_{i=1} \frac{(M_i - O_i)^2}{\epsilon^2_i}. 
    \end{aligned}
\end{equation}

Here, $N$ is the number of observational bins, $M_i$ is the model transit depth in a bin, $O_i$ is the observed transit depth in that bin, and $\epsilon_i$ is the transit depth error in the observation. During a retrieval, given a set of priors, \texttt{SANSAR} utilizes \texttt{PyMultiNest} to explore the posterior probability distribution of the requested parameters, which are constrained by the observations.

In the current version of \texttt{SANSAR} free retrieval, the free parameters can be radius (at a reference pressure) or pressure at the observed radius, chemical abundances of individual species, temperature profile (isothermal, Guillot, Madhusudhan), haze, and cloud parameters. The chemical abundances are assumed to be constant throughout each layer in the free retrieval for all the molecules. An offset between multiple instruments could also be a free parameter if needed to perform retrievals across multiple instruments, as shown in further sections. Currently, during retrieval, model spectra are binned into observation bins by simply taking the average of all the model points within an observation bin, since we did not find much difference between instrument-specific binning and simple averaging in each bin. We plan to implement instrument-specific binning in future updates of \texttt{SANSAR}.

\subsection{Grid Retrieval}
\label{sec:grid-theory}
In this work, we also interpret the observations of HD~209458~b using a self-consistent model atmosphere grid and the corresponding grid of simulated transmission spectra. We detail our implementation of grid-retrieval in this section. 

We generated a self-consistent model atmosphere grid for HD~209458~b, with Radiative-Convective-Thermochemical-Equilibrium (RCTE) $P$-$T$ profiles consistent with equilibrium chemistry using ATMO \citep{Tremblin2015, Drummond2016, Goyal2018, Goyal2020} for a range of parameters. This model atmosphere grid is similar to the one presented in \citet{Arora2024} for different sets of exoplanets. The model atmospheres are computed for 9 recirculation factors (0.2 -- 1.2), 20 metallicity values ($0.025\times$ -- $100\times$), 13 C/O ratios (0.1 -- 2), and 4 T$_{\rm int}$ values (100\,K -- 400\,K). We included H$_2$O, CO$_2$, CO, CH$_4$, NH$_3$, Na, K, Li, Rb, Cs, TiO, VO, FeH, CrH, PH$_3$, HCN, C$_2$H$_2$, H$_2$S, SO$_2$, H$^-$, Fe, H$_2$-H$_2$, and H$_2$-He CIA correlated-k opacities for computing these $P$-$T$ profiles, detailed in \citet{Goyal2020}. The self-consistent $P$-$T$ profiles from this model atmosphere grid, along with the corresponding equilibrium chemical abundances, are further used to generate a grid of transmission spectra using \texttt{SANSAR}. In addition to the model atmosphere grid parameters, we generate a transmission spectra grid with five cloud top pressure values (10$^{-5}$ - 10$^{-1}$ bar), six cloud fraction values (0.3 - 1), six Rayleigh-enhancement factor (1 - 10$^{5}$) and three scattering exponent values (-7, -4, -1). Thus, our self-consistent transmission spectra grid to perform grid-retrieval has more than two million spectra, obtained through systematic narrowing of the parameter space. We also note that we have used all the opacities used in the model atmosphere grid to generate the transmission spectra as well (except H$^-$ as its abundance is quite low) at a higher spectral resolution of R$\sim$3000 with the correlated-k method. This transmission spectra grid was then used to perform grid-retrieval. 

For grid-retrieval, the model transmission spectra grid generated with self-consistent model atmospheres of HD~209458~b is fitted to observations using Bayesian sampling. Here, the model spectra are sampled from the grid. This is beneficial as we do not need to run the self-consistent models again, which are computationally very expensive. For grid-retrieval, too, we use \texttt{PyMultiNest} as the Bayesian sampler. Since the self-consistent grids are generated for a discrete set of parameters, the spectra have to be interpolated for the parameter values within these discrete values. Therefore, following various recent exoplanet studies applying grid-retrieval \citep[for e.g.][]{Gagnebin2024}, we use the \texttt{RegularGridInterpolator} (RGI) for interpolation from \texttt{scipy.interpolate} \citep{2020SciPy-NMeth}. The free parameters used in our grid-retrieval setup for HD~209458~b are recirculation factor $(\text{rcf})$, $\log(Z)$, C/O, $T_{\text{int}}$, $\log(a)$, $\gamma$, $\log(P_{\text{cloud}})$, $\phi$, and two offsets $\delta_1$ and $\delta_2$, where $\delta_1$ is the offset used to move whole model spectra (as a reference radius proxy) and $\delta_2$ is used as an offset between HST and NIRCam observations.

RGI uses a rectilinear grid with models defined for each combination of grid points. There were some combinations of the parameters for which there were no converged RCTE $P$-$T$, and hence, the grid did not have those $P$-$T$ and abundance files. Most of these missing points were near the edges of each parameter range. We used the nearest neighbor method to generate files for these missing grid points, resulting in a complete set of $P$-$T$ and abundance profiles across the grid. We use these files to generate a regular grid consisting of models for each parameter combination and then use RGI for interpolation over the entire grid.
This interpolated spectrum for a set of sampled parameters is then binned to the same wavelength grid as the observations, and the log-likelihood is calculated by comparing the model spectrum with the observed spectrum. This is repeated for each combination of sampled parameter values from the prior space. The nested sampler (\texttt{PyMultiNest}) tries to increase the log-likelihood in each iteration until it converges, and we get the highest log-likelihood value to obtain robust constraints on the parameter values. Finally, after convergence, we generate the corner plot containing posterior probability distributions for the selected samples by the nested sampler. Equally weighted samples ($\sim$5000 samples) are used to get the interpolated models to generate best-fit models. The median model is computed using the $50^{\text{th}}$ percentile, and the confidence interval is shown using $84.1^{\text{th}}$ and $15.9^{\text{th}}$ percentile for $\pm1\sigma$, and $97.7^{\text{th}}$ and $2.3^{\text{rd}}$ percentile for the $\pm2\sigma$ bounds.
The RCTE $P$-$T$ profiles have different pressure grids for each model in the grid, as ATMO uses optical depth rather than pressure for fixing each atmospheric layer vertically. The temperature for each $P$-$T$ profile was interpolated to a common log-spaced pressure grid using linear interpolation. \texttt{RBFInterpolator} with the linear kernel is then used to interpolate the temperatures for the equally weighted selected samples, which are then used to plot the median $P$-$T$ along with their $\pm1\sigma$ bounds.

\section{Benchmarking}
\label{sec:benchmark}
In this section, we detail the benchmarking of the \texttt{SANSAR} forward model transmission spectra and retrieval with various published works. Since we have included two opacity treatment (opacity sampling and correlated-k) methods in \texttt{SANSAR}, we present benchmarking for each of them. We also show the computational performance benchmarking of \texttt{SANSAR} in this section. 

\subsection{Forward Model Benchmarking}
\subsubsection{Opacity Sampling Benchmarking}
The forward model transmission spectra computation with opacity sampling in \texttt{SANSAR} is benchmarked with POSEIDON \citep{Macdonald2017, MacDonald_12023}, an open-source retrieval code. For this comparison, we generated a cloud-haze free forward model of WASP-96 b with abundances from \citet{Taylor2023}, at a resolution of R $\sim$ 20,000. To first benchmark the transmission spectra computation, without any influence of opacities, we use the POSEIDON opacity database for comparison. The POSEIDON opacity files used for this test are H$_2$O \citep{Polyansky2018}, CO \citep{Li2015}, CO$_2$ \citep{Tashkun2011}, Na \citep{Ryabchikova2015} and K \citep{Ryabchikova2015}. The log abundances of H$_2$, He, H$_2$O, CO, CO$_2$, Na, and K used for the benchmark test are -0.072, -0.820, -3.586, -3.246, -4.376, -6.846, and -8.036, respectively. We note that we have turned off CIA opacities in both models for this comparison, as the format of CIA opacities used in POSEIDON is different from that in \texttt{SANSAR}. The atmosphere is defined from $10^2$ bar to $10^{-6}$ bar as the POSEIDON opacity database has a minimum pressure of $10^{-6}$ bar. The isothermal temperature is assumed at 1200 K. Other parameters like the radius of the planet (1.2 $R_{\text{J}}$), gravitational acceleration of the planet (8.263 ms$^{-2}$), and radius of WASP-96 (1.05 $R_{\odot}$) are same. Figure \ref{fig:poseidon-bench-main} in Appendix \ref{app: benchmarking} shows the comparison between both \texttt{SANSAR} and POSEIDON spectra. The maximum difference between both spectra is 32 ppm, with an absolute mean difference of 10 ppm, which is well below the observational errors. The differences are mainly due to different methods of interpolation, radial profile calculation, and spectra generation. 

\begin{figure}
    \includegraphics[width=8cm]{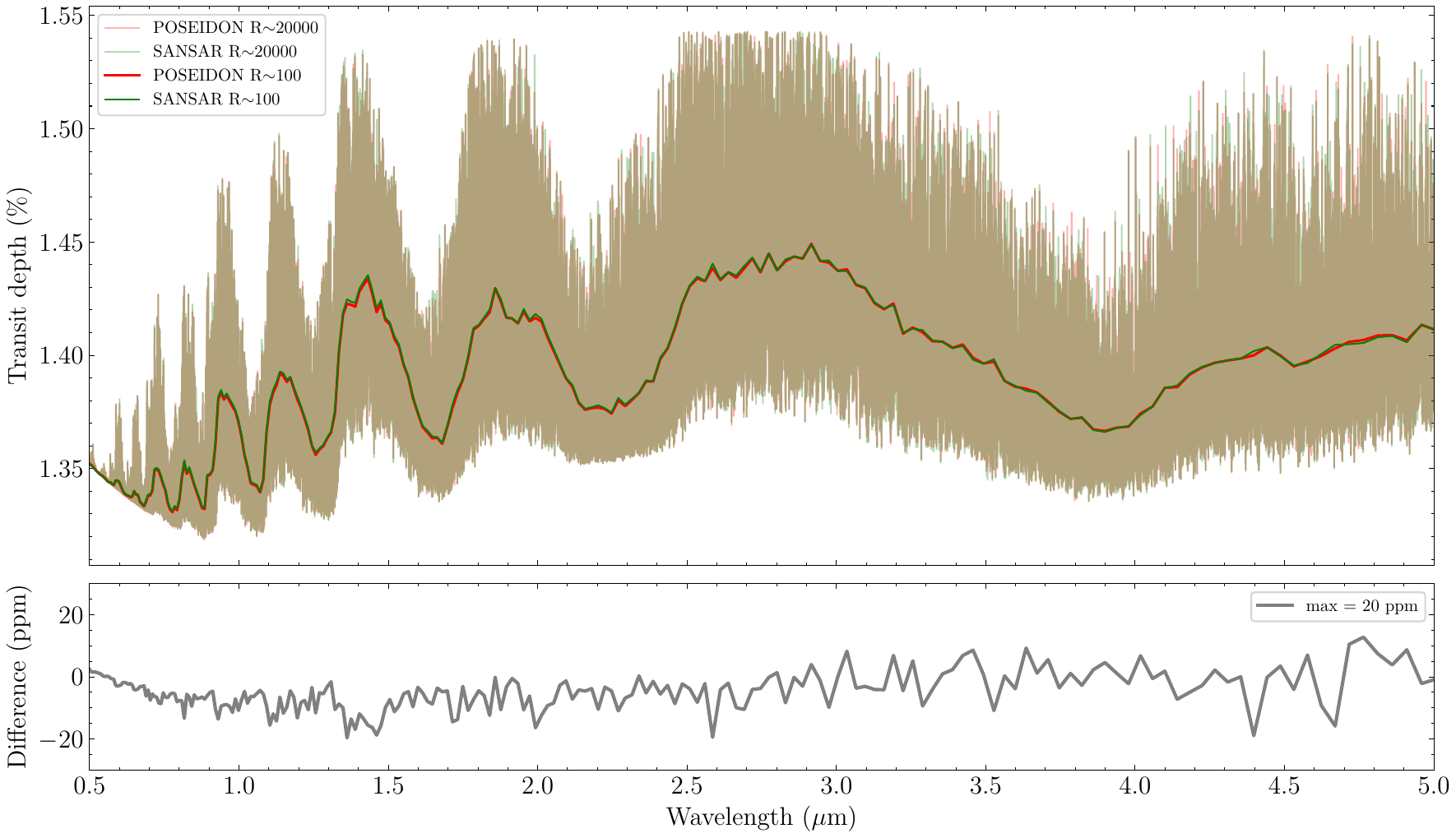}
    \caption{Comparison of \texttt{SANSAR} forward model with POSEIDON for a WASP-96 b type planet. Both \texttt{SANSAR} and POSEIDON use the POKAZATEL line list here. \texttt{SANSAR} samples lines from a 0.001 cm$^{-1}$ opacity file contrary to POSEIDON, which uses an opacity file with a resolution of 0.01 cm$^{-1}$. }
   
    \label{fig:poseidon-bench-2}
\end{figure}

We also performed another forward model comparison where we directly generated spectrum from both POSEIDON and \texttt{SANSAR}, using their respective opacity databases for a H$_2$ and He dominated atmosphere for WASP-96 b with only \ce{H2O} as trace gas, where log abundances of H2, He and \ce{H2O} are -0.07179, -0.81998, and -2.996, respectively. Comparison is shown in Figure \ref{fig:poseidon-bench-2}. Here, opacity files for both POSEIDON and \texttt{SANSAR} are generated using the POKAZATEL line list \citep{Polyansky2018}; the only difference is that POSEIDON LBL resolution is 0.01 cm$^{-1}$, whereas for LBL, the opacity file used in \texttt{SANSAR}, is 0.001 cm$^{-1}$. This comparison also assumes an isothermal atmosphere with a temperature of 1200 K, and other planetary parameters are the same as before. In this test, H$_2$-H$_2$ and H$_2$-He CIA are also used in both models. In this comparison, we get a maximum difference between POSEIDON and \texttt{SANSAR} to be 20 ppm with an absolute mean difference of 6 ppm, as seen in Figure \ref{fig:poseidon-bench-2}.

\subsubsection{Correlated-k Benchmarking}
\begin{figure}
    \includegraphics[width=8cm]{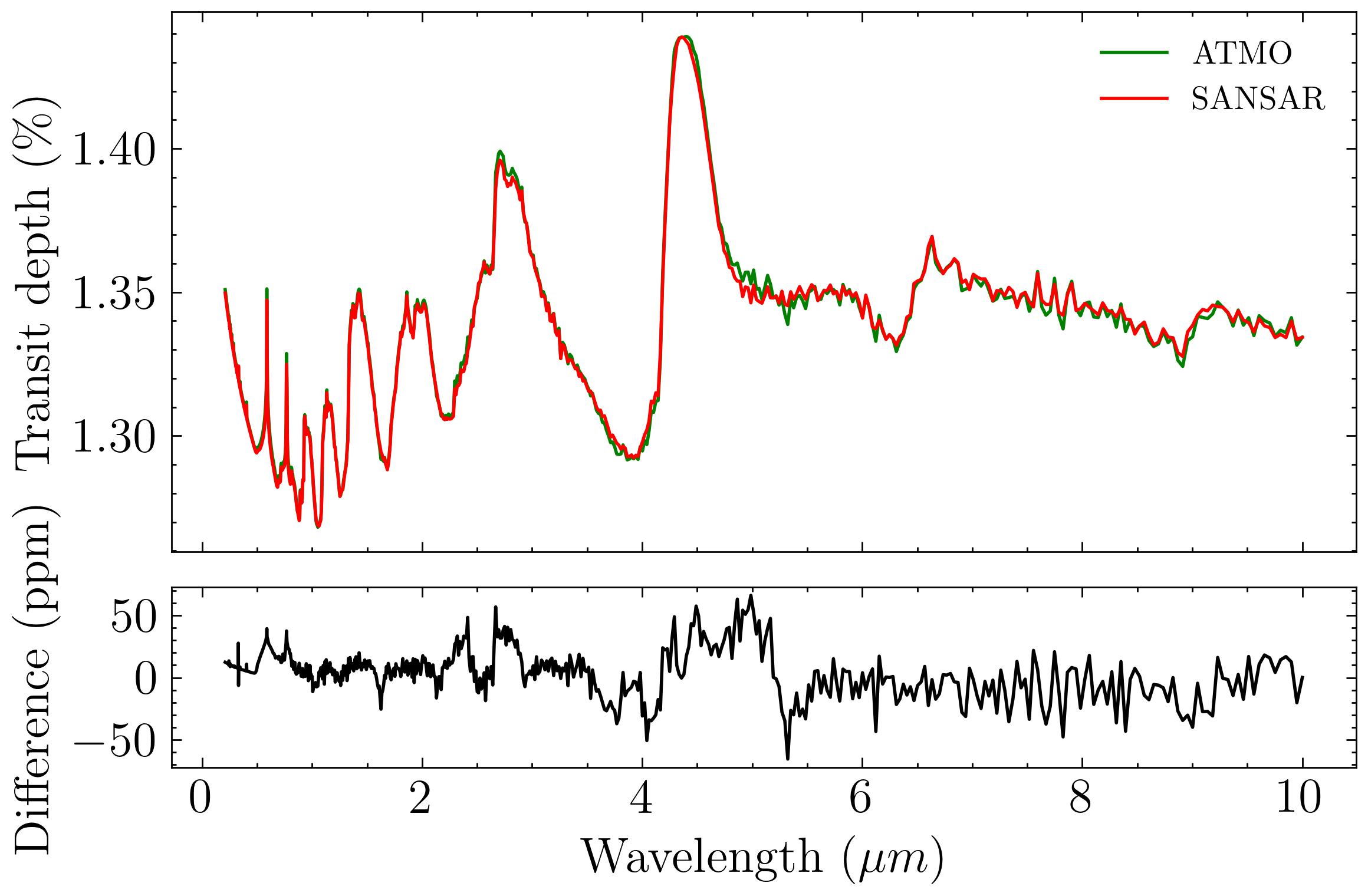}
    \caption{Comparison of forward models generated with \texttt{SANSAR} and ATMO for a WASP-96 b type planet. Both models were generated with the correlated-k method at a resolution of R$\sim$1000 and binned to R$\sim$200 for comparison.}
    \label{fig:corrk_bench}
\end{figure}

For benchmarking \texttt{SANSAR} forward model transmission spectra with the correlated-k method, we use ATMO \citep{Amundsen2014,Tremblin2015,Goyal2018} as it computes transmission spectra using the correlated-k method. For this comparison also, we generated transmission spectra for WASP-96 b with the same set of planetary parameters and opacity species as mentioned in the previous section using both ATMO and \texttt{SANSAR}, at a spectral resolution of R $\sim$ 1000. We have precomputed k-tables for all the molecules to use in \texttt{SANSAR} as detailed in Section \ref{sec:corrk}. The LBL database for \texttt{SANSAR} and ATMO is the same. However, the k-table computation methodology is different. Figure \ref{fig:corrk_bench} shows the comparison of the transmission spectra between \texttt{SANSAR} and ATMO. The maximum discrepancy observed here is approximately 50 ppm, primarily arising from differences in how \texttt{SANSAR} and \texttt{ATMO} handle the computation of k-coefficients for mixed atmospheric compositions. While \texttt{ATMO} employs the random overlap method with resorting and rebinning, \texttt{SANSAR} adopts random overlap without these additional steps. Additional sources of variation include differences in the underlying k-tables, atmospheric profiles, and spectral computation methodologies.

\subsection{Inverse Model Benchmarking}
\label{sub:inverse-benchmark}

\begin{figure*}
    \includegraphics[width=\textwidth]{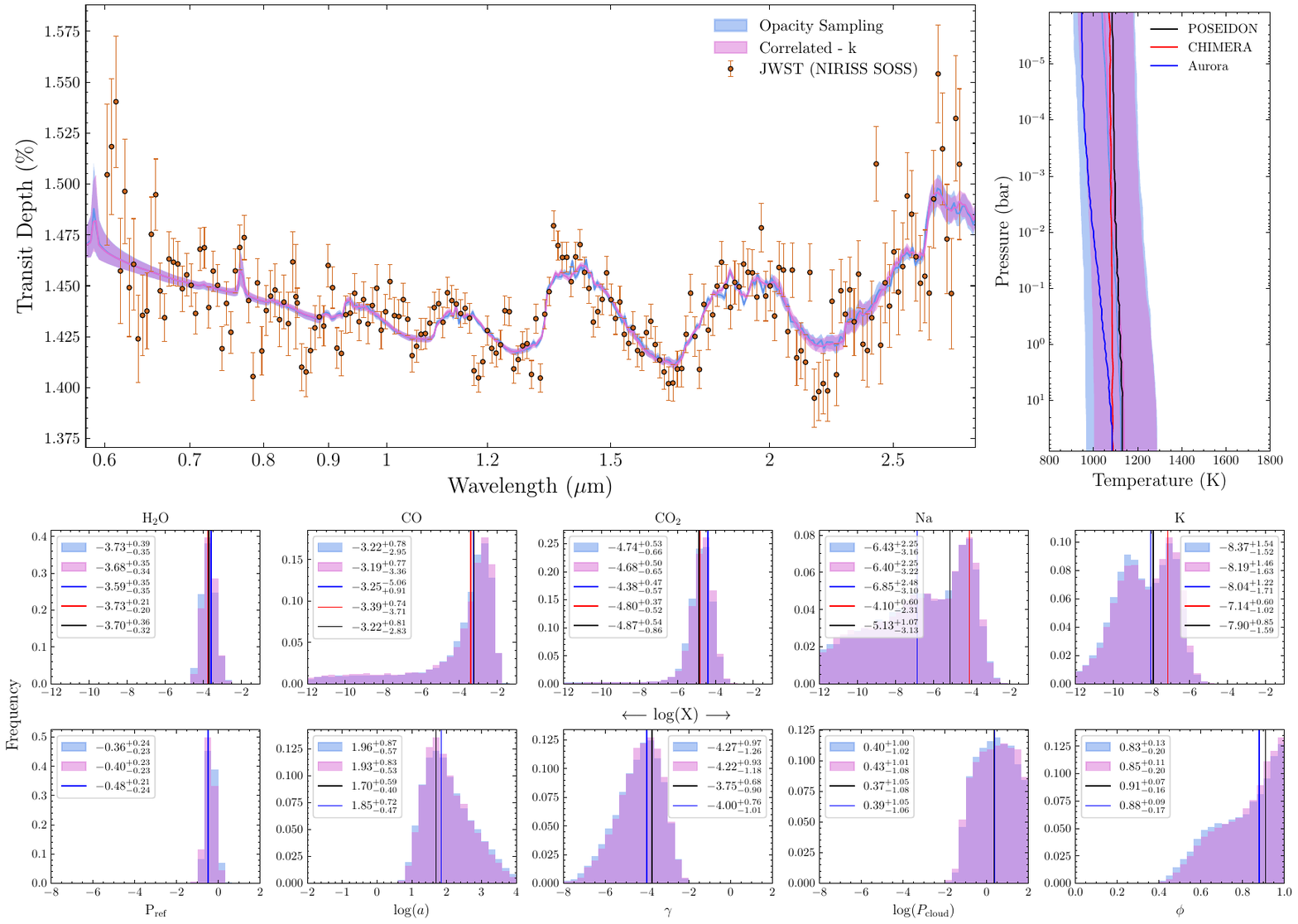}
    \caption{Figure showing benchmarking of \texttt{SANSAR} retrieval with other widely used retrieval models using WASP-96 b observations from \citet{Radica2023} and model results from \citet{Taylor2023}. Opacity sampling and correlated-k median transmission spectra are represented by light blue and violet colors, respectively, with $1\sigma$ uncertainty. The temperature profile is shown for both opacity sampling and correlated-k along with the $1\sigma$ error bar, whereas the median retrieved temperature profiles from POSEIDON (black), CHIMERA (red), and Aurora (Blue) are also plotted on top. The bottom panel shows the posteriors of different chemical species as well as cloud and haze parameters, along with the values retrieved by other models. The median values of the posteriors from POSEIDON, CHIMERA, and Aurora are shown as solid lines.}
    \label{fig:wasp96final}
\end{figure*}

As mentioned before, we benchmark the \texttt{SANSAR} retrieval using WASP-96 b JWST observations from \citet{Radica2023} and the multi-retrieval study of WASP-96 b from \citet{Taylor2023}. The WASP-96 b retrieval in \texttt{SANSAR} is set up assuming 84\% H$_2$ and 15\% He atmosphere. The atmosphere is in hydrostatic equilibrium and has 100 layers from $10^2$ bar to $10^{-8}$ bar with a logarithmically equidistant pressure difference between each layer. We also include clouds and haze in the retrievals using \texttt{SANSAR}. Similar to forward model benchmarking, we perform inverse model benchmarking for both opacity sampling as well as correlated-k method. We have used similar priors in opacity sampling as well as correlated-k inverse model benchmarking as given in Table \ref{tab:wasp96-b retr table}. The results of this inverse model benchmarking are summarized in Table \ref{tab:wasp96-b retr table} and Figure \ref{fig:wasp96final}.

\subsubsection{Opacity sampling inverse model benchmarking}
We benchmark the \texttt{SANSAR} inverse model with the opacity sampling method by comparing \texttt{SANSAR} retrieval results with POSEIDON and Aurora retrieval studies for WASP-96 b from \citet{Taylor2023}. For this comparison, we have assumed constant (with pressure) chemical abundances (\ce{H2O, CO, CO2, Na, K}) and Madhusudhan $P$-$T$ profile as used in \citet{Taylor2023}.
In the \texttt{SANSAR} retrieval setup, we have considered a model wavelength grid from 0.2 to 3 $\mu$m at a resolution of R $\sim$ 20000 for opacity sampling, as used for POSEIDON and Aurora. The results of this benchmarking study are shown in Table \ref{tab:wasp96-b retr table}. The abundances of \ce{H2O}, \ce{CO}, and \ce{CO2} obtained by \texttt{SANSAR} are in excellent agreement with those derived by POSEIDON and Aurora. The Na absorption feature at 0.589 $\mu$m falls just outside the observed wavelength range, resulting in a poor constraint but in agreement with others. For K, the abundances retrieved using \texttt{SANSAR} are consistent within $1\sigma$ of those derived by other retrieval models. However, the Na and K line lists and absorption cross-sections are quite different in the three models, which can lead to discrepancies.  A similar trend is observed for the cloud, haze, and reference pressure parameters as shown in Table \ref{tab:wasp96-b retr table}, where the values retrieved by \texttt{SANSAR} are well within the $1\sigma$ range of those obtained by other retrieval models.

\subsubsection{Correlated-k Inverse Model Benchmarking}
We benchmark the \texttt{SANSAR} inverse model with the correlated-k method by comparing \texttt{SANSAR} retrieval results with the CHIMERA \citep{Line2013} retrieval study for WASP-96 b from \citet{Taylor2023}, as CHIMERA also used the correlated-k method. Here again, we have used all the chemical species described in the previous section as free parameters with constant abundances along the pressure. For this retrieval, we have used k-tables generated using \texttt{Exo\_k} as described in Section \ref{sec:corrk} with a resolution of R $\sim$ 1000. The results of this benchmarking study are shown in Table \ref{tab:wasp96-b retr table}. The retrieved abundances of all chemical species from \texttt{SANSAR} agree within 1$\sigma$ uncertainty of that of CHIMERA, thus benchmarking the \texttt{SANSAR} inverse model with the correlated-k method. 

As demonstrated in Section \ref{sec:corrk}, a correlated-k forward model generated at low resolution provides accuracy comparable to a forward model with opacity sampling at much higher spectral resolution. Therefore, we anticipate that retrievals performed using correlated-k at R$\sim$1000 should yield similar results to those obtained from opacity sampling retrievals at R$\sim$20000 for WASP-96 b with JWST NIRISS SOSS, assuming all other conditions are the same. This is indeed confirmed by our retrievals, where both methods produce very similar results, as seen in Table \ref{tab:wasp96-b retr table}. This agreement also serves as a benchmark of our retrieval approach with correlated-k.

\begin{deluxetable*}{lcccccc}
    \label{tab:wasp96-b retr table}

    \renewcommand{\arraystretch}{1.5}
    \tabletypesize{\footnotesize}
    \tablecolumns{7} 
    \tablecaption{Table showing inverse model (Retrieval) benchmarking of \texttt{SANSAR} with Aurora, CHIMERA and POSEIDON results obtained from \citet{Taylor2023}, for WASP-96 b. Median value along with their 1$\sigma$ uncertainty is shown for each parameter.}
    \tablehead{Parameters & Prior & Aurora &  POSEIDON & \texttt{SANSAR} & CHIMERA & \texttt{SANSAR} \\
    & & (Opacity sampling)& (Opacity sampling)&(Opacity sampling)&(Correlated-k)& (Correlated-k)
    }
    \startdata
    log(\ce{H2O}) & $\mathcal{U}$(-12,-1) &$-3.59_{-0.35}^{+0.35}$ &  $-3.70^{+0.36}_{-0.32}$ & $-3.73^{+0.39}_{-0.35}$ &$-3.73^{+0.21}_{-0.20}$ & $-3.70^{+0.35}_{-0.34}$\\
    log(\ce{CO})  & $\mathcal{U}$(-12,-1)&$-3.25_{+0.91}^{-5.06}$& $-3.22^{+0.81}_{-2.83}$ & $-3.22^{+0.78}_{-2.95}$ &$-3.39^{+0.74}_{-3.71}$  &  $-3.20^{+0.77}_{-3.34}$ \\
    log(\ce{CO2}) & $\mathcal{U}$(-12,-1)&  $-4.38^{+0.47}_{-0.57}$ & $-4.87^{+0.54}_{-0.86}$ &  $-4.74^{+0.53}_{-0.66}$ & $-4.80^{+0.37}_{-0.52}$  &$-4.70^{+0.50}_{-0.63}$ \\
    log(\ce{Na}) & $\mathcal{U}$(-12,-1)&$-6.85^{+2.48}_{-3.10}$&  $-5.13^{+1.07}_{-3.13}$ &  $-6.43^{+2.25}_{-3.16}$ &$-4.10^{+0.60}_{-2.31}$ & $-6.42^{+2.25}_{-3.22}$ \\
    log(\ce{K}) & $\mathcal{U}$(-12,-1)& $-8.04^{+1.22}_{-1.71}$ &  $-7.90^{+0.85}_{-1.59}$ & $-8.37^{+1.54}_{-1.52}$ &$-7.14^{+0.60}_{-1.02}$  & $-8.21^{+1.46}_{-1.63}$ \\
    $\log(a)$ & $\mathcal{U}$(-2,4)& $1.85^{+0.72}_{-0.47}$  &  $1.7^{+0.59}_{-0.40}$  & $1.96^{+0.87}_{-0.57}$ & $\cdots$ &$1.93^{+0.83}_{-0.53}$ \\
    $\gamma$ & $\mathcal{U}$(-8 , 2)& $-4.00^{+0.76}_{-1.01}$ & $-3.75^{+0.68}_{-0.90}$ &  $-4.27^{+0.97}_{-1.26}$ & $\cdots$ & $-4.22^{+0.93}_{-1.18}$\\
    $\log(P_{\text{cloud}})$ &$\mathcal{U}$(-8, 2)&  $0.39^{+1.05}_{-1.06}$  & $0.37^{+1.05}_{-1.08}$ & $0.40^{+1.00}_{-1.02}$ &$\cdots$ &$0.43^{+1.01}_{-1.08}$ \\
    $\phi$ &$\mathcal{U}$(0, 1)& $0.88^{+0.09}_{-0.17}$ & $0.91^{+0.07}_{-0.16}$ & $0.83^{+0.13}_{-0.20}$&$\cdots$ & $0.85^{+0.11}_{-0.20}$ \\
    $\text{P}_{\text{ref}}$ & $\mathcal{U}$(-8, 2)&$-0.48^{+0.21}_{-0.24}$& $\cdots$ &$-0.36^{+0.24}_{0.23}$ &$\cdots$& $-0.40^{+0.23}_{-0.23}$
    \enddata
\end{deluxetable*}

\subsection{Model Computation Time}\label{sec:perf bench}
Exoplanet atmosphere retrieval studies require millions of forward model runs to obtain robust constraints, and this has to be done for multiple instrument combinations, different prescriptions of physical processes, etc., making these retrieval studies computationally very intensive. Therefore, in this section, we report the Python-based {\texttt{SANSAR} model computation time for different commonly used retrieval setups. Computing \texttt{SANSAR} model transmission spectra with opacity sampling 
requires the loading of LBL opacities, and these opacities are generated at a resolution of $\delta \nu$ = 0.001 cm$^{-1}$ with 800 $P$-$T$ pairs for \texttt{SANSAR}. Therefore, the size of these files for each molecule is very large; for example, the H$_2$O cross-section file generated with the BT2 line list is 105 GB. Hence, it is currently not possible to run \texttt{SANSAR} with the opacity sampling method directly on a typical desktop due to RAM limitations.  
However, \texttt{SANSAR} with correlated-k can be run directly on a typical desktop with 16 GB RAM at R $\sim$ 1000 with ease. We report \texttt{SANSAR} performance on a Windows machine (inside Windows Subsystem for Linux) with Intel Core i7 10700 CPU and 64 GB RAM in Table \ref{tab:perf-bench} of Appendix \ref{app: benchmarking}. This table presents two types of tests. The first, referred to as otf (on-the-fly), involves interpolating cross sections based on the atmospheric $P$-$T$ profile at each iteration during the retrieval process. The second, labeled grid, corresponds to selecting the nearest pre-interpolated cross-sections from a densely sampled grid based on the atmospheric $P$-$T$. Comparison of accuracy for otf and grid based interpolated forward model transmission spectra is shown in Figure \ref{fig:otfvsgrid} in Appendix \ref{app: benchmarking}. All simulations are conducted at a spectral resolution of R$\sim$20000 for opacity sampling (corresponding to a total of 100213 wavelength points) and R$\sim$1000 for the correlated-k method (corresponding to 5011 wavelength points). In the fast grid mode, \texttt{SANSAR} achieves computation time of $\sim$1 second for opacity sampling and 610 milliseconds for correlated-k. Running \texttt{SANSAR} for a typical hot-Jupiter requires $\sim$4 GB of RAM in correlated-k (at R$\sim$1000) mode with otf, while running in grid mode requires $\sim$30 GB of memory. Thus, while otf mode is slower than grid, it requires very less RAM compared to grid. We note that running retrieval models with opacity sampling at R $<$ 20000 is not recommended to interpret JWST observations, as it is highly inaccurate.  For opacity sampling due to the limitations mentioned above, we first load the LBL cross section on the server and then convert it to a binary file with R$\sim$20000 (thereby reducing size by an order of magnitude) and transfer it to a desktop. The benchmark shown in Table \ref{tab:perf-bench} only represents the time for spectra calculation, and it does not include any opacity initialization time and parameter setup, as these steps are done only once during initialization.  These models were generated for an \ce{H2}, and \ce{He} dominated planet with opacities from \ce{H2}-He CIA, \ce{H2}-\ce{H2} CIA, \ce{H2O}, \ce{CO}, \ce{CO2}, Na, and K considered, and the temperature was chosen to be isothermal. Other parameters are the same as WASP-96 b.

\section{Results}
\label{sec:results}

As described earlier, \citet{Xue2024} presented the JWST NIRCam transmission spectra observations of HD~209458~b from 2.3 to 5.1 $\mu$m. They performed retrievals with equilibrium chemistry to place constraints on the C/O ratio and metallicity. They also performed equilibrium chemistry retrievals on JWST NIRCam plus HST WFC3 \citep{demingwfc3} transmission spectra observations. In this work, we show a more detailed investigation of HD~209458~b transmission spectra observations with the \texttt{SANSAR} model. We have divided this section into three subsections. First, we present the results that we obtain with free retrieval analysis, followed by equilibrium chemistry retrieval, and finally using grid-retrieval with self-consistent radiative-convective equilibrium $P$-$T$ profiles detailed in \citet{Goyal2020}. We also perform all these retrieval analyses on three instrument combinations: JWST NIRCam, JWST NIRCam plus HST WFC3, and JWST NIRCam plus HST WFC3 plus HST STIS. The primary HST WFC3 and HST STIS observations for all these simulations are adopted from \citet{Sing2016}. However, in the equilibrium chemistry retrieval analysis where we compare our results with \citet{Xue2024}, we also utilize HST WFC3 observations of HD~209458~b from \citet{demingwfc3}, since they used these observations in their combined retrieval analysis. We note that \citet{Xue2024} did not include HST STIS data in their retrieval analysis.

For all our retrieval analysis, we have used the planetary radius of HD~209458~b to be 1.39 R$_\text{J}$ whereas stellar radius and effective temperature are taken to be 1.19 R$_{\odot}$ and 6091\,K respectively \citep{Stassun_2017}. The gravitational acceleration is taken as 9.365 ms$^{-2}$. The distance of HD~209458~b from its host star is 0.047 AU \citep{Guillermo2008}, and the internal temperature of the planet is taken to be 100\, K.

\begin{figure*}
    
    \includegraphics[width=\textwidth]{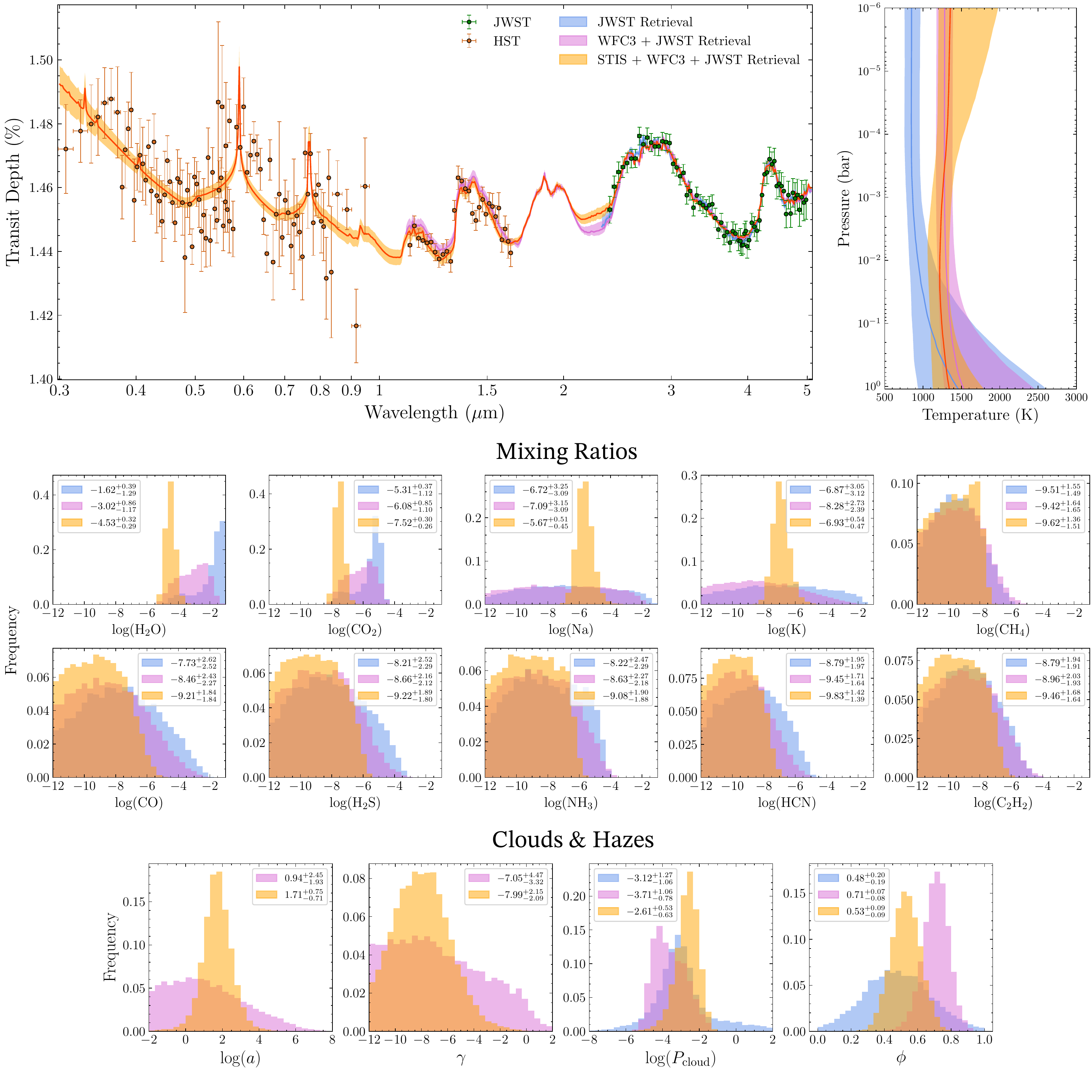}
    \caption{Retrieved transmission spectra, abundance constraints, and $P$-$T$ profile with free chemistry retrievals. The top panel consists of median transmission spectra and $P$-$T$ profiles of HD 209458~b for the three retrievals, JWST (NIRCam), HST WFC3 + JWST NIRCam, and HST STIS + HST WFC3 + JWST NIRCam. All the spectra shown here are generated with 5000 random equal-weighted samples from \texttt{PyMultiNest} and binned at a resolution of R$\sim$200. The shaded region indicates the 1-$\sigma$ uncertainty in the median fit. JWST observations are represented in green, while combined STIS and WFC3 observations are shown in brown. A retrieved median offset of 204 ppm is applied to the HST observations, whereas the offset for WFC3 in the case of the WFC3+NIRCam combination is not shown here. For all the retrievals, the corresponding $P$-$T$ profiles are shown beside the median spectra. The middle panel shows the posteriors of the logarithm of mixing ratios of the ten molecules for three retrievals. The bottom panel shows the posteriors of the clouds and hazes parameters.}
    \label{fig:full_retrieval}
\end{figure*}

\begin{deluxetable*}{lcccc}
    \renewcommand{\arraystretch}{1.8}
    \tabletypesize{\footnotesize}
    \tablecolumns{5} 
    \tablecaption{Table showing the results of free chemistry retrievals for HD~209458~b with different observational configurations: JWST NIRCam, HST WFC3 + JWST NIRCam, and HST STIS + HST WFC3 + JWST NIRCam, as discussed in Section \ref{subsec:free}. Median value along with its 1$\sigma$ uncertainty is shown for each parameter.}
    \tablehead{Parameters & Priors & NIRCam & WFC3 + NIRCam & STIS + WFC3 + NIRCam}
    \startdata
    \hline
     log(\ce{H2O}) & $\mathcal{U}$(-12, -1)& $-1.62^{+0.39}_{-1.29}$ & $-3.02^{+0.86}_{-1.17}$ & $-4.53^{+0.32}_{-0.29}$ \\ 
        log(\ce{CO2}) & $\mathcal{U}$(-12, -1)& $-5.31^{+0.37}_{-1.12}$ & $-6.08^{+0.85}_{-1.10}$  & $-7.52^{+0.30}_{-0.26}$ \\ 
        log(\ce{Na}) & $\mathcal{U}$(-12, -1)& $-6.72^{+3.25}_{-3.09}$  & $-7.09^{+3.15}_{-3.09}$ & $-5.67^{+0.51}_{-0.45}$  \\ 
        log(\ce{K}) & $\mathcal{U}$(-12, -1)& $-6.87^{+3.05}_{-3.12}$  & $-8.28^{+2.43}_{-2.27}$ & $-6.93^{+0.54}_{-0.47}$ \\ 
        log(\ce{CH4}) & $\mathcal{U}$(-12, -1)& $-9.51^{+1.55}_{-1.49}$  & $-9.42^{+1.64}_{-1.65}$ & $-9.62^{+1.36}_{-1.51}$  \\ 
        log(\ce{CO}) & $\mathcal{U}$(-12, -1)& $-7.73^{+2.62}_{-2.52}$ & $-8.46^{+2.43}_{-2.27}$  & $-9.21^{+1.84}_{-1.84}$  \\ 
        log(\ce{H2S}) & $\mathcal{U}$(-12, -1)& $-8.21^{+2.52}_{-2.29}$ & $-8.66^{+2.16}_{-2.12}$  & $-9.22^{+1.89}_{-1.80}$  \\ 
        log(\ce{NH3}) & $\mathcal{U}$(-12, -1)& $-8.22^{+2.47}_{-2.29}$ & $-8.63^{+2.27}_{-2.18}$  & $-9.08^{+1.90}_{-1.88}$ \\ 
        log(\ce{HCN}) & $\mathcal{U}$(-12, -1)& $-8.77^{+2.01}_{-1.89}$& $-9.45^{+1.71}_{-1.64}$  & $-9.83^{+1.42}_{-1.39}$ \\ 
        log(\ce{C2H2}) & $\mathcal{U}$(-12, -1)& $-8.79^{+1.94}_{-1.91}$ & $-8.96^{+2.03}_{-1.93}$  & $-9.46^{+1.68}_{-1.64}$ \\
        R$_{\text{p},\text{ref}}$ (R$_\text{J}$) & $\mathcal{U}$(1, 1.5) &$1.387^{+0.004}_{-0.005}$ & $1.375^{+0.015}_{-0.011}$  & $1.395^{+0.004}_{-0.005}$ \\ 
        $\kappa_{\text{IR}}$ & $\mathcal{U}$(-5, 5)& $-2.03^{+1.53}_{-1.73}$ & $-3.05^{+1.39}_{-1.26}$  & $-2.53^{+1.83}_{-1.62}$ \\ 
        $\gamma_1$ & $\mathcal{U}$(-5, 3)& $-2.31^{+1.74}_{-1.63}$  & $-2.27^{+1.76}_{-1.76}$ & $-0.65^{+1.92}_{-2.80}$ \\ 
        $\gamma_2$ & $\mathcal{U}$(-5, 3)& $-2.33^{+1.76}_{-1.64}$  & $-2.30^{+1.73}_{-1.73}$ & $-0.82^{+2.00}_{-2.71}$ \\ 
        $\beta$ & $\mathcal{U}$(0, 2)& $0.66^{+0.09}_{-0.08}$ & $1.00^{+0.09}_{-0.11}$  & $0.90^{+0.13}_{-0.15}$\\ 
        $\alpha$ & $\mathcal{U}$(0, 1)& $0.50^{+0.31}_{-0.31}$  & $0.50^{+0.33}_{-0.33}$ & $0.50^{+0.33}_{-0.33}$ \\ 
        $\log(a)$ & $\mathcal{U}$(-2, 8)&$\cdots$ & $0.94^{+2.45}_{-1.93}$  & $1.71^{+0.75}_{-0.71}$  \\ 
        $\gamma$ & $\mathcal{U}$(-12, 2)& $\cdots$  & $-7.05^{+4.47}_{-3.32}$  & $-7.99^{+2.15}_{-2.09}$\\
        $\log(P_{\text{cloud}})$ & $\mathcal{U}$(-8, 2)&$-3.12^{+1.27}_{-1.06}$  & $-3.71^{+1.06}_{-0.78}$  & $-2.61^{+0.53}_{-0.63}$ \\ 
        $\phi$ & $\mathcal{U}$(0, 1)& $0.48^{+0.20}_{-0.19}$  & $0.71^{+0.07}_{-0.08}$ & $0.53^{+0.09}_{-0.09}$\\ 
        $\delta$ & $\mathcal{U}$(-500, 500) &$\cdots$ & $181^{+18}_{-17}$  & $204^{+16}_{-15}$ \\
        $\chi^2_{\text{red}}$  & $\cdots$ & $0.57$ & $0.97$ & $1.21$ 
    \enddata
    \label{tab:free-retrieval}
        \tablecomments{R$_{\text{J}}$ = 7.1492 $\times 10^7$ m, $\delta$ is in ppm, $\log(P_{\text{cloud}})$ is in bar}
\end{deluxetable*}

\begin{figure*}
    \includegraphics[width=\textwidth]{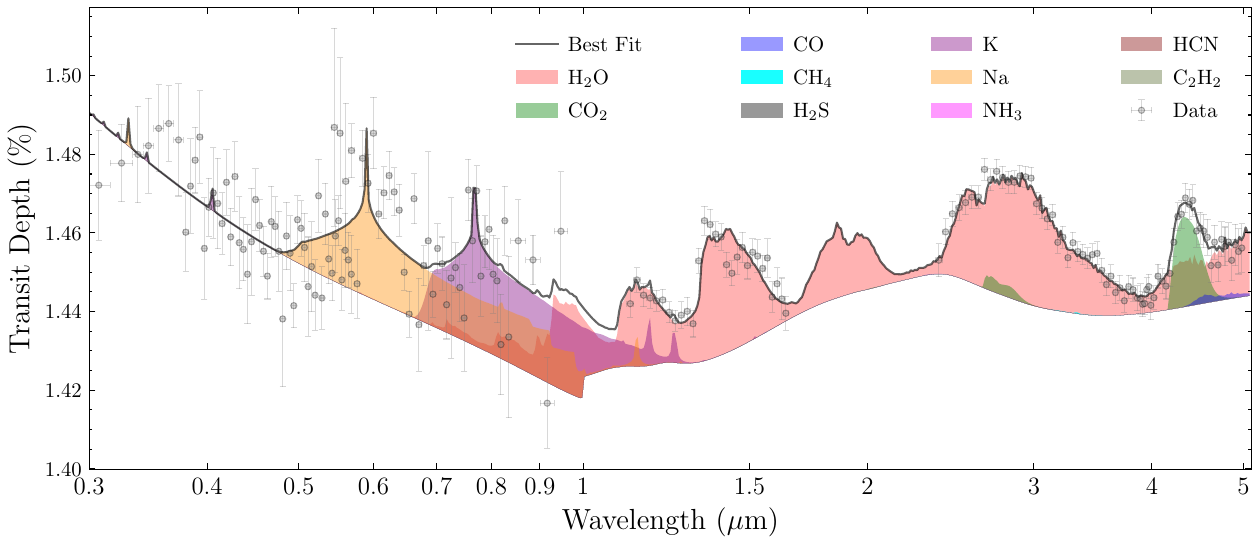}
    \caption{Figure showing the contribution of different gases to the best-fit spectrum obtained from the free chemistry retrieval binned down to R$\sim$200, for STIS+WFC3+NIRCam observations.}
    
    \label{fig:contribution}
\end{figure*}

\begin{deluxetable*}{lccccccc}
    \label{tab:stats}
    \renewcommand{\arraystretch}{1.5}
    \tabletypesize{\footnotesize}
    \tablecolumns{4} 
    \tablecaption{Retrieval statistics for STIS+WFC3+NIRCam observations. This table shows all major test statistics of retrievals with different model assumptions. }
    \tablehead{Retrieval Type & Model &  Bayesian Evidence & Bayes Factor & $\chi^2_{\text{red}}$ & $\chi^2$ & DOF & Detection Significance }
    \startdata
    Free chemistry&  All species & 1560.60 & Reference & 1.21 & 205.57& 170& Reference \\
    Free chemistry&  No \ce{H2O} & 1483.69 & $2.51\times10^{33}$ & 2.06 & 352.95& 171& 12.6  $\sigma$ \\
    Free chemistry&  No \ce{CO2} & 1532.60 & $1.44\times10^{12}$ & 1.56 & 266.54& 171& 7.8 $\sigma$ \\
    Free chemistry&  No \ce{Na} & 1539.02 & $2.35\times10^{9}$ & 1.47 & 250.76& 171& 6.9 $\sigma$ \\
    Free chemistry&  No \ce{K} & 1550.60 & $2.21\times10^{4}$ & 1.33 & 227.11& 171& 4.9 $\sigma$ \\
    Free chemistry&  No \ce{NH3} & 1560.86 & $\cdots$ & 1.20 & 206.00& 171& $\cdots$ \\
    Free chemistry&  No \ce{HCN} & 1561.39 & $\cdots$ & 1.21 & 206.20& 171& $\cdots$ \\
    Free chemistry&  Homogeneous cloud & 1557.87 & $1.53\times10^{1}$ & 1.24 & 211.90& 171& 2.8 $\sigma$ \\
    Equilibrium chemistry &  Change C & 1556.74 & $\cdots$ & 1.26 & 225.20 & 178 & $\cdots$ \\
    Equilibrium chemistry &  Change O & 1556.56 & $\cdots$ & 1.27 & 225.94 & 178 & $\cdots$ \\
    Self-consistent chemistry &  Change O & 1561.06 & $\cdots$ & 1.30 & 236.14 & 181 & $\cdots$ 
    \enddata
\end{deluxetable*}

\subsection{Free Chemistry Retrieval}
\label{subsec:free}

In this section, we present the results of the free-chemistry retrieval analysis of HD 209458~b using the \texttt{SANSAR} model. For all the free-chemistry retrieval analyses, the atmosphere ranges from a pressure of 100 bar at the base to \(10^{-8}\) bar at the upper boundary. Our free retrieval setup assumes a \ce{H2}-\ce{He} dominated atmosphere, comprising $\sim$84\% hydrogen and $\sim$15\% helium by number density. The abundances (mixing ratios) of other trace gases are treated as free parameters, with the selection of gases included in the retrieval analysis guided by findings from previous studies. We included \ce{NH3} and \ce{HCN} based on their detection claims by \citet{Macdonald2017}, alongside \ce{H2O}, \ce{CO}, \ce{C2H2}, and \ce{CH4}, as reported by \citet{Giacobbe2021}. Additionally, \ce{H2S} and \ce{CO2} were incorporated following \citet{Xue2024}. Furthermore, we included \ce{Na} and \ce{K} following \citet{Sing2016}. Our model considers a fractional gray cloud coverage on the terminator region along with haze, as described in Section \ref{sec:cloud haze} (except for NIRCam only analysis, where we do not consider haze). For each of the three instrument combinations, we performed two separate retrievals with different parametrized pressure-temperature profiles: one assuming the Guillot temperature structure and the other considering the Madhusudhan temperature structure as described in Section \ref{sec:atm pt}. All the free retrievals are performed using the correlated-k method at R $\sim$1000, using 2000 live points and an evidence tolerance of 0.5. The retrieved radius is defined at the reference pressure of 10 mbar for all the free retrieval analyses. All the results and priors from these analyses are shown in Table \ref{tab:free-retrieval} and Figure \ref{fig:full_retrieval}. We note that we only show results with the Guillot $P$-$T$ profile in this table because we do not find any significant difference in our results between the two $P$-$T$ parameterizations. We also note that our results with STIS+WFC3+NIRCam observations are the main results. 
 
\subsubsection{JWST Only} \label{subsub:free_jwst}

For this retrieval, JWST NIRCam instrument observations of HD~209458~b from \citet{Xue2024} reduced using \texttt{Eureka!} reduction pipeline \citep{Taylor2022} were used. These observations were taken from 2.3-5.1 $\mu$m with F444W and F322W2 filters. Considering ten chemical species described earlier, two cloud parameters ($\log(P_{\text{cloud}})$, $\phi$), five parameters for the Guillot $P$-$T$ profile, and one reference radius, the retrieval involves a total of 18 parameters. The median retrieved values for all the parameters using \texttt{SANSAR} for HD~209458~b are shown in Table \ref{tab:free-retrieval} and Figure \ref{fig:full_retrieval}. We detect H$_2$O and CO$_2$ robustly similar to \citet{Xue2024} and do not find robust evidence of any other species. With only JWST observations we constrain the H$_2$O and CO$_2$ abundances to be $-1.62^{+0.39}_{-1.29}$ and $-5.31^{+0.37}_{-1.12}$, respectively. It can also be noted from Figure \ref{fig:full_retrieval} that the retrieved median $P$-$T$ profile is almost isothermal with the temperature of $\sim$900\,K in the region probed with transmission spectra, which is quite less than what is retrieved with other instrument combinations, as discussed in further sections. Our retrieved cloud parameters are $-3.12^{+1.27}_{-1.06}$ and $0.48^{+0.20}_{-0.19}$ for $\log(P_{\text{cloud}})$ and $\phi$, respectively.

\subsubsection{HST WFC3 + JWST}
\label{subsub:free_jwst_hst}
In this section, we show the results obtained by performing the retrieval analysis on combined HST WFC3+NIRCam transmission spectra observations.  This provides a combined wavelength coverage from $1.1-5.1$ 
 $\mu$m. HST WFC3 observations are obtained from \citet{Sing2016}. This retrieval analysis includes all the same parameters as used for the JWST analysis discussed in the previous section, along with two haze parameters and an offset between the WFC3 and JWST observations. Thus, the total number of retrieved parameters for this analysis is 21. The results from this retrieval are summarized in Table \ref{tab:free-retrieval} and Figure \ref{fig:full_retrieval}. With WFC3+NIRCam observations, we constrain the H$_2$O abundance to be $-3.02^{+0.86}_{-1.17}$, almost two orders of magnitude less than what we constrain with JWST NIRCam observations. The CO$_2$ abundance is constrained to $-6.08^{+0.85}_{-1.10}$, almost an order of magnitude less than what we constrain with JWST NIRCam observations. The $\log(P_{\text{cloud}})$ and $\phi$ are constrained to  $-3.71^{+1.06}_{-0.78}$ and $0.71^{+0.07}_{-0.08}$, respectively, which is much higher and well-constrained than the cloud parameters constrained with NIRCam observations. The best-fit spectra in this retrieval have a $\chi^2_{\text{red}}$ value of 0.97. It can also be noted from Figure \ref{fig:full_retrieval} that the retrieved median $P$-$T$ profile is almost isothermal with the temperature of $\sim$1300\,K in the region probed with transmission spectra which is $\sim$ 400\,K higher than the isothermal temperature (at 1 mbar) retrieved with just NIRCam observations. 

The median spectra in retrieved with WFC3+NIRCam and just NIRCam observations both match the spectral features very well in the NIRCam wavelength region. However, the median spectra with just NIRCam observations lead to larger \ce{H2O} spectral feature in the HST WFC3 bandpass (at 1.4 $\mu$m) compared to WFC3 observations. The high abundance of \ce{H2O} retrieved with just NIRCam observations is unable to fit the 1.4 $\mu$m \ce{H2O} feature in WFC3+NIRCam observations. Therefore, to reduce the \ce{H2O} spectral feature size in the WFC3 bandpass, the abundance of \ce{H2O} is reduced (compared to just NIRCam observations), as can be noticed from the \ce{H2O} abundance constraint with WFC3+NIRCam observations. However, decreasing the \ce{H2O} abundance will decrease the \ce{H2O} spectral feature in the NIRCam bandpass as well. Therefore, in the WFC3+NIRCam retrieval, the temperature is ramped up as can be noticed from \ref{fig:full_retrieval} where the $P$-$T$ profile with WFC3+NIRCam observations is $\sim$ 400\,K hotter than the $P$-$T$ profile with just NIRCam observations. This hotter $P$-$T$ profile will also lead to an increase in the \ce{CO2} spectral feature size in the NIRCam bandpass. Therefore, even \ce{CO2} abundance is reduced in the WFC3+NIRCam retrievals compared to retrievals with just NIRCam observations. The haze and cloud parameters further alter the model spectra to fit observations, which is detailed further in the discussion section.  

We also find that the constraints on the abundance of \ce{H2O} do not improve substantially (large uncertainty) in retrievals with WFC3+NIRCam observations compared to just JWST NIRCam, even though the observation wavelength range is now increased. We attribute this to the large uncertainty in the HST WFC3 observations themselves, as \ce{H2O} abundance is mainly dictated by the WFC3 bandpass. It can be noticed from Table \ref{tab:free-retrieval} that the constraints on haze and cloud parameters are highly uncertain, governed by a lack of optical observations. The large uncertainty in H$_2$O abundance leads to high uncertainty in \ce{CO2} abundance as well, as they are highly correlated. 
The comparatively high cloud fraction that we obtain with WFC3+NIRCam observations compensates for the low radius that we constrain with WFC3+NIRCam observations. This also highlights the degeneracy between radius, \ce{H2O} abundance, and cloud/haze parameters, leading to larger uncertainties for WFC3+NIRCam observations.

\subsubsection{HST STIS + HST WFC3 + JWST NIRCam}
\label{subsub:free_full}
In this section, we show the results obtained by performing the retrieval analysis on combined STIS+WFC3+NIRCam transmission spectra observations. These observations provide a total wavelength coverage from 0.3-5.1 $\mu$m. In this retrieval analysis too, we include all the parameters described in the previous section with WFC3+NIRCam observations, however the offset is between STIS+WFC3 and NIRCam. Table \ref{tab:free-retrieval} and Figure \ref{fig:full_retrieval} show the result of this combined retrieval, while a corner plot is provided in the Appendix (Figure \ref{fig:SWN_corner}). We obtain a $\chi^2_{\text{red}}$ value of 1.21 for our best-fit model with STIS+WFC3+NIRCam observations. Figure \ref{fig:contribution} shows the contribution of different gases to the best-fit spectrum obtained from this retrieval. Since this retrieval analysis is our main result, we report detection significance only for this. The detection significance of \ce{H2O}, \ce{CO2}, \ce{Na} and \ce{K} are $12.6\sigma$, $7.8\sigma$, $6.9\sigma$ and $4.9\sigma$, respectively and other stastical parameters are provided in Table \ref{tab:stats}. This detection significance has been calculated by comparing the Bayesian evidence from the retrieval run with all the parameters and that without the respective molecule. As expected, due to the additional H$_2$O feature in the NIRCam bandpass, our H$_2$O detection significance with STIS+WFC3+NIRCam observations is comparatively higher than that of the 9.1$\sigma$ reported by \citet{Macdonald2017} with HST STIS and WFC3 observations. \citet{Macdonald2017} also reported detection of HCN and/or NH$_3$ with more than 3$\sigma$ significance with the same HST observations. Furthermore, \citet{Giacobbe2021} reported detection of \ce{CO}, \ce{NH3}, \ce{C2H2}, \ce{CH4}, and \ce{HCN} with more than 5$\sigma$ detection significance for each molecule with high-resolution observations utilizing the cross-correlation technique. However, we do not detect any of these molecules in the \texttt{SANSAR} free retrievals at significance greater than 3$\sigma$.

We constrain the abundances of the H$_2$O, CO$_2$, Na and K to $-4.53^{+0.32}_{-0.29}$, $-7.52^{+0.30}_{-0.26}$, $-5.67^{+0.51}_{-0.45}$ and $-6.93^{+0.54}_{-0.47}$ respectively. Our retrieved value of H$_2$O abundance with STIS+WFC3+NIRCam observations is consistent with $-4.66^{+0.39}_{-0.30}$ and $-4.46^{+0.46}_{-0.35}$, reported by \citet{Pinhas2019} and \citet{fairman2024}, respectively, using HST+Spitzer observations. \citet{Xue2024} did not report any H$_2$O or \ce{CO2} abundance constraint because of the chemical equilibrium assumption in their retrievals for H$_2$O and \ce{CO2}.
We did not find significant absorption by other species included in our retrievals, as seen in Figure \ref{fig:contribution}. However, we are able to place upper limits on them. The 3$\sigma$ upper limits on the log(X) that we obtain for \ce{CO}, \ce{NH3}, \ce{C2H2}, \ce{CH4}, \ce{H2S}, and \ce{HCN} are -5.35, -5.79, -6.17, -7.30, -5.69, and -6.97, respectively. 

The $\log(P_{\text{cloud}})$ is robustly constrained to $-2.61^{+0.53}_{-0.63}$ and cloud fraction is constrained to $0.53^{+0.09}_{-0.09}$. We also find evidence of patchy clouds with a detection significance of 2.8$\sigma$, which is lower than 4.8$\sigma$ reported by \citet{Macdonald2017} using HST observations. The median retrieved $P$-$T$ profile is quasi-isothermal in this case as well, with a temperature of $\sim$1300\,K, similar to that obtained from WFC3+NIRCam but higher by $\sim$400\,K compared to NIRCam. The haze enhancement factor, $\log(a)$, is constrained to $\sim$10 -- 288, which is very low compared to the value obtained by \citet{Macdonald2017} (5000 -- 100000) with HST STIS plus WFC3 observations. However, we note that we also get very high $\log(a)$ values similar to \citet{Macdonald2017} when we consider only HST data.

\begin{figure*}
    \includegraphics[width=\textwidth]{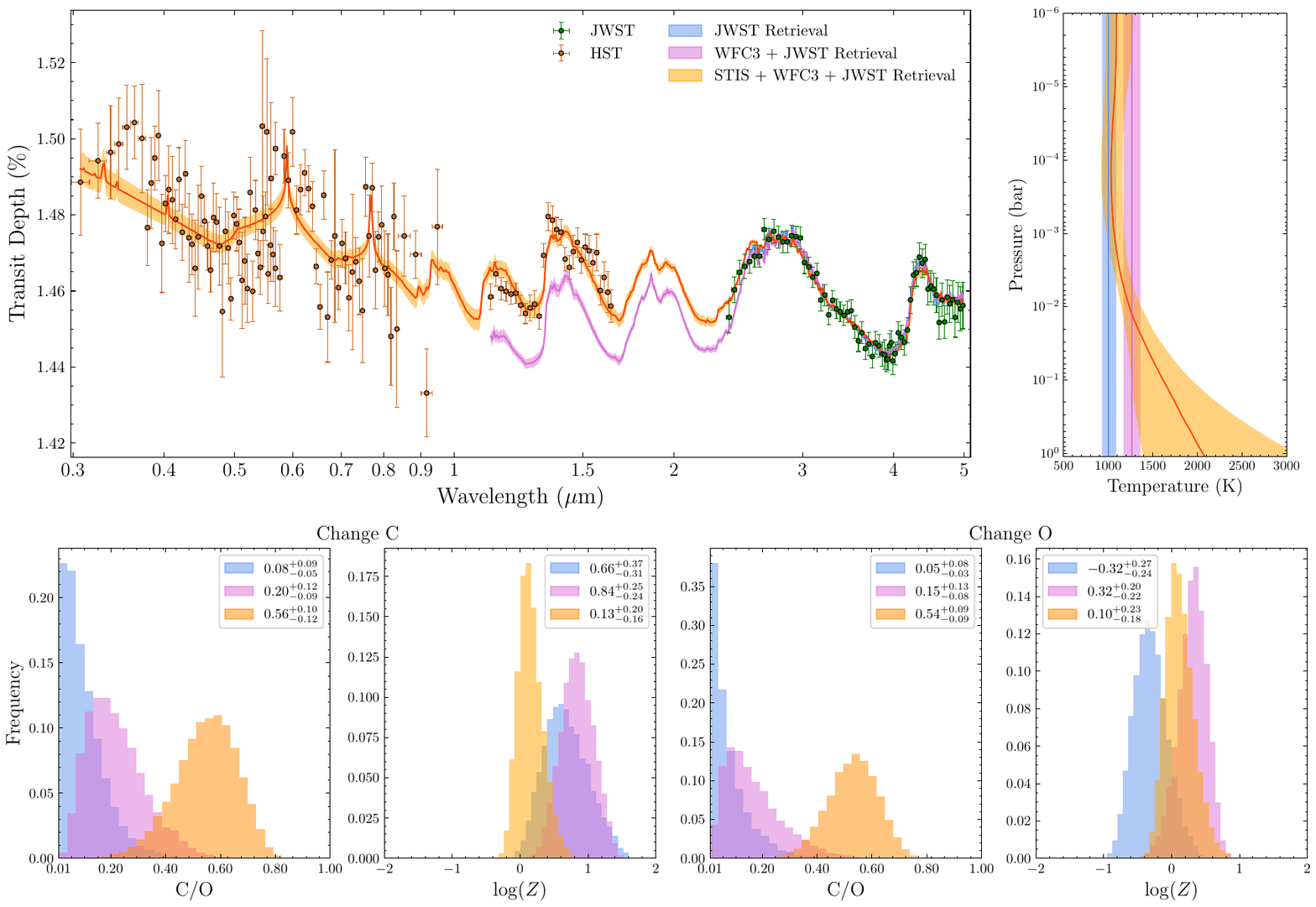}
    \caption{Retrieved transmission spectra, $P$-$T$ profile, C/O and metallicity posteriors of HD 209458 b with equilibrium chemistry.
    The top panel consists of median transmission spectra and $P$-$T$ profile of HD 209458 b for the three retrievals, JWST (NIRCam), WFC3 + NIRCam, and STIS + WFC3 + NIRCam. All the spectra shown here are generated with 5000 random equal-weighted samples from \texttt{PyMultiNest} and binned at a resolution of R$\sim$200. The shaded region indicates the 1$\sigma$ uncertainty in the median fit. JWST observations are represented in green, while combined STIS and WFC3 observations are shown in brown. A retrieved median offset of 39 ppm is applied to the STIS + WFC3 observations, whereas the offset for WFC3 in the case of the WFC3 + JWST combination is not shown here. For all the retrievals, the corresponding $P$-$T$ profiles are shown beside the median spectra. The bottom panel shows the posteriors of C/O and $\log(Z)$ by varying C and varying O for the C/O ratio. The posterior values shown in this panel correspond to the results shown in Table \ref{tab:eqbm_chem_table}.}
    \label{fig:eqbm}
\end{figure*}

\begin{table*}
    \renewcommand{\arraystretch}{1.5}
	\centering
    \caption{Table showing \texttt{SANSAR} retrieval results  in comparison to a previous study of HD-209458 b by \citet{Xue2024}. For the results shown in this table, we also perform retrieval for the combined observations from HST WFC3 \citep{demingwfc3} and JWST NIRCam observations as done by \citet{Xue2024}. $\delta$ is the instrumental offset between HST WFC3 and JWST NIRCam observations in ppm.}
		\begin{tabular}{@{}lccccc@{}}
			\hline
                \hline
			Parameters & Prior &
			\multicolumn{2}{c}{JWST NIRCam} &
			\multicolumn{2}{c}{HST (WFC3) + JWST (NIRCam)} \\ \cmidrule(l){3-6} 
			&  & \citet{Xue2024}                       & \texttt{SANSAR}                    & \citet{Xue2024}                       & \texttt{SANSAR}                                     \\ \cmidrule(r){1-6}
			T~(K)                    & $\mathcal{U}$(400, 1600) & $1088^{+103}_{-88}$       & $1062^{+91}_{-76}$        & $1290^{+83}_{-81}$        & $1281^{+79}_{-76}$              \\
			R$_{\text{p},\text{ref}}$ (R$_\text{J}$) & $\mathcal{U}$(0.8, 1.5) & $1.353^{+0.006}_{-0.007}$ & $1.354^{+0.006}_{-0.007}$ & $1.339^{+0.007}_{-0.006}$ & $1.341^{+0.006}_{-0.006}$  \\
			$\log(Z)$                   & $\mathcal{U}$(-2, 2) & $0.54^{+0.30}_{-0.23}$    & $0.66^{+0.34}_{-0.31}$    & $0.69^{+0.34}_{-0.25}$    & $0.85^{+0.27}_{-0.27}$        \\
			C/O                        & $\mathcal{U}$(0.01, 1) & $0.11^{+0.12}_{-0.06}$    & $0.08^{+0.09}_{-0.05}$    & $0.23^{+0.12}_{-0.15}$    & $0.18^{+0.13}_{-0.08}$       \\
			$\log$(P$_{\text{cloud}}$)    & $\mathcal{U}$(-8, 2) & $-3.31^{+0.50}_{-0.68}$   & $-3.61^{+0.63}_{-1.40}$   & $-3.68^{+0.45}_{-0.44}$   & $-3.84^{+0.39}_{-0.36}$      \\
			$\phi$ & $\mathcal{U}$(0, 1) & $0.68^{+0.19}_{-0.20}$    & $0.60^{+0.24}_{-0.24}$    & $0.82^{+0.09}_{-0.09}$    & $0.84^{+0.09}_{-0.09}$        \\
			$\delta$   & $\mathcal{U}$(-500, 500) & $\cdots$                         & $\cdots$                         & $126^{+12}_{-12}$         & $126^{+11}_{-11}$                  
                           \\ \bottomrule
		\end{tabular}
	\tablecommentsnormal{Here, the C/O ratio is varied by changing the elemental abundance of carbon, and the radius is retrieved at 1 bar.}
        \label{tab: eq chem comp}
    
\end{table*}

\begin{deluxetable*}{lccccccc}
    \label{tab:eqbm_chem_table}
    \renewcommand{\arraystretch}{1.9}
    \tabletypesize{\footnotesize}
    \tablecolumns{8} 
    \tablecaption{Table showing comparison of \texttt{SANSAR} equilibrium chemistry retrieval results for HD~209458~b with different instrument combinations, by changing elemental abundance of carbon as well as oxygen for varying C/O ratio. The retrievals were performed on JWST (NIRCam) observations from \citet{Xue2024}, HST (WFC3), and HST (STIS) from \citet{Sing2016}. \citet{Guillot2010} $P$-$T$ parametrization is included only for HST STIS + HST WFC3 + JWST NIRCam retrieval. Instrumental offset ($\delta$) is applied to WFC3 and WFC3+STIS observations, keeping JWST observations fixed. Reduced chi-squared value ($\chi^2_{\text{red}}$) is also shown for all the retrievals.}
    \tablehead{ &  & \multicolumn{2}{c}{JWST (NIRCam)} & \multicolumn{2}{c}{HST (WFC3) + JWST (NIRCam)} & \multicolumn{2}{c}{HST (STIS) +  HST (WFC3) + JWST (NIRCam)}}
    \startdata
    Parameter \phantom{space} & Prior & Change C & Change O & Change C & Change O & Change C & Change O \\ \hline
    $T$ (K) & $\mathcal{U}$(400, 1600) & $1062^{+94}_{-81}$ &$998^{+92}_{-71}$ & $1306^{+83}_{-81}$ &  $1266^{+95}_{-95}$ & $\cdots$& $\cdots$ \\
    R$_{\text{p},\text{ref}}$ (R$_\text{J}$) & $\mathcal{U}$(0.8,1.5) & $1.378^{+0.006}_{-0.006}$ &$1.381^{+0.005}_{-0.006}$ & $1.367^{+0.005}_{-0.005}$ &  $1.366^{+0.005}_{-0.006}$& $1.384^{+0.004}_{-0.006}$ & $1.384^{+0.004}_{-0.009}$ \\
    $\log(Z)$ & $\mathcal{U}$(-2, 2) & $0.66^{+0.37}_{-0.31}$ &$-0.32^{+0.27}_{-0.24}$ & $0.84^{+0.25}_{-0.24}$ &  $0.32^{+0.20}_{-0.22}$ & $0.13^{+0.20}_{-0.16}$ & $0.10^{+0.23}_{-0.18}$ \\
    C/O & $\mathcal{U}$(0.01, 1) & $0.08^{+0.09}_{-0.05}$ &$0.05^{+0.08}_{-0.03}$ & $0.20^{+0.12}_{-0.09}$ &  $0.15^{+0.13}_{-0.08}$ & $0.56^{+0.10}_{-0.12}$ & $0.54^{+0.09}_{-0.09}$ \\
    $\log$(P$_{\text{cloud}}$) & $\mathcal{U}$(-8, 2) & $-3.63^{+0.64}_{-1.66}$ & $-3.29^{+0.53}_{-0.68}$ & $-4.00^{+0.38}_{-0.35}$ &  $-4.19^{+0.42}_{-0.38}$ & $-0.49^{+1.64}_{-1.69}$ & $-0.50^{+1.68}_{-1.77}$ \\
    $\phi$ & $\mathcal{U}$(0, 1) & $0.60^{+0.24}_{-0.26}$ & $0.71^{+0.16}_{-0.17}$ & $0.80^{+0.09}_{-0.08}$ &  $0.78^{+0.09}_{-0.08}$ & $0.56^{+0.05}_{-0.05}$ & $0.56^{+0.06}_{-0.05}$ \\
    $\delta$ & $\mathcal{U}$(-500, -500) & $\cdots$ & $\cdots$ & $190^{+11}_{-11}$ & $181^{+13}_{-13}$ & $41^{+20}_{-20}$ & $39^{+21}_{-21}$ \\
    $\log(a)$ & $\mathcal{U}$(-2, 8) & $\cdots$ & $\cdots$ & $\cdots$ &  $\cdots$ & $3.81^{+0.49}_{-0.47}$ & $3.82^{+0.50}_{-0.48}$  \\
    $\gamma$ & $\mathcal{U}$(-12, 2) & $\cdots$ & $\cdots$ & $\cdots$ &  $\cdots$ & $-2.50^{+0.44}_{-0.48}$ & $-2.53^{+0.44}_{-0.50}$ \\
    $\kappa_{\text{IR}}$ & $\mathcal{U}$(-5, 5) & $\cdots$ & $\cdots$ & $\cdots$ &  $\cdots$ & $-0.91^{+1.40}_{-1.54}$ & $-1.05^{+1.01}_{-1.66}$ \\
    $\gamma_1$ & $\mathcal{U}$(-5, 3) & $\cdots$ & $\cdots$ & $\cdots$ &  $\cdots$ & $-1.45^{+2.29}_{-2.21}$ & $-1.66^{+1.52}_{-2.13}$ \\
    $\gamma_2$ & $\mathcal{U}$(-5, 3) & $\cdots$ & $\cdots$ & $\cdots$ &  $\cdots$ & $-1.35^{+2.35}_{-2.26}$ & $-1.66^{+1.56}_{-2.16}$ \\
    $\beta$ & $\mathcal{U}$(0, 2) & $\cdots$ & $\cdots$ & $\cdots$ &  $\cdots$ & $0.81^{+0.12}_{-0.12}$ & $0.80^{+0.12}_{-0.11}$ \\
    $\alpha$ & $\mathcal{U}$(0, 1) & $\cdots$ & $\cdots$ & $\cdots$ &  $\cdots$ & $0.50^{+0.33}_{-0.32}$ & $0.50^{+0.33}_{-0.33}$ \\
    $\chi^2_{\text{red}}$ & $\cdots$ & 0.55 & 0.48 & 0.94 & 0.92 & 1.26 & 1.27 
    \enddata
    
\end{deluxetable*}

\subsection{Equilibrium chemistry Retrieval}
\label{subsec:eq}

 In this section, we detail the equilibrium chemistry retrieval results of HD~209458~b  with \texttt{SANSAR}. These retrievals were also performed for three instrument combinations similar to free retrieval: JWST NIRCam observations from \citet{Xue2024} described in Section \ref{eqbm_jwst}, HST WFC3 \citep{demingwfc3, Sing2016} plus JWST NIRCam observations detailed in Section \ref{eqbm_wfc3jwst}, and by combining HST STIS and HST WFC3 \citep{Sing2016} with JWST NIRCam observations as described in Section \ref{eqbm_stiswfc3jwst}. We note that we perform equilibrium chemistry retrieval with WFC3+NIRCam observations using two separate datasets, one from \citet{demingwfc3} with \citet{Xue2024} and the second from \citet{Sing2016} as used for free retrieval. The retrievals are performed using the correlated-k method at R$\sim$1000 using \texttt{SANSAR}. Unlike the mixing ratios, which are free parameters in free chemistry retrieval, in the equilibrium chemistry retrievals, the C/O ratio and metallicity, $\log(Z)$, are free parameters. We note that the equilibrium chemistry retrievals that we perform in this work are without condensation. We adopt the simple gray-cloud prescription parameterized by cloud-top pressure, $\log(P_{\text{cloud}})$, and cloud fraction, $\phi$, as for free retrievals. The pressure grid ranges from $10^{2}$ bar to $10^{-8}$ bar with 100 layers. We chose 10 mbar as our reference pressure (R$_{\text{p},\text{ref}}$) for all our equilibrium chemistry retrievals, similar to free retrievals, except for the retrievals that were run to compare with \citet{Xue2024}, in which case it is taken similar to that adopted by them i.e. 1 bar. All the retrievals use the correlated-k method at R$\sim$1000 with 2000 live points. The median spectra along with the $P$-$T$ profile and posteriors for the retrievals in Section \ref{eqbm_jwst}, \ref{eqbm_wfc3jwst}, and \ref{eqbm_stiswfc3jwst} are shown in Figure \ref{fig:eqbm}.

\subsubsection{JWST NIRCam}\label{eqbm_jwst}
In the equilibrium chemistry retrieval with just JWST NIRCam observations, the free parameters are $T$, R$_{\text{p,ref}}$, C/O, $\log(Z)$, $\log(P_{\text{cloud}})$, and $\phi$. We perform two separate retrievals to vary the C/O ratio, one by varying the elemental abundance of C (C/H) since that was adopted in \citet{Xue2024} and the second by varying the elemental abundance of O (O/H). We use the same opacities of molecules as considered by \citet{Xue2024}, namely \ce{CH4}, \ce{CO}, \ce{CO2}, \ce{H2O}, \ce{H2S}, \ce{C2H2}, \ce{HCN}, and \ce{SO2}. Table \ref{tab: eq chem comp} shows the comparison of our retrieval results with \citet{Xue2024}, and we see that our retrieval results with 1$\sigma$ errors for all the parameters agree quite well. In summary, similar to \citet{Xue2024}, we also obtain super-solar metallicity and a very low C/O ratio for HD~209458~b with an equilibrium chemistry retrieval with just JWST NIRCam observations and changing C/H to vary the C/O ratio. 

We further investigated the constraints that we obtain on the retrieved parameters when O/H is changed to vary the C/O ratio. These constraints are shown in Table \ref{tab:eqbm_chem_table}. For retrieval on JWST NIRCam observations, we notice a stark difference in $\log(Z)$ which changes from super-solar ($0.66^{+0.37}_{-0.31}$) to sub-solar metallicity ($-0.32^{+0.27}_{-0.24}$) when compared to our previous results where the elemental abundance of C was varied for varying C/O ratio. The constraint on the C/O ratio does not change and is still highly sub-solar. Varying C/H and O/H affects the abundances of \ce{H2O} and \ce{CO2} very differently at different C/O ratios as highlighted by \citet{Drummond2019}. With just NIRCam observations the required \ce{H2O} and \ce{CO2} abundance to match the observations is quite high (super-solar) as we noticed with free retrievals. High abundances of \ce{H2O} and \ce{CO2} require a high amount of O/H. When we vary C/H to vary the C/O ratio, O/H is fixed, and therefore, O/H can only be increased by an increase in metallicity, thus leading to a high metallicity constraint when we change C/H. On the contrary, when we vary O/H to vary the C/O ratio, O/H can increase by decreasing the C/O ratio to produce the required \ce{H2O} and \ce{CO2} abundances. This can also be noticed by the low C/O ratio constraint that we obtain. 

\subsubsection{HST WFC3 + JWST NIRCam}\label{eqbm_wfc3jwst}
To compare our results for WFC3+NIRCam observations from \texttt{SANSAR} with \citet{Xue2024}, we use HST WFC3 data from \citet{demingwfc3} along with the JWST NIRCam observations. We use the same set of opacities as described in the previous section. Since observations, in this case, are from two different instruments, we also include offset between NIRCam and WFC3 observations as a free parameter, as done for the free retrievals. Table \ref{tab: eq chem comp} shows the comparison of our retrieval results with \citet{Xue2024}, and here also we see that our retrieval results with 1$\sigma$ errors for all the parameters agree quite well.

We further perform an equilibrium retrieval with WFC3+NIRCam observations with HST WFC3 data from \citet{Sing2016} along with the JWST NIRCam observations as we have done for free retrievals. For this, we perform retrieval with two cases, one by changing C/H and the other by changing O/H to change C/O. The results for these retrieval runs are shown in Table \ref{tab:eqbm_chem_table}. The major difference that we see with this test is also for metallicity. By varying C/H we constrain the metallicity to be $0.84^{+0.25}_{-0.24}$ and by varying O/H we constrain it to be $0.32^{+0.20}_{-0.22}$. However, we obtain a higher metallicity when we use C/H compared to when we use O/H to vary the C/O ratio. We also find a similar trend with just NIRCam observations, as discussed in Section \ref{eqbm_jwst}, where the difference was much larger, with C/H variation giving super-solar metallicity while O/H variation led to sub-solar metallicity. We also see a minor increase in the constrained value of the C/O ratio in both cases (C/H and O/H) with WFC3+NIRCam observations. 

\subsubsection{HST STIS + HST WFC3 + JWST NIRCam}\label{eqbm_stiswfc3jwst}
Similar to free retrieval, we perform combined retrieval with STIS+WFC3+NIRCam observations. The STIS and WFC3 observations are from \citet{Sing2016} while JWST observations are from \citet{Xue2024}. For this retrieval, in addition to the opacities considered before in sections \ref{eqbm_jwst} and \ref{eqbm_wfc3jwst}, we also include the opacities of \ce{Na} and \ce{K}, as they have major absorption lines in the optical region. An offset is added for combined STIS+WFC3 observations while fixing NIRCam data. We include haze in our study for this configuration as described in Section \ref{sec:cloud haze}, resulting in two additional free parameters $\log(a)$ and $\gamma$. We use Guillot $P$-$T$ parametrization for this retrieval to consider the physically consistent atmospheric conditions. The retrievals were performed by changing C/H as well as by changing O/H to vary C/O, and both the results are shown in Table \ref{tab:eqbm_chem_table}. Figure \ref{fig:eqbm} shows the median model and the corresponding 1$\sigma$ bounds of the retrieved spectra for all three scenarios, along with the retrieved $P$-$T$ profile and posteriors. Detection significances and other statistical parameters for these retrievals are listed in Table \ref{tab:stats}

We find that with STIS+WFC3+NIRCam observations, metallicity is constrained to $0.13^{+0.20}_{-0.16}$ when varying C/H and $0.10^{+0.23}_{-0.18}$ when varying O/H. It can be noted that the metallicity constraint is consistent with solar value in both cases. However, metallicity constraints in both cases are almost similar, unlike NIRCam and WFC3+NIRCam observations discussed in previous sections. The similarity in metallicity constraint in both cases again highlights the importance of a broader wavelength range of observations for the retrievals. We also find the C/O ratios in both cases are constrained to values close to solar C/O ratio, i.e. $0.56^{+0.10}_{-0.12}$ with C/H and $0.54^{+0.09}_{-0.09}$ with O/H. This is substantially different from the sub-solar C/O ratio constraint that we obtain with just NIRCam and WFC3+NIRCam observations.  

The retrieved $P$-$T$ profile with STIS+WFC3+NIRCam observations shows an isothermal structure with a temperature of $\sim$1100\,K from top of the atmosphere until $\sim$1 mbar after which it starts increasing as shown in Figure \ref{fig:eqbm}. The retrieved quasi-isothermal temperature with STIS+WFC3+NIRCam observations is $\sim$100\,K higher than the isothermal temperature retrieved with just NIRCam observations, but cooler by about 200\,K compared to WFC3+NIRCam observations at 1 mbar pressure level. 

\begin{figure*}
    \centering
    \includegraphics[width=\textwidth]{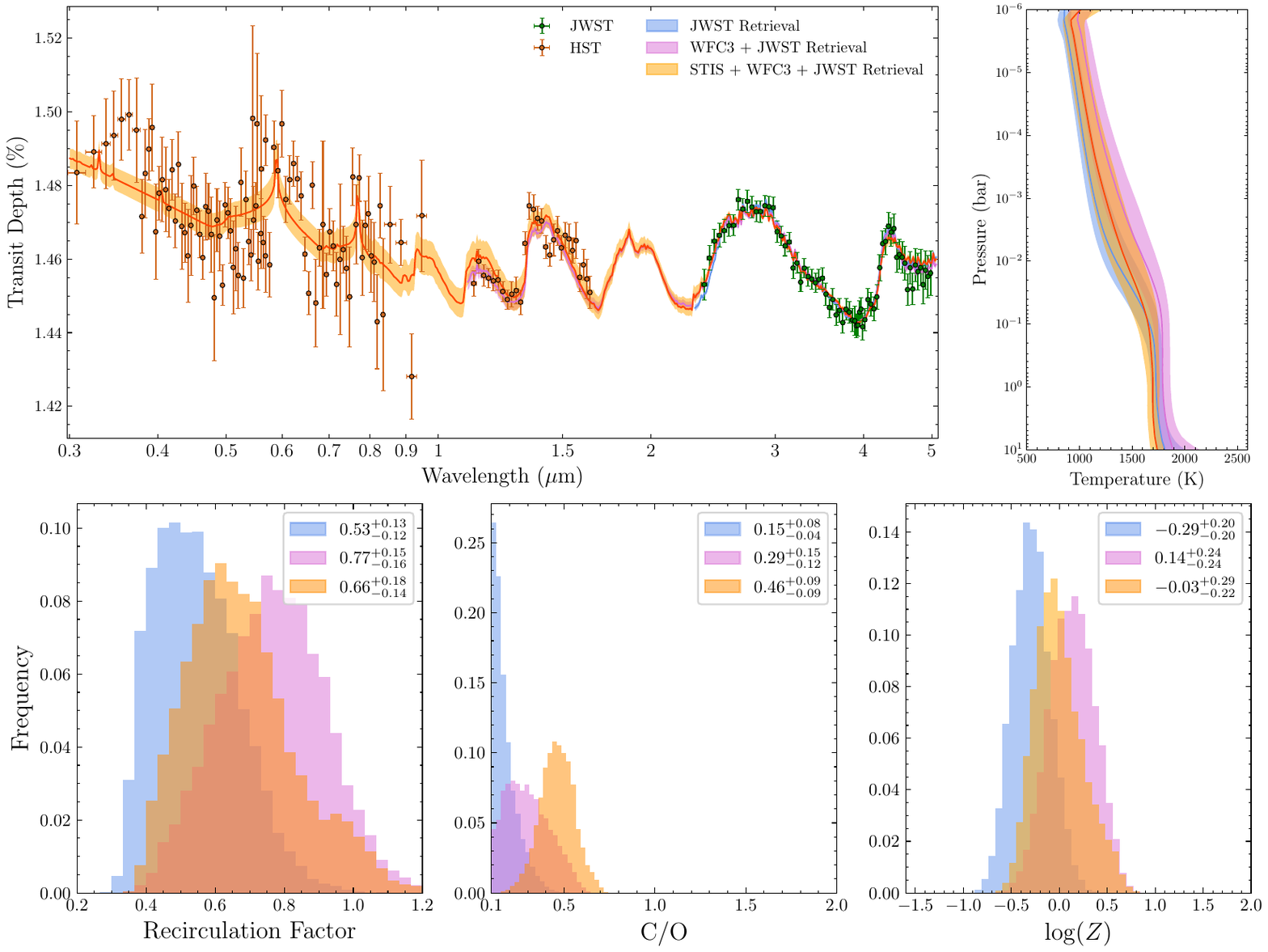}
    
    \caption{Retrieved transmission spectra, $P$-$T$ profile along with rcf, C/O ratio and metallicity posteriors with grid-retrieval.
    The top panel consists of median transmission spectra and $P$-$T$ profile of HD~209458~b for the three grid-retrievals, with just JWST (NIRCam), WFC3 + NIRCam, and STIS + WFC3 + NIRCam observations. All the spectra shown here are generated with 5000 random equal-weighted samples from \texttt{PyMultiNest} and binned at a resolution of R$\sim$200. The shaded region indicates the 1$\sigma$ uncertainty in the median fit. JWST observations are represented in green, while combined STIS and WFC3 observations are shown in brown. A retrieved median offset of 90 ppm is applied to the STIS + WFC3 observations, whereas the offset for WFC3 in the case of the WFC3 + JWST combination is not shown here. For all the retrievals, the corresponding $P$-$T$ profiles are shown beside the median spectra. The bottom panel shows the posteriors of the recirculation factor, C/O ratio, and $\log(Z)$.}
    \label{fig:grid spectra}
\end{figure*}

\begin{deluxetable*}{lcccc}
    \label{tab:grid_ret}
    \renewcommand{\arraystretch}{1.5}
    \tabletypesize{\footnotesize}
    \tablecolumns{4} 
    \tablecaption{Table showing the results of self-consistent grid-retrievals performed on JWST NIRCam observations from \citet{Xue2024}, HST (WFC3), and HST (STIS) observations from \citet{Sing2016}. Instrumental offset ($\delta_2$) is applied to WFC3 and WFC3+STIS observations, keeping JWST observations fixed.}
    \tablehead{Params & Prior & NIRCam & WFC3+NIRCam & STIS+WFC3+NIRCam }
    \startdata
    rcf & $\mathcal{U}$(0.2 , 1.2) &$0.53^{+0.13}_{-0.12}$ & $0.77^{+0.15}_{-0.16}$ & $0.66_{-0.14}^{+0.18}$ \\
    log (Z)  & $\mathcal{U}$(-1.6, 2.0)&$-0.29^{+0.20}_{-0.20}$ &$0.14^{+0.24}_{-0.24}$ & $-0.03_{-0.22}^{+0.29}$\\
    C/O & $\mathcal{U}$(0.1, 2.0)& $0.15^{+0.08}_{-0.04}$ &$0.29^{+0.15}_{-0.12}$ & $0.46_{-0.09}^{+0.09}$\\
    $T_{\text{int}}$ & $\mathcal{U}$(100, 400)&$255^{+98}_{-104}$&$236^{+110}_{-94}$ & $228_{-89}^{+110}$\\
    $\log (a)$ & $\mathcal{U}$(1, 5)& $\cdots$ & $2.66_{-0.54}^{+0.45}$ & $3.53_{-0.51}^{+0.58}$\\
    $\gamma$ & $\mathcal{U}$(-7 , -1)& $\cdots$& $-1.38_{-0.52}^{+0.27}$ & $-2.81_{-0.49}^{+0.43}$\\
    $\log(P_{\text{cloud}})$ &$\mathcal{U}$(-4, -1)& $-2.89^{+0.38}_{-0.35}$ &$-2.21^{+0.82}_{-0.78}$ & $-2.27_{-0.56}^{+0.70}$\\
    $\phi$ &$\mathcal{U}$(0.3, 1)& $0.76^{+0.16}_{-0.20}$ &$0.76^{+0.10}_{-0.08}$ & $0.64_{-0.07}^{+0.09}$\\
    $\delta_2$ & $\mathcal{U}$(-500, 500)& $\cdots$& $108^{+15}_{-17}$ & $90^{+40}_{-35}$ \\
    $\chi_{\text{red}}^2$ & $\cdots$& 0.54& 0.90 & 1.30
    \enddata
\end{deluxetable*}

\subsection{Grid Retrieval}
\label{subsec:grid}

We also perform grid-retrievals on HD~209458~b observations with the transmission spectra grid generated using a grid of self-consistent radiative-convective equilibrium $P$-$T$ profiles and the corresponding chemical abundances for HD~209458~b, described in Section \ref{sec:grid-theory}. The variables in the self-consistent model atmosphere grid are the re-circulation factor (rcf), metallicity, C/O ratio, and internal temperature. Additionally, the transmission spectra grid includes cloud and haze parameters, as discussed in previous sections. 
We note that in this self-consistent atmosphere model grid, the C/O ratio is varied by varying O/H. Similar to free and equilibrium chemistry retrievals, we also perform three distinct grid-retrievals with different HST and JWST instrument combinations, which are just JWST NIRCam, WFC3+NIRCam, and STIS+WFC3+NIRCam observations. All these grid-retrievals have been performed with 2000 live points with an evidence tolerance of 0.5.  The grid-retrieval results with observations from different instrument combinations are shown in Table \ref{tab:grid_ret} and Figure \ref{fig:grid spectra}. The table shows the median values of each parameter along with their $\pm1\sigma$ values.

\subsubsection{JWST NIRCam only}

The results of the grid-retrieval with just JWST NIRCam observations ranging from 2.3 - 5.1 $\mu$m are shown in Table \ref{tab:grid_ret} and Figure \ref{fig:grid spectra}. We do not use haze parameters in the grid-retrieval with just NIRCam observations, as it does not affect the spectrum covered by NIRCam. We find that the metallicity, $\log(Z)$, is constrained to $-0.29^{+0.20}_{-0.20}$, indicating a sub-solar value. This result is consistent with the $-0.32^{+0.27}_{-0.24}$ obtained from equilibrium chemistry retrievals using only NIRCam observations, where O/H is varied to adjust the C/O ratio, following the approach used in our self-consistent grid model. The C/O ratio is constrained to  $0.15^{+0.08}_{-0.04}$, which is highly sub-solar and similar to equilibrium chemistry retrieval. Figure \ref{fig:grid spectra} shows the median $P$-$T$ profile and its 1$\sigma$ bounds, that we obtain with our self-consistent grid-retrieval. Our retrieved temperature is hotter than the retrieved $P$-$T$ profile with free and equilibrium chemistry retrievals for NIRCam observations at 1 mbar pressure level. We do not obtain any robust constraints on the internal temperature of the planet.

\subsubsection{HST WFC3 + JWST NIRCam}

The results of the grid-retrieval with WFC3+NIRCam observations ranging from 1.1 - 5.1 $\mu$m are shown in Table \ref{tab:grid_ret} and Figure \ref{fig:grid spectra}. With WFC3+NIRCam observations, metallicity is constrained to $0.14^{+0.24}_{-0.24}$ which is consistent with equilibrium chemistry retrievals ($0.32^{+0.20}_{-0.22}$) for WFC3+NIRCam observations. The C/O ratio is constrained to be $0.29^{+0.15}_{-0.12}$, which is sub-solar but higher than the value that we obtain with NIRCam observations. We obtain a reduced chi-square $\chi^{2}_\text{red}$ value of 0.90 for the best-fit model spectra. The cloud fraction, $\phi$, is constrained to $0.76^{+0.10}_{-0.08}$, which is very similar to what we obtain with NIRCam observations

\subsubsection{HST STIS + HST WFC3 + JWST NIRCam}

The results of the grid-retrieval with STIS+WFC3+NIRCam observations ranging from $0.3-5.1$ $\mu$m are shown in Table \ref{tab:grid_ret} and Figure \ref{fig:grid spectra}. The log(Z) and C/O ratio is constrained to $-0.03^{+0.29}_{-0.22}$ and $0.46^{+0.09}_{-0.09}$, respectively, consistent with solar value and equilibrium chemistry retrieval for STIS+WFC3+NIRCam. We obtain reduced $\chi^{2} = 1.30$ for the best-fit model spectrum in this grid-retrieval.
We see a sequential increase in the C/O ratio constraints, going from 0.15 with just NIRCam observations to 0.46 with full STIS+WFC3+NIRCam observations. We notice the same trend with equilibrium chemistry retrievals as well. Including WFC3 and STIS leads to more robust constraints on the cloud fraction. The cloud fraction is also much lower with STIS+WFC3+NIRCam observations. Therefore, since the overall cloudiness is reduced, the abundance of \ce{H2O} and \ce{CO2} required to match the observed spectral features is reduced, leading to higher C/O ratio constraints with STIS+WFC3+NIRCam observations. We note that the abundance of \ce{H2O} and \ce{CO2} both decrease with an increase in the C/O ratio from sub-solar to solar values.  Here again, we do not obtain a good constraint on the internal temperature of the planet. The median $P$-$T$ profile does not show any temperature inversion.

\section{Discussion}
\label{sec:discussion}
The quantity and quality of exoplanet observations that we have been receiving with recent JWST observations have necessitated robust testing of our models used to interpret exoplanet atmosphere observations. In this work, we presented a new planetary atmosphere modeling suite \texttt{SANSAR}, particularly focusing on the computation of the model transmission spectra. We interpret the HST and JWST observations of HD~209458~b using \texttt{SANSAR}, with three distinct retrieval approaches: free retrievals, equilibrium chemistry retrievals, and self-consistent grid-based retrievals. Each approach is applied to three different instrument configurations: (i) NIRCam-only observations, (ii) combined HST(WFC3) and NIRCam observations, and (iii) a complete data set comprising HST STIS, HST WFC3, and NIRCam observations. In this section, we discuss in detail how these model choices with different instrument combinations can result in different constraints on the atmospheric properties of HD~209458~b. 

\subsection{On Opacity Treatment Method}
The choice of the opacity treatment method adopted in retrievals to constrain exoplanet atmosphere properties is crucial. In \texttt{SANSAR}, we include both the opacity sampling and correlated-k methods for opacity treatment. We performed robust benchmarking tests by comparing line-by-line model spectra with opacity sampling and correlated-k model spectra at different resolutions. We find that using opacity sampling at less than R$\sim$20000 can lead to substantial errors compared with LBL spectra for observations binned down to R$\sim$100. We find that using the correlated-k method at R$\sim$1000 gives a maximum error of the same order as opacity sampling at R$\sim$100000, while reducing the model computation time by a factor of $\sim$10.

\subsection{On Wavelength Coverage}
With free retrievals for STIS+WFC3+NIRCam observations, the \ce{H2O} abundance is constrained to $-4.53^{+0.32}_{-0.29}$ compared to $-1.62^{+0.39}_{-1.29}$ using just NIRCam observations and $-3.02^{+0.86}_{-1.17}$ using WFC3+NIRCam observations. Furthermore, the \ce{CO2} abundance is constrained to $-7.52^{+0.30}_{-0.26}$ with STIS+WFC3+NIRCam observations compared to $-5.31^{+0.37}_{-1.12}$ using just NIRCam observations and $-6.08^{+0.85}_{-1.10}$ using WFC3+NIRCam observations. Thus, we find a systematic decrease in the abundance of both the species \ce{H2O} and \ce{CO2} as we go from shorter wavelength coverage of just NIRCam to large wavelength coverage of STIS+WFC3+NIRCam observations. 

The NIRCam only retrieval yields abnormally high abundances of \ce{H2O} and \ce{CO2} with large lower bound uncertainties and a $\sim$400 K lower temperature than retrievals performed using broader wavelength coverage. A low $\chi_{\text{red}}^2$ value of 0.57 for NIRCam retrieval also suggests overfitting, likely due to the absence of a strong anchor for temperature as well as cloud-haze, resulting in a very high abundance of \ce{H2O} and \ce{CO2}. However, when HST(WFC3) observations are included for the retrieval, the prominent 1.4 $\mu$m \ce{H2O} absorption feature provides a crucial anchor for the temperature profile, thus increasing it. This higher temperature leads to a larger scale height and enhances absorption cross-sections, thereby requiring lower \ce{H2O} and \ce{CO2} abundances to reproduce the same spectral features in the NIRCam bandpass, thus making the overall fit better with the $\chi^2_{\text{red}}$ value of 0.97 for WFC3+NIRCam observations. However, a degeneracy between the molecular abundances and cloud-haze opacity remains, as seen in the posterior distributions from the WFC3+NIRCam retrieval shown in the Appendix (Figure \ref{fig:WN_corner}), since WFC3 bandpass cannot completely anchor cloud-haze properties. The STIS data captures the \ce{Na} and \ce{K} absorption features, which are highly sensitive to cloud opacity, as well as the scattering slope in the optical, which constrains haze properties \citep{Welbanks2019}. These additional constraints enable the retrieval to disentangle the effects of clouds and abundances, resulting in significantly more robust estimates of \ce{H2O} and \ce{CO2}, compared to WFC3+NIRCam and just NIRCam observations. Such estimates show the critical importance of broad wavelength coverage spanning optical to infrared for accurately constraining atmospheric properties and chemical composition of exoplanetary atmospheres in the JWST era, highlighted by many previous studies using only HST observations \citep{Wakeford2018, Welbanks2019, fairman2024}. 

Even with equilibrium chemistry and grid retrievals, utilizing just NIRCam observations leads to overfitting. Similar to free retrievals, this is driven by the required high abundance of \ce{H2O} and \ce{CO2}, thus leading to overestimation of metallicity depending on model choices discussed in detail in the next Section (\ref{sec:model_choice}) and underestimation of the C/O ratio. However, when STIS+WFC3+NIRCam observations are used, the abundance of \ce{H2O} and \ce{CO2} required to explain observed features reduces, thus leading to solar metallicity and solar C/O ratio constraint that we obtain with equilibrium chemistry and grid retrievals. Similar to free retrievals as detailed earlier, even in equilibrium chemistry and grid retrievals, the inclusion of HST STIS and WFC3 observations leads to better fits compared to just NIRCam observations, as indicated by their $\chi^2_{\text{red}}$ values shown in Table \ref{tab:eqbm_chem_table} and \ref{tab:grid_ret}. 

In summary, utilizing only JWST NIRCam observations can lead to an overestimation of abundances by free retrievals. It also results in misleading constraints on metallicity and C/O ratio with equilibrium chemistry and grid retrievals due to overfitting. Even though we can detect \ce{CO2} features with just NIRCam observations, we cannot constrain its abundances robustly without the optical baseline provided by HST STIS observations. If we extend the implications of this finding, we also predict that the abundances of \ce{CO2}, \ce{H2O}, and many other species constrained by retrievals that only utilize JWST NIRSpec G395M/H and G235M/H observations may also be overestimated, compared to retrievals that include UV/optical observations.

\begin{figure}
    \centering
    \includegraphics[width=\columnwidth]{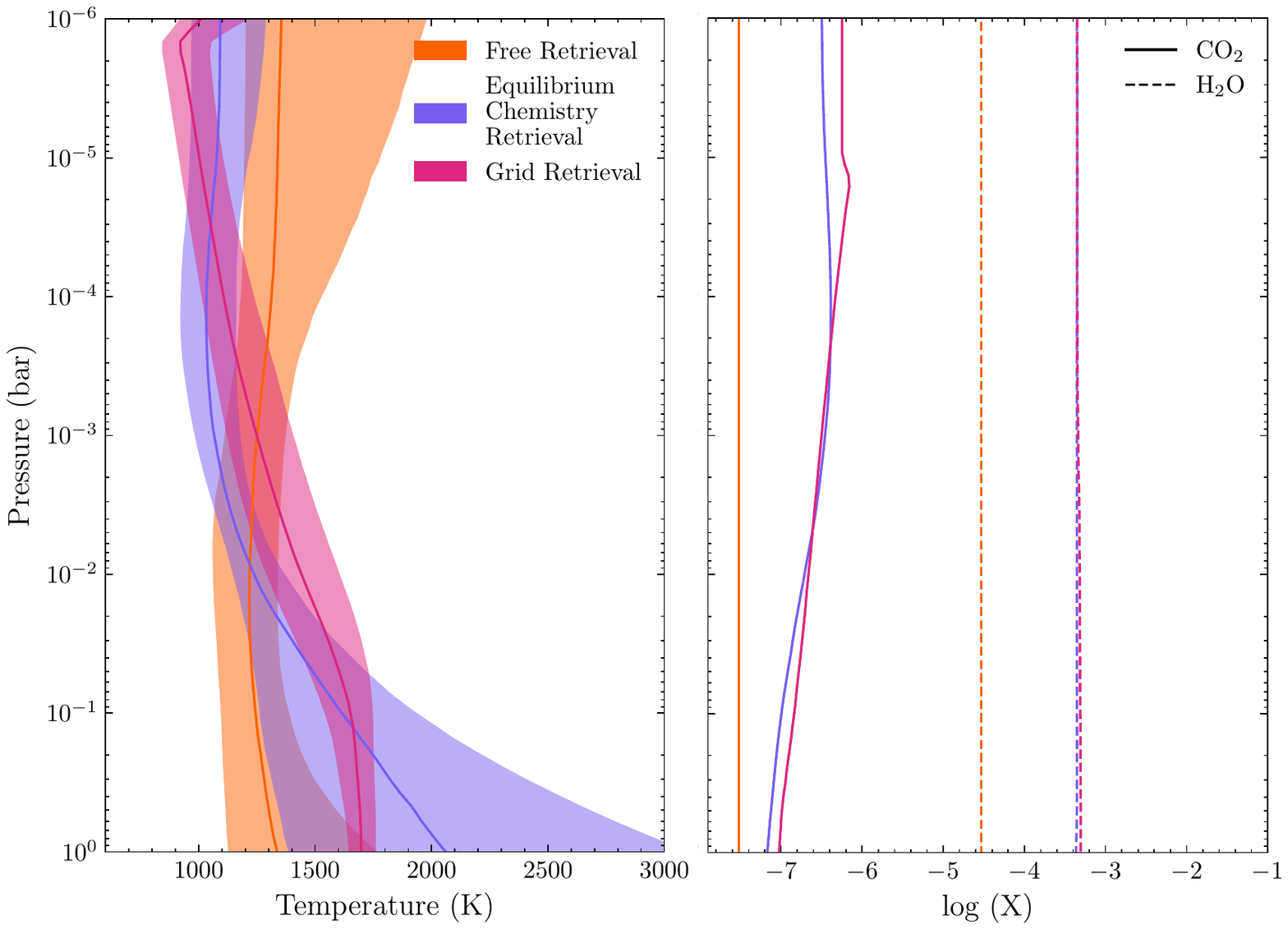}
    \caption{Left: Figure showing median $P$-$T$ profile of HD 209458 b for the three retrievals, free chemistry (orange), equilibrium chemistry (violet), and grid retrieval (pink), performed using STIS+WFC3+NIRCam observations. The shaded region shows 1$\sigma$ uncertainty on the $P$-$T$ profile. Right: The corresponding mixing ratios of \ce{CO2} (solid) and \ce{H2O} (dashed) obtained from the median values of all three retrievals are shown.}
    \label{fig:swn_disc}
\end{figure}

\begin{figure}
    \centering
    \includegraphics[width=\columnwidth]{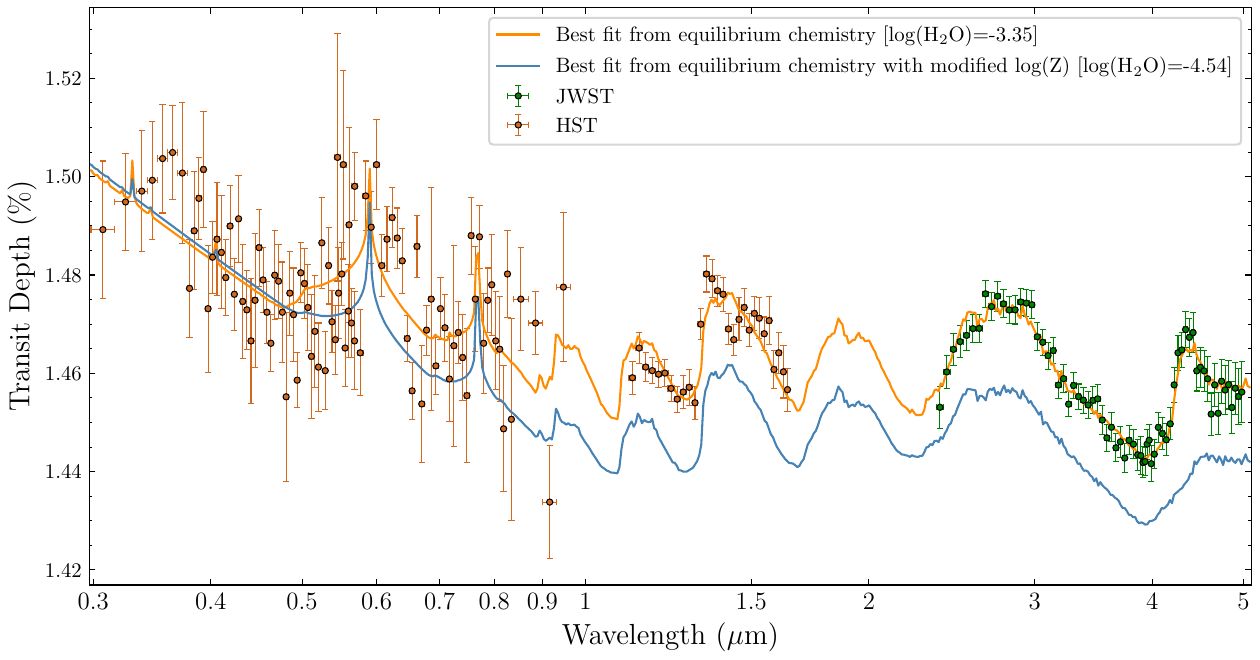}
    \caption{Best fit transmission spectrum (orange) from equilibrium chemistry retrieval for STIS+WFC3+NIRCam observations is compared to the spectrum where the best-fit metallicity value is changed to match the retrieved \ce{H2O} abundance from the free chemistry retrieval shown in blue.}
    \label{fig:eqbm_to_free}
\end{figure}

\subsection{On Model/Retrieval Choices}\label{sec:model_choice}

We find that in the case of free retrieval, abundances of \ce{H2O} and \ce{CO2} are highly sub-solar compared to equilibrium chemistry and grid retrieval.
The equilbrium chemistry abundances of \ce{H2O} and \ce{CO2} are $-3.35^{+0.20}_{-0.17}$ and $-6.46^{+0.27}_{-0.32}$, respectively, at 1 millibar pressure level with STIS+WFC3+NIRCam observations, similar to grid retrieval abundances as seen from Figure \ref{fig:swn_disc}. Equilibrium chemistry and grid retrieval constrained the atmospheric metallicity for HD~209458~b to be $\sim$ $0.8$ -- $2.1$ $\times$ and $0.6$ -- $1.8$$\times$ solar metallicity, respectively, with STIS+WFC3+NIRCam observations, which are essentially consistent with the solar value. To explain why equilibrium chemistry retrievals do not give as low H$_2$O abundance as free retrievals, we compare the best-fit equilibrium chemistry spectrum and the same best-fit spectrum with lower metallicity such that we obtain the same H$_2$O abundance as the free retrieval, in Figure \ref{fig:eqbm_to_free}. We find that in the equilibrium chemistry retrieval, as the H$_2$O abundance decreases with a decrease in metallicity, the abundance of CO$_2$ also reduces substantially, as it can be noticed in Figure \ref{fig:eqbm_to_free} at 4.2 microns, showing diminished CO$_2$ feature for low metallicity model. However, the observed CO$_2$ feature for HD~209458~b with JWST NIRCam observations does not allow CO$_2$ abundance to drop as low as log(X) = -8.9, which is the value expected at this lower metallicity. Therefore, the abundance of the \ce{H2O} cannot be as low as that obtained with free retrievals. One might argue that if low metallicity is unable to form enough \ce{CO2} to match the spectral feature, then maybe changing C/O could provide enough carbon or oxygen to form the \ce{CO2} at low metallicity, but this is not the case as we see from our predicted equilibrium chemical abundances heatmap shown in Figure \ref{Change_C} and Figure \ref{Change_O} of the Appendix. Changing C/O also does not increase the \ce{CO2} abundance to the level required to match the observations. Both \ce{H2O} and \ce{CO2} in equilibrium chemistry retrieval are constrained together by metallicity. On the contrary, in free chemistry retrievals, CO$_2$ and H$_2$O abundances are independent of each other, and therefore, they can be independently changed to match observed spectral features. Since in grid retrievals also, the abundances are obtained using self-consistent $P$-$T$ profiles and equilibrium chemistry, \ce{CO2} and \ce{H2O} are constrained together by metallicity here as well. Therefore, the constrained metallicity with grid-retrieval is also solar. 

As \ce{H2O} and \ce{CO2} are the driving species of metallicity and C/O ratio in equilibrium chemistry and grid retrieval, it can be valuable to infer metallicity and C/O ratio values, using free retrieval abundances, for fair comparison. Therefore, we also compute the metallicity using free retrieval abundances with our STIS+WFC3+NIRCam retrieval result by summing up all the metal atoms (in our case, these metal atoms are C,O, and S) and then dividing it by all the H-atoms following \citet{zhang2024}. We then compare this value to the solar reference computed using our equilibrium chemistry model at the 1 millibar pressure level, where the solar (C+O+S)/H ratio is 7.37$\times$10$^{-4}$. Following this procedure, we computed the metallicity from 5,000 randomly sampled, equal-weight posterior draws from our free retrieval, yielding a value of $-1.62^{+0.32}_{-0.29}$, which is significantly sub-solar.. We also compute free retrieval C/O ratio using this method, which we obtain as $0.0017^{+0.0036}_{-0.0006}$, which is also highly sub-solar.

In equilibrium chemistry retrievals, when C/H is used to vary the C/O ratio, we constrain metallicity to be super-solar (2.2 -- 10.7$\times$ solar) with just NIRCam observations, as reported by \citet{Xue2024} as well. However, when O/H is changed, we constrain the metallicity to be sub-solar (0.3 -- 0.9$\times$ solar). As noticed in free retrieval, the abundances of \ce{CO2} and \ce{H2O}, required to explain the observations in just the NIRCam bandpass, is very high due to low temperature. A similar trend is observed in equilibrium chemistry retrievals; however, in this case, the abundances are governed by the metallicity and the C/O ratio. Therefore, when C/H is varied to vary the C/O ratio, the oxygen inventory remains the same, and the metallicity is increased in the retrieval to increase the abundance of oxygen to produce required \ce{CO2} and \ce{H2O} for NIRCam observations. Due to this, we obtain super-solar metallicity with just NIRCam observations, similar to \citet{Xue2024}. Conversely, when O/H is varied, the decrease in C/O is achieved by increasing oxygen inventory, and therefore, metallicity does not have to increase to produce the required \ce{CO2} and \ce{H2O} for NIRCam observations. Hence, we obtain sub-solar metallicity with just NIRCam observations when we vary O/H to vary C/O ratio. We also explain the origin of this difference in metallicity more robustly using heat maps for predicted equilibrium chemistry \ce{H2O} abundance as a function of metallicity and C/O ratio, shown in Figure \ref{Change_C} in the case of change C and Figure \ref{Change_O} for the change O case. Our retrieved metallicity and C/O ratio, with just NIRCam observations and equilibrium chemistry retrieval, correspond to a \ce{H2O} log abundance of approximately -2.5. While such a water abundance is achievable when varying C/H at high metallicities, shown in Figure \ref{Change_C}, it is not supported when varying O/H, as shown in Figure~\ref{Change_O}.  Although similar water and \ce{CO2} abundance can be obtained at high metallicity in the change O case, it would require a C/O ratio of $\sim$0.4. This solution, however, is physically disfavored due to the excessively high \ce{CO} abundance it requires (see Figure~\ref{Change_O}), which is more than the upper limit of CO constrained by our free retrievals.  As a result, when we vary O/H, the retrieval is forced toward sub-solar metallicity and low C/O values in order to simultaneously reproduce the observed \ce{H2O} and \ce{CO2} features while minimizing the spectral contribution from \ce{CO}. Even for grid retrieval, where we vary O/H to vary C/O, the metallicity and C/O ratio are constrained to sub-solar values, with just NIRCam observations, similar to O/H equilibrium chemistry retrievals.

Interestingly, varying C/H or O/H leads to very close constraints, consistent with solar metallicity with combined STIS+WFC3+NIRCam observations. The constrained C/O ratio is also solar in both cases. We observe a trend where the C/O ratio constraint changes from highly sub-solar to solar value for both cases (varying C/H and O/H) as the wavelength coverage increases from just NIRCam to combined STIS+WFC3+NIRCam observations. As in free retrievals, the abundance of \ce{CO2} and \ce{H2O}, required to explain the combined STIS+WFC3+NIRCam observations, is comparatively lower than just NIRCam observations. Therefore, we obtain solar metallicity and solar C/O ratio with combined STIS+WFC3+NIRCam observations, irrespective of varying C/H or O/H, as they are at their solar value. This is clearly evident from the abundance heat map Figures \ref{Change_C} and \ref{Change_O} as well. For both change C/H and O/H, the solar metallicity and solar C/O ratio yield the same \ce{H2O} and \ce{CO2} abundances, approximately $-3.4$ and $-6.7$, respectively, as required by equilibrium chemistry retrieval to fit STIS+WFC3+NIRCam observations.
For grid retrievals too,  metallicity and C/O ratio are constrained to solar values, with STIS+WFC3+NIRCam observations.

In free retrievals, with STIS+WFC3+NIRCam observations, we constrain the haze enhancement factor to be $1.71^{+0.75}_{-0.71}$, while with equilibrium chemistry retrieval, we constrain it to be $3.82^{+0.50}_{-0.48}$. We note that the difference between the haze enhancement factor for free and equilibrium chemistry retrievals is more than two orders of magnitude. Additionally, the instrument offset between HST and JWST observations that we obtain with free retrieval is $204^{+16}_{-15}$ ppm, while with equilibrium chemistry retrieval, this is $39^{+21}_{-21}$ ppm, both with STIS+WFC3+NIRCam observations. In free retrieval, the large retrieved offset value compensates for the difference between the mean transit depth of HST and JWST observations. However, in the equilibrium chemistry retrieval, this is achieved by enhancing the haze, which can increase the transit depth in HST band-passes. The exact reason why both retrieval methods choose different parameters to compensate for the offset between the instruments remains elusive. We also find surprising uniformity in the retrieved cloud coverage fraction ($\phi$) across the retrieval methods. $\phi$ ranges from 0.44 -- 0.73 for STIS+WFC3+NIRCam observations with different model assumptions. Given this consistent value of cloud fraction, we also investigated the detection significance of inhomogeneous cloud and found it to be 2.8$\sigma$. We find that the median retrieved $P$-$T$ profile with free retrieval is hotter by $\sim$150-200\,K than the median retrieved $P$-$T$ profile with equilibrium chemistry retrieval, in the region probed by transmission spectra as shown in Figure \ref{fig:swn_disc}. However, the self-consistent RCTE $P$-$T$ profile spans the entire temperature range covered by the other two retrievals, as seen in the same figure. 

In this work, we constrain the C/O ratio of HD 209458 b to be consistent with solar C/O ratio with chemical equilibrium and grid retrievals. This is in contrast to the sub-solar C/O ratio constrained by \citet{Xue2024}. Moreover, we constrain the metallicity of HD 209458 b to be solar with chemical equilibrium and grid retrievals and highly sub-solar with free retrievals, in contrast to super-solar metallicity constrained by \citet{Xue2024}. This difference in constraints can substantially affect the interpretation of the formation location of HD~209458~b. \cite{Polanski2022} constrained the C/O ratio of the host star HD~209458 to be $0.42^{+0.08}_{-0.08}$. Therefore, the C/O ratio constraint that we obtain with our chemical equilibrium retrievals ($0.54^{+0.09}_{-0.09}$) is consistent with the stellar C/O ratio, indicating that the C/O ratio of HD~209458~b is similar to that of its host star.  It is important to note that, including the uncertainties, the C/O ratio of HD~209458~b is consistent with solar and stellar values since the C/O ratio of HD~209458 is quite close to the solar value.  The stellar (solar) C/O ratio and solar/sub-solar metallicity that we constrain for HD~209458~b implies that the atmosphere of HD~209458~b is not polluted i.e. enriched by metal-rich solid materials during/after its formation and formed interior to both the water snowline and the carbon-grain evaporation line \citep{Oberg2011, Espinoza2017}.

\section{Conclusions}
\label{sec:conclusions}
In this work, we have presented \texttt{SANSAR} (Suite of Adaptable plaNetary atmoSphere model And Retrieval), particularly the transmission spectra model of the suite. We discussed the implementation of opacity sampling and correlated-k methodology to treat opacities in \texttt{SANSAR} and compared both with each other and line-by-line opacities at the resolution of 0.001 cm$^{-1}$. We benchmark \texttt{SANSAR} forward model transmission spectra using the opacity sampling method with POSEIDON and the correlated-k method with ATMO. We benchmark the \texttt{SANSAR} retrieval model using both the opacity sampling and correlated-k methods by performing retrievals on the recent JWST NIRISS SOSS observations of WASP-96~b. For opacity sampling, we compare our results with POSEIDON and Aurora; for correlated-k, we compare with CHIMERA. We further interpret the HST and JWST observations of HD~209458~b using \texttt{SANSAR} with different instrument combinations, that is, just JWST NIRCam, HST WFC3 and JWST NIRCam, and finally combining HST STIS and WFC3 with JWST NIRCam observations, all with free chemistry, equilibrium chemistry, and self-consistent grid-retrieval. The most important findings of this work are as follows:

\begin{enumerate}

\item The accuracy of model transmission spectra with the correlated-k method at R$\sim$1000 is equivalent to that of the opacity sampling method at R$\sim$100000, highlighting the benefits of transmission spectra retrieval with a correlated-k method (for low-resolution observations) for speed and accuracy. The opacity sampling method should be used with caution at R $<$ 20000, especially for interpreting JWST observations. 
 
\item We detect H$_2$O, CO$_2$, Na and K in the atmosphere of HD~209458~b with detection significance of 12.6$\sigma$, 7.8$\sigma$, 6.9$\sigma$ and 4.9$\sigma$, respectively. The detection of H$_2$O and CO$_2$ is in agreement with \citet{Xue2024}, but they did not report any detection significances. The detection of Na and K is in line with several previous HST studies.  

\item We constrain the abundances (log of mixing ratio) of H$_2$O, CO$_2$, Na, and K to $-4.53^{+0.32}_{-0.29}$, $-7.52^{+0.30}_{-0.26}$, $-5.67^{+0.51}_{-0.45}$ and $-6.93^{+0.54}_{-0.47}$ respectively, with STIS+WFC3+NIRCam observations for free chemistry retrieval. The retrieved abundance of H$_2$O and CO$_2$ are highly sub-solar with STIS+WFC3+NIRCam observations, almost three orders of magnitude lower, compared to the highly super-solar abundance of H$_2$O ($-1.62^{+0.39}_{-1.29}$) and  CO$_2$ ($-5.31^{+0.37}_{-1.12}$) with just JWST NIRCam observations. The significant differences in the retrieved abundances between STIS+WFC3+NIRCam observations and just NIRCam observations highlight that NIRCam observations alone can lead to an overestimation of abundances from free retrievals for exoplanet atmospheres and should be used in combination with UV/Optical and near-infrared observations. In particular, even though we can detect \ce{CO2} feature with just NIRCam observations, we cannot constrain its abundances robustly without the optical baseline. We also predict that the abundances of \ce{CO2} and \ce{H2O} constrained by retrievals that only utilize JWST NIRSpec G395M/H and G235M/H observations may also be overestimated, compared to retrievals that include optical data.

\item We constrain the atmospheric metallicity of HD~209458~b to be $\sim$0.8--2.1 $\times$solar and 0.6--1.8 $\times$solar, with equilibrium chemistry retrieval and self-consistent grid-retrieval, respectively, with combined STIS+WFC3+NIRCam observations. The retrieved metallicity with both methods is consistent with solar and the host star metallicity. However, it is comparatively lower than the metallicity we obtain with just JWST NIRCam observations or WFC3+NIRCam observations by changing C/H to vary the C/O ratio. 

\item We find varying C/H to vary C/O ratio leads to super-solar metallicity ($\sim$2.2 -- 10.7 $\times$solar) with just NIRCam observations, as found by \citet{Xue2024}, however, when O/H is used, we constrain the metallicity to be sub-solar ($\sim$0.3--0.9 $\times$solar) with equilibrium chemistry retrievals. Varying either C/H or O/H yields similar constraints with combined STIS+WFC3+NIRCam data, resulting in near-solar metallicities ($\sim$0.8--2.1 $\times$solar) and C/O ratios (0.56 and 0.54, respectively).

\item The retrieved abundance of H$_2$O ($-3.35^{+0.20}_{-0.17}$) and \ce{CO2} ($-6.46^{+0.27}_{-0.32}$) with the equilibrium chemistry retrieval at one millibar, is higher by more than an order of magnitude, compared to that with free retrieval with STIS+WFC3+NIRCam observations. We find that the dependence of H$_2$O abundance on CO$_2$ abundance via metallicity in equilibrium chemistry retrieval does not allow H$_2$O abundance to go as low as in free retrieval, anchored by \ce{CO2} feature in NIRCam observations. This is also the reason why we constrain sub-solar metallicity with free retrievals, while it is solar with equilibrium chemistry and grid retrievals. 

\item We do not find any evidence of \ce{CO}, \ce{NH3}, \ce{C2H2}, \ce{CH4}, \ce{H2S}, and \ce{HCN}, which is consistent with \citet{Xue2024}, however, we are able to place upper limits on their abundances. The 3$\sigma$ upper limits on the abundances, log(X), that we obtain for \ce{CO}, \ce{NH3}, \ce{C2H2}, \ce{CH4}, \ce{H2S} and \ce{HCN} with STIS+WFC3+NIRCam observations and free retrievals are -5.35, -5.79, -6.17, -7.30, -5.69, and -6.97, respectively. 

\item We constrain the atmospheric C/O ratio of HD~209458~b to be $0.54^{+0.09}_{-0.09}$ and $0.46^{+0.09}_{-0.09}$, with equilibrium chemistry retrieval and self-consistent grid-retrieval, respectively, for combined STIS+WFC3+NIRCam observations. The retrieved C/O ratio from both methods is very close to the solar value. However, it is substantially higher than the C/O ratio we obtain with just JWST NIRCam observations or WFC3+NIRCam observations and that constrained by \citet{Xue2024}. 

\item The reduced $\chi^2$ values for the best-fit model with free, equilibrium, and grid retrievals with just NIRCam observations are lower than 0.6, while those with STIS+WFC3+NIRCam observations are comparatively closer to 1, thus highlighting that just NIRCam observations lead to overfitting by models and therefore leads to unphysical constraints.

\item We constrain the $P$-$T$ structure of HD~209458~b's atmosphere to be without a temperature inversion with free chemistry, equilibrium chemistry, and self-consistent grid retrievals with STIS+WFC3+NIRCam observations. The retrieved $P$-$T$  profile is cooler by about $\sim$400\,K with just NIRCam observations, as compared to WFC3+NIRCam and STIS+WFC3+NIRCam observations, with free chemistry retrievals, in the region probed by transmission spectra. 

\end{enumerate}

 Although, by using combined HST and JWST observations that provide broad wavelength coverage we are able to place constraints on the H$_2$O and CO$_2$ abundances in the atmosphere of HD~209458~b, we find that the choice of the methodology to interpret the observations can lead to different constraints on the H$_2$O and CO$_2$ abundances and therefore, its metallicity and C/O ratio. The future observations of HD~209458~b with other JWST instruments and their modes, offering broad wavelength coverage and high spectral resolution, may reconcile these differences. In our current work, we did not consider the 2D/3D nature of HD~2029458~b's atmosphere as detailed in previous works \citep[for e.g.][]{Lines2018} and also did not include any disequilibrium processes like vertical mixing and photochemistry\citep[for e.g.][]{Drummond2016}. Including these effects might lead to more robust constraints in the future and may reconcile the differences we see.

\section{Acknowledgments}
We would like to thank Ryan MacDonald for the extremely thorough referee report and very productive comments that have improved the quality of this manuscript substantially. AV and JG acknowledge funding from the SERB SRG Grant SRG/2022/000727-G of the Department of Science and Technology (DST), Government of India. 

%

\vspace{5mm}
\facilities{HST(STIS, WFC3), JWST(NIRCam)}

 
\software{  
\texttt{NumPy} \citep{numpy},
\texttt{Numba} \citep{numba}, 
\texttt{SciPy} \citep{2020SciPy-NMeth},
\texttt{Matplotlib} \citep{Hunter:2007matplotlib},
\texttt{SciencePlots} \citep{SciencePlots},
\texttt{SpectRes} \citep{spectres},
\texttt{PyMultiNest} \citep{pymultinest}, 
\textsc{PyFastChem} \citep{stock2022fastchem}, 
\texttt{Exo\_k} \citep{Leconte_2020}
}



\newpage
\bibliographystyle{aasjournal}

\appendix

\section{transit depth equation}
\label{ap:trans-derive}

\begin{figure*}
    \centering
    \includegraphics[width=\textwidth]{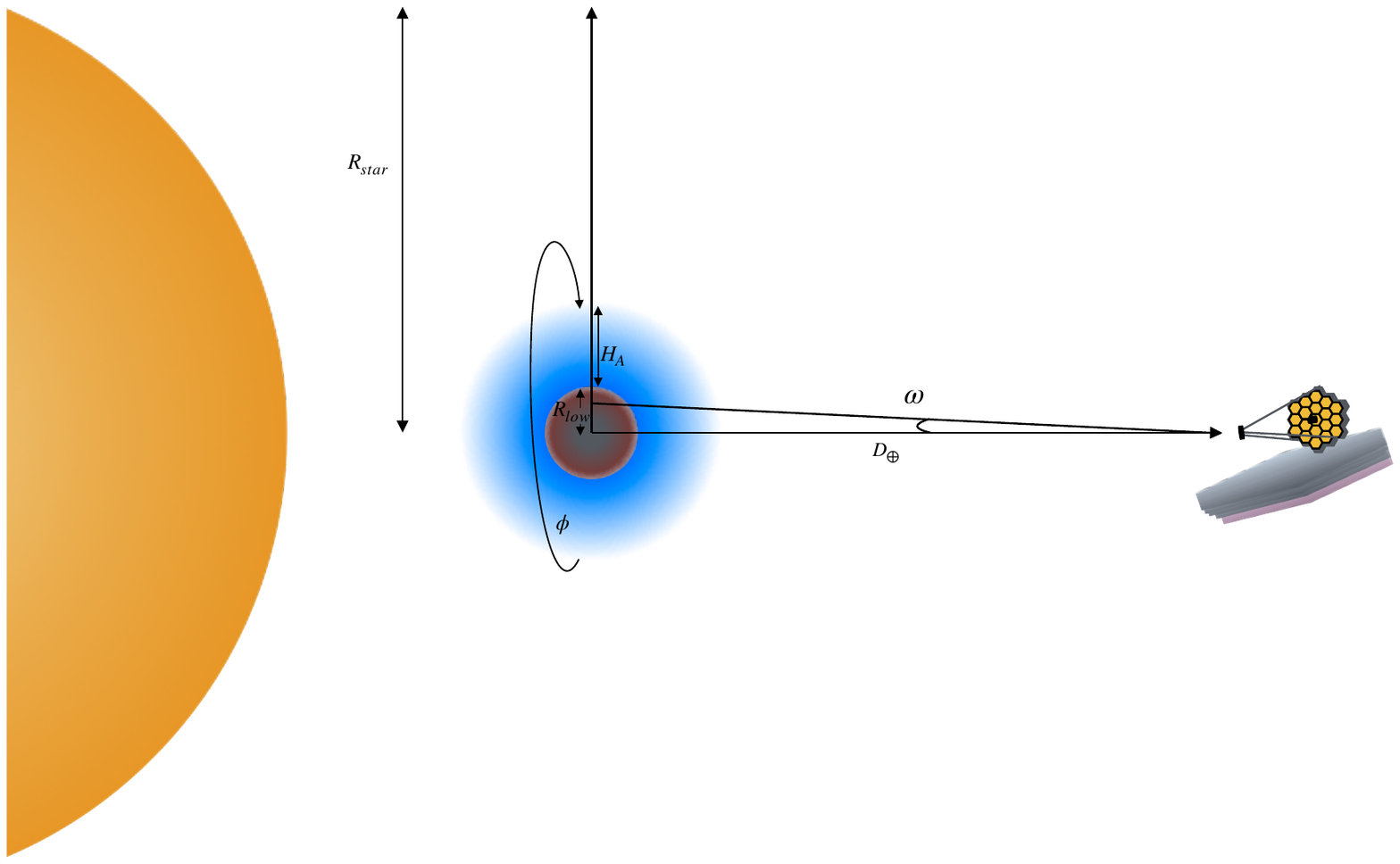}
    \caption{Side view schematic of a planet in transit being observed by a telescope (JWST).}
    \label{fig:transit-planet}

\end{figure*}

We show the complete derivation of the model transmission spectrum that we started in Section \ref{subsec:method_trans} here. We cannot spatially resolve exoplanets currently, therefore all the radiation that we receive from the planet-star hemisphere system is averaged into a single value of the flux. This wavelength ($\lambda$) and time ($t$) dependent flux from the planet/star received at the detector ($\mathcal{F}_{\oplus}(\lambda, t)$) is given by
\begin{equation}
\mathcal{F}_{\oplus}(\lambda, t)=\int_0^{2 \pi} \int_0^{R_{\mathrm{P}} / D_{\oplus}}
I_{\mathrm{S}}(\vartheta, \phi, \lambda, t) \cos \omega \sin \omega~\text{d} \omega~\text{d} \phi
\label{eq:flux-earth}
\end{equation}
where, $I_{\mathrm{S}}(\vartheta, \phi, \lambda, t)$ is the surface intensity of radiation of planet/star system, $\omega$ is the angle subtended by planet on the detector, $\phi$ is the azimuthal angle and $\vartheta$ is the planetary polar angle related to $\omega$. This equation integrates the intensity of radiation over the solid angle subtended on the detector. In the limit of $R_p \ll D_{\oplus}$, equation \ref{eq:flux-earth} simplifies as,
\begin{equation}
\mathcal{F}_{\oplus}(\lambda, t)=\left(\frac{R_{\mathrm{p}}}{D_{\oplus}}\right)^2 F_{\mathrm{S}}(\lambda, t)
\label{eq:rel-flux}
\end{equation}
where, $F_{\mathrm{S}}(\lambda, t)$ is the surface flux of the planet/star. A more detailed solution can be found in \citet{Seager2010}

During a transit, the planet is in the line of sight of the detector and the star as shown in Figure \ref{fig:transit-planet}, due to which the flux received by the detector reduces as compared to when the planet is out of transit. In-transit flux received by the detector can be divided into three parts: planet flux, atmosphere flux, and stellar flux. Here, planet flux refers to the planetary emission, atmosphere flux is the stellar radiation that gets transmitted through the planet's atmosphere, and stellar flux is the unobstructed stellar radiation reaching the detector. Therefore,

\begin{equation}
\begin{aligned}
\mathcal{F}_{\text {in trans }, \oplus}(\lambda) & =\int_0^{2 \pi} \int_{\left(R_{\mathrm{low}}+H_A\right) / D_{\oplus}}^{R_{\star} / D_{\oplus}} \hspace{-25pt} I_{\star}(\vartheta, \phi, \lambda) \cos \omega \sin \omega d \omega d \phi~+\\
& \int_0^{2 \pi} \int_0^{R_{\mathrm{low}} / D_{\oplus}} I_{\mathrm{p}}(\vartheta, \phi, \lambda) \cos \omega \sin \omega d \omega d \phi~+ \\
& \int_0^{2 \pi} \int_{R_{\mathrm{low}} / D_{\oplus}}^{\left(R_{\mathrm{low}}+H_A\right) / D_{\oplus}} \hspace{-23pt} I_{\mathrm{atm}}(\vartheta, \phi, \lambda) \cos \omega \sin \omega d \omega d \phi.
\end{aligned}
\end{equation}

Here, $\mathcal{F}_{\text {in trans }, \oplus}(\lambda)$ is the flux received by the detector during transit, $I_{\star}$ is the stellar intensity, $I_p$ is the intensity of radiation emitted by the planet, $I_{\text{atm}}$ is the intensity of the radiation transmitted through the planet's atmosphere, $R_{\text{low}}$ is the lowest radius below which no stellar radiation can pass through the planet and $H_A$ is the height of the atmosphere extending from $R_{\text{low}}$. Since the distance between the planet and the host star is much less than that of the distance of the star-planet system from the detector near Earth, distance of the planet and star from the detector is assumed to be the same. Since planetary emission is comparatively low as compared to stars in the wavelength range of our interest (0.2 $\mu$m to 5 $\mu$m),  we can set $I_p = 0$ and utilizing equation \ref{eq:rel-flux}, we get

\begin{equation}
    \begin{aligned}
\mathcal{F}_{\text {in trans }, \oplus}(\lambda) & =\frac{R_{\star}^2 - (R_{\text{low}}+H_A)^2}{D_{\oplus}^2} \mathcal{F}_{S,\star}\\
& +\frac{(R_{\text{low}}+H_A)^2 - R_{\text{low}}^2}{D_{\oplus}^2} \mathcal{F}_{S,\text{atm}}
\end{aligned}
\end{equation}

Where $\mathcal{F}_{S,\text{atm}}$ is the flux received from the atmosphere and $\mathcal{F}_{S,\star}$ is the stellar flux received.

As shown in equation \ref{eq:transit depth}, transit depth is given by,
\begin{equation}
    \delta(\lambda) = \frac{\mathcal{F}_{\text {out trans }, \oplus}(\lambda) - \mathcal{F}_{\text {in trans }, \oplus}(\lambda)}{\mathcal{F}_{\text {out trans }, \oplus}(\lambda)}. 
    \label{eq:td-eq-main}
\end{equation}

We know that $\mathcal{F}_{\text {out trans }, \oplus}(\lambda)$ is the flux received by the detector outside of transit, hence this is equal to the stellar flux received. From equation \ref{eq:rel-flux}, we can relate stellar flux at the surface of the star with the stellar flux received by detector as $$\mathcal{F}_{\text {out trans }, \oplus}(\lambda) = \frac{R_{\star}^2}{D_{\oplus}^2}\mathcal{F}_{S,\star}$$

Substituting values of $\mathcal{F}_{\text {out trans }, \oplus}(\lambda)$ and $\mathcal{F}_{\text {in trans }, \oplus}(\lambda)$ in equation \ref{eq:td-eq-main} we obtain, 

\begin{equation}
    \begin{aligned}
        \delta(\lambda) &=\frac{\frac{R_{\star}^2}{D_{\oplus}^2}\mathcal{F}_{S,\star} - \frac{R_{\star}^2 - (R_{\text{low}}+H_A)^2}{D_{\oplus}^2} \mathcal{F}_{S,\star} - \frac{(R_\text{low}+H_A)^2 - R_{\text{low}}^2}{D_{\oplus}^2} \mathcal{F}_{S,\text{atm}}}{\frac{R_{\star}^2}{D_{\oplus}^2}\mathcal{F}_{S,\star}}\\
        &= \frac{R_\text{low}^2+H_A^2+2R_{\text{low}}H_A - (H_A^2+2R_{\text{low}}H_A)\frac{\mathcal{F}_{S,\text{atm}}}{\mathcal{F}_{S,\star}}}{R^2_{\star}}\\
        &=\frac{R_{\text{low}}^2+H_A^2-H_A^2(\frac{\mathcal{F}_{S,\text{atm}}}{\mathcal{F}_{S,\star}})+2R_{\text{low}}H_A(1-\frac{\mathcal{F}_{S,\text{atm}}}{\mathcal{F}_{S,\star}})}{R_{\star}^2}\\
        &=\frac{R_{\text{low}}^2 + 2R_{\text{low}}H_A(1-\frac{\mathcal{F}_{S,\text{atm}}}{\mathcal{F}_{S,\star}})}{R_{\star^2}}. 
    \end{aligned}
\end{equation}

Here, since $\frac{H_A^2}{R_{\star}^2} \approx 0$, $H_A^2$ terms were neglected. $\mathcal{F}_{S,\text{atm}}$ is the stellar radiation coming through the planet's atmosphere after absorption (as we are neglecting emission by the planet itself, and also no multiple-scattering processes add rays in the line of sight). Therefore, by applying Beer's law in this plane-parallel geometry, we obtain

\begin{equation}
\frac{\mathcal{F}_{S,\text{atm}}}{\mathcal{F}_{S,\star}} = e^{-\tau(\lambda)} 
   \label{eq:td-beer}
\end{equation}

where, $\tau(\lambda)$ is the wavelength dependent optical depth. Therefore,

\begin{equation}
    \delta(\lambda) = \frac{R_{\text{low}}^2 + 2R_{\text{low}}H_A(1-e^{\tau(\lambda)})}{R^2_\star}
    \label{eq:td-simple}
\end{equation}

This solution is for a simplified case where the whole atmosphere is represented with a single optical depth (single layer). This is not true in reality, as different portion of the atmosphere absorbs radiation differently due to pressure, temperature, chemical abundances, and path length variations. Thus, to accommodate these variables, we divide the atmosphere into several layers, and the above equation can be written in an integral form as,
\begin{equation}
\delta (\lambda) = \frac{R_{\text{low}}^2 + 2\int_{R_{\text{low}}}^{R_{\text{low}}+H_{\text{A}}} b (1-e^{-\tau(\lambda, b)}) \text{d} b}{R^2_\star}
\label{eq:td main}
\end{equation}
\eject
In comparison to equation \ref{eq:td-simple}, we now have multiple atmospheric layers, each represented by impact parameter $b$, with the size of each layer represented by d$b$ governed by hydrostatic equilibrium as described in Section \ref{sec:model atmosphere}. 








\section{Benchmarking} \label{app: benchmarking}
In this section we show the Figures and Tables that
demonstrate the benchmarking and performance of \texttt{SANSAR}
with regards to speed and accuracy. Table \ref{tab:perf-bench} summarizes the
time required by \texttt{SANSAR} to generate forward models under
different configurations. The accuracy of bicubic interpolation
is demonstrated in Figure \ref{fig:A bicubic comp}. Figure \ref{fig:poseidon-bench-main} compares the forward
models produced by \texttt{SANSAR} and \texttt{POSEIDON}, while Figure
\ref{fig:otfvsgrid} illustrates the difference between on-the-fly interpolation
and grid-based interpolation in generating forward models.

\begin{deluxetable}{lcccc}[h!]
    \label{tab:perf-bench}
    \renewcommand{\arraystretch}{1.1}
    \tabletypesize{\footnotesize}
    \tablecolumns{4} 
    \tablecaption{Performance benchmarking of \texttt{SANSAR} on a typical Desktop. Here, otf and grid represent on-the-fly and grid-based interpolation respectively.}
    \tablehead{Method & Cloud-Haze& Wavelength range ($\mu$m) & Time (s)}
    \startdata
         Sampling (otf)&Yes  & $0.2 - 30$  & $ 3.5$ \\
         Sampling (grid)& Yes& $0.2 - 30$&  $1$\\
         Correlated-k (otf)&  Yes& $0.2 - 30$ & $1.93$ \\
         Correlated-k (grid)& Yes & $0.2 - 30$ & $0.61$\\
    \enddata
\end{deluxetable}

\begin{figure}[h!]
    \includegraphics[width=0.5\textwidth]{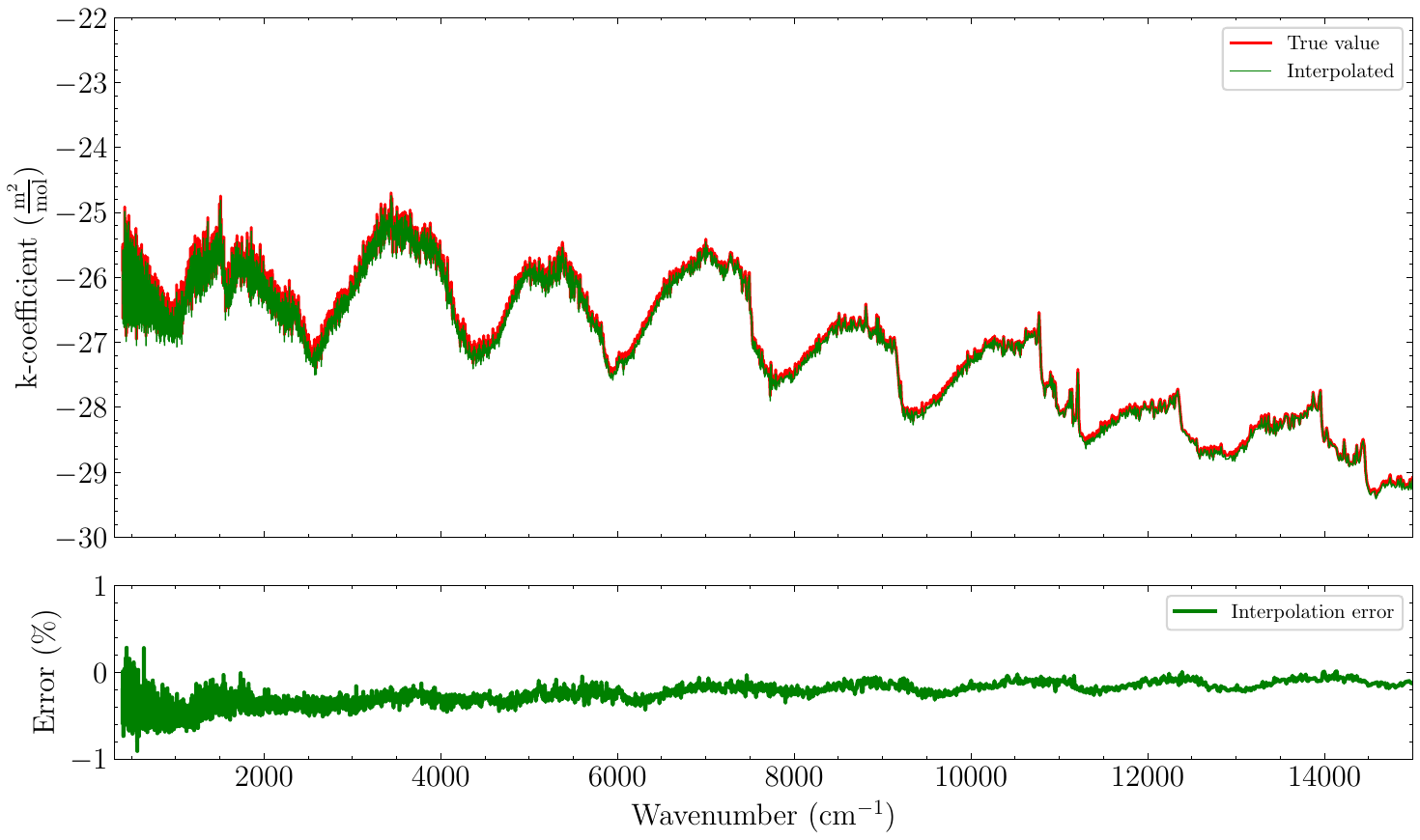}
    \caption{This figure shows the accuracy of bicubic convolution interpolation. The red curve represents the true k-coefficient at 
    $\sim$ 1mbar and 1360 K pressure and temperature respectively, whereas the green curve is the interpolated. The bottom plot shows the percentage error of interpolated values with respect to the true value.}
    \label{fig:A bicubic comp}
\end{figure}

\begin{figure}[h!]
    \includegraphics[width=0.48\textwidth]{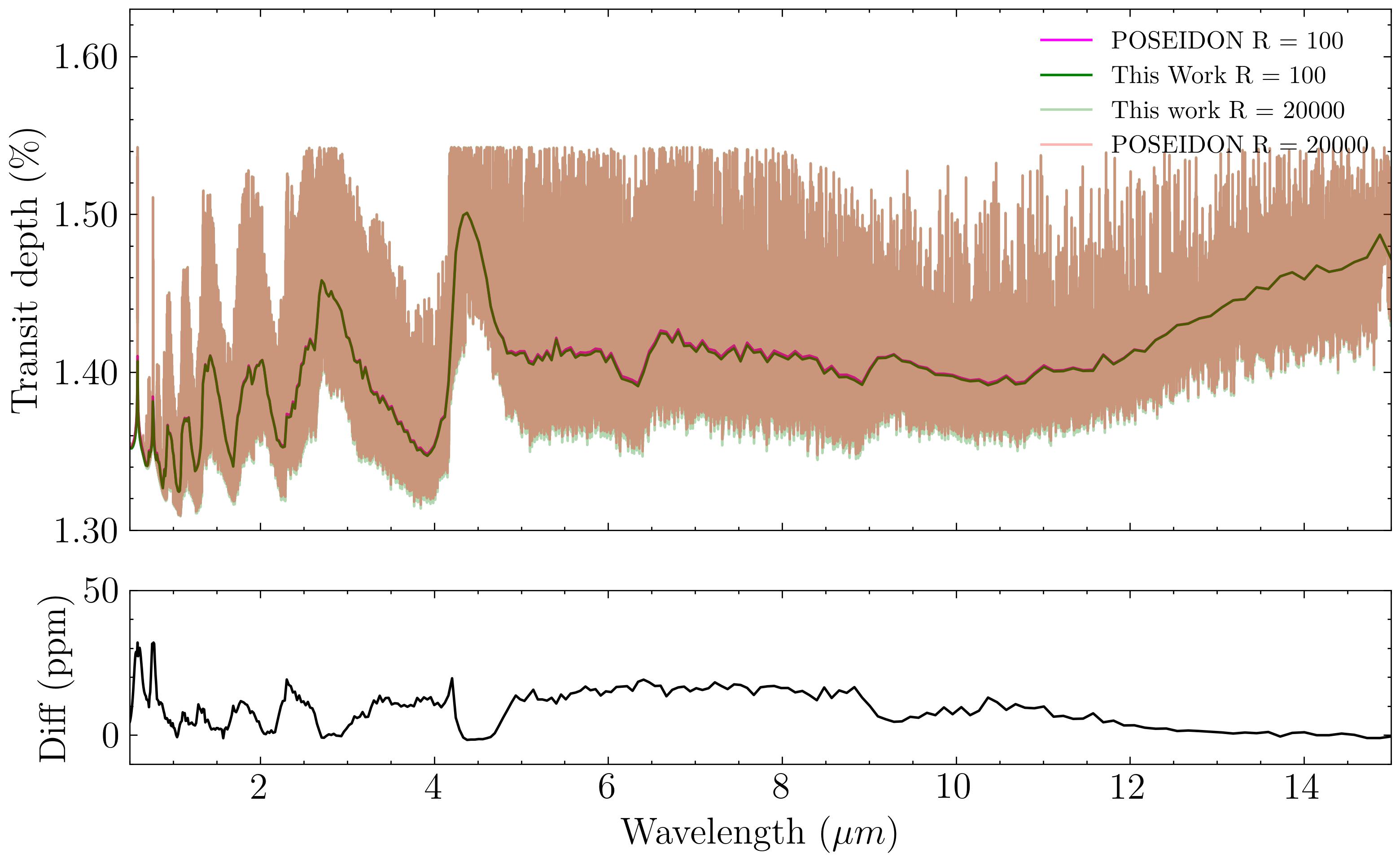}
    \caption{Comparison of forward models from \texttt{SANSAR} and POSEIDON. Here, we have used the opacity files of POSEIDON in \texttt{SANSAR}. The bottom panel shows the difference between SANSAR and POSEIDON forward model binned at R$\sim$100.}
    \label{fig:poseidon-bench-main}
\end{figure}

\begin{figure}[h!]
    \includegraphics[width=0.48\textwidth]{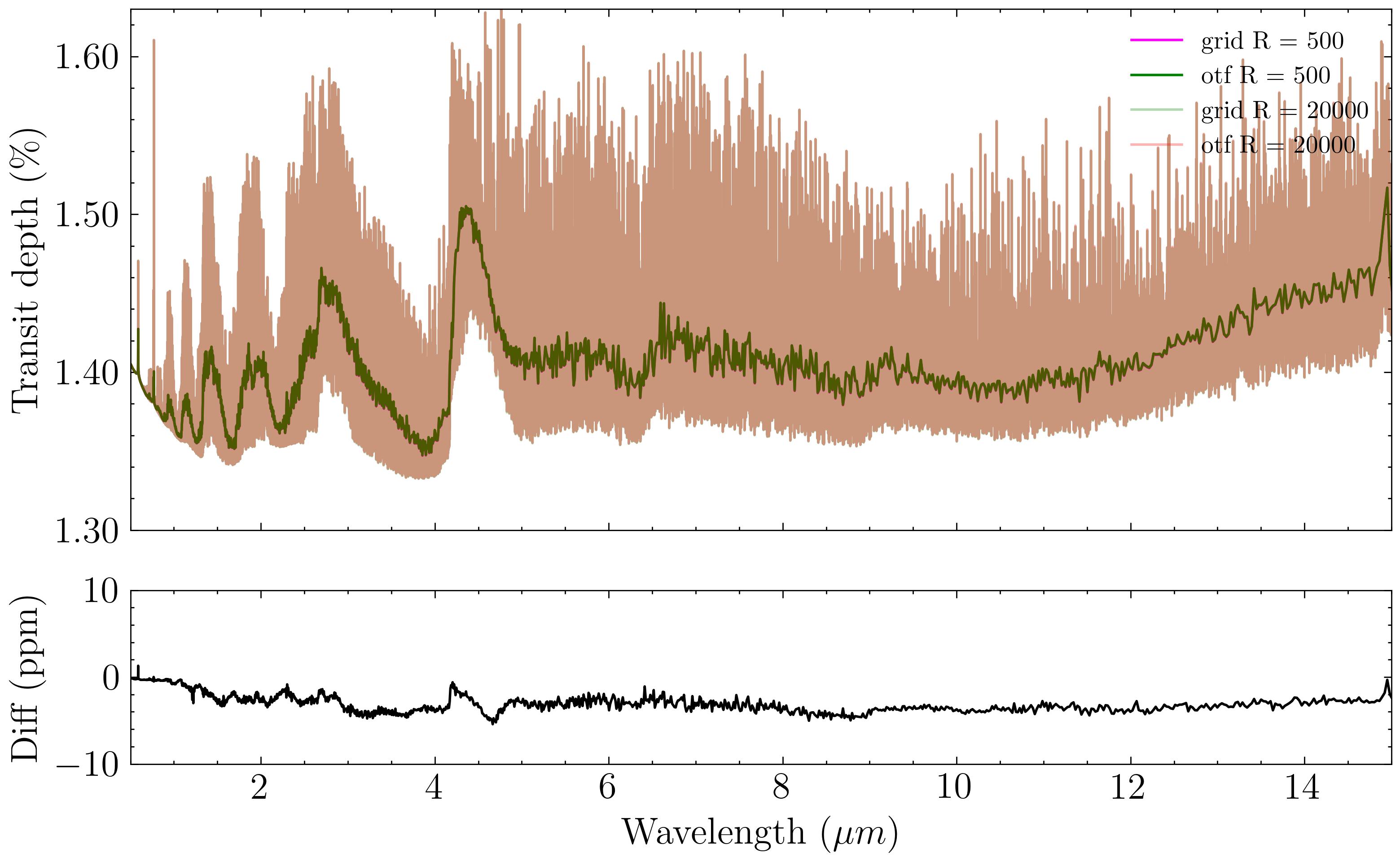}
    \caption{Comparison of grid-based interpolation and on-the-fly interpolation when generating the forward model. This comparison was performed for a WASP-96 b type planet with an isothermal temperature of 1205\,K with abundances provided by \citet{Taylor2023}.}
    \label{fig:otfvsgrid}
\end{figure}


\section{Corner plots}
In this section we show the corner plots from different retrieval
analyses. Figure \ref{fig:SWN_corner} shows the posterior distributions of various
parameters from the free chemistry retrieval using STIS+WFC3
+NIRCam observations, while Figure \ref{fig:WN_corner} shows the posteriors
obtained using only WFC3+NIRCam observations.

\begin{figure*}[h!]
    \includegraphics[width=\textwidth]{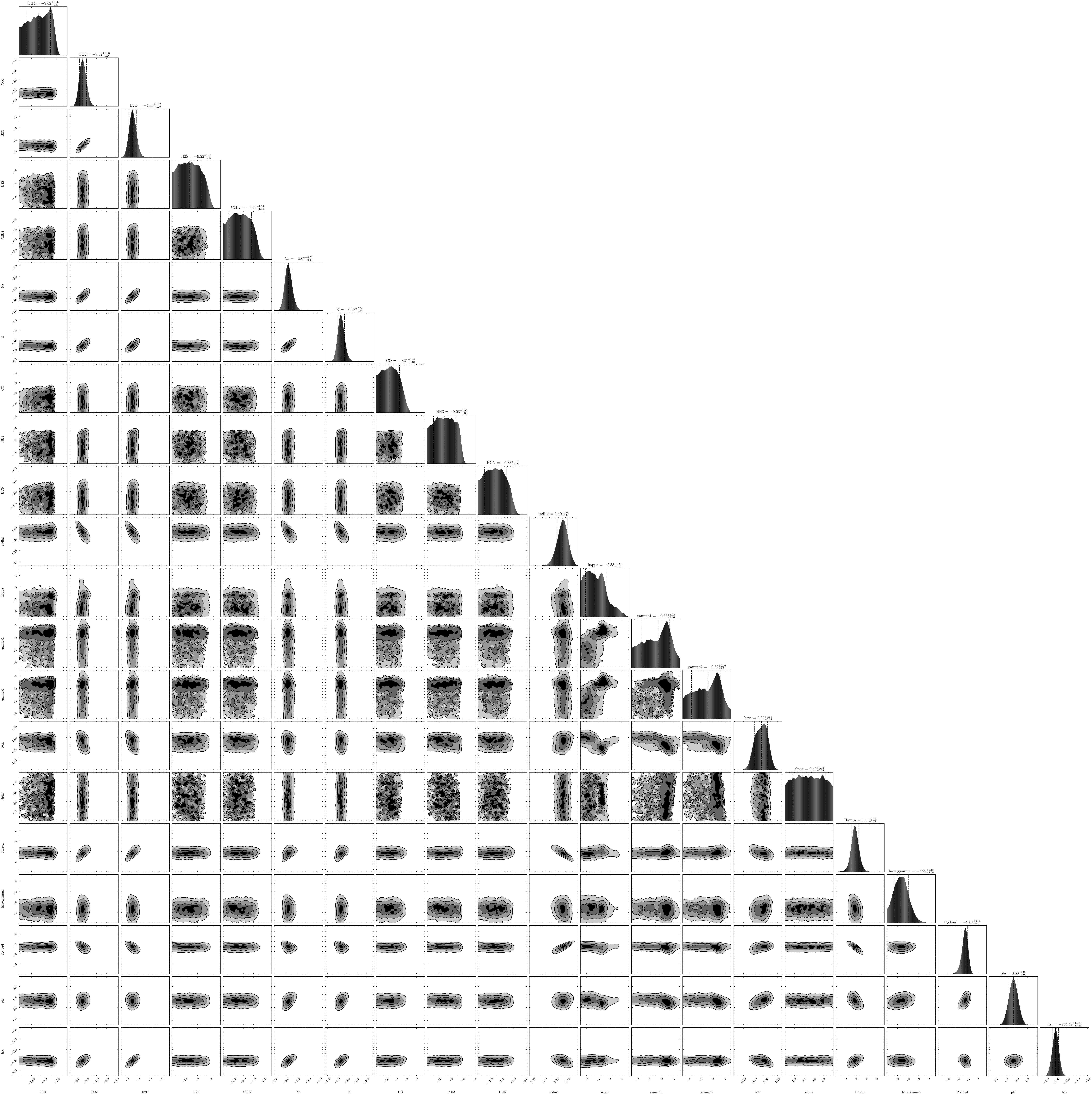}
    \caption{Figure showing the retrieved posterior of different parameters with free chemistry retrieval obtained with STIS+WFC3+NIRCam observations. Here, retrieval was performed with the correlated-k method.}
    \label{fig:SWN_corner}
\end{figure*}

\begin{figure*}[h!]
    \includegraphics[width=\textwidth]{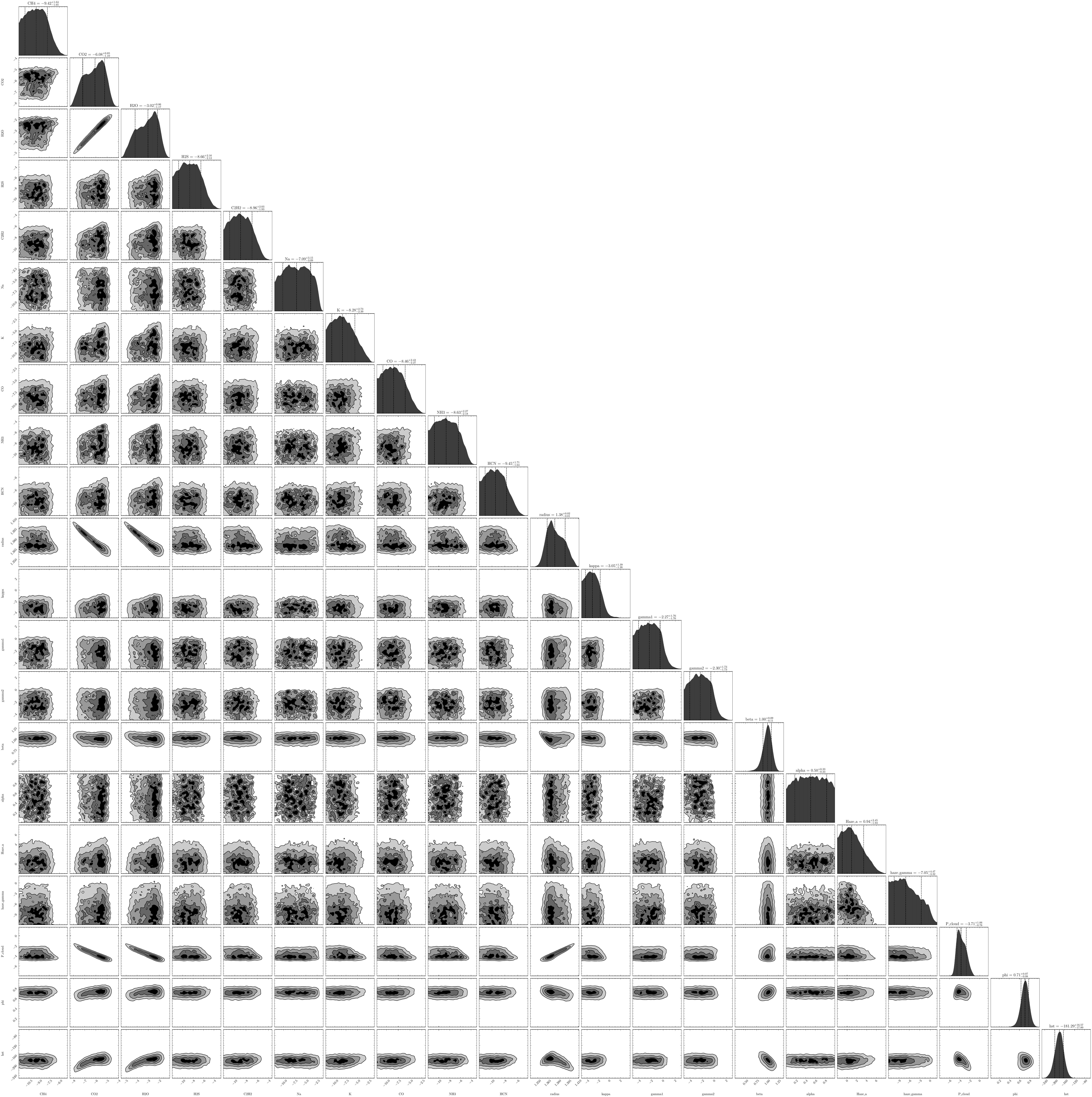}
    \caption{Figure showing the retrieved posterior of different parameters with free chemistry assumption obtained with WFC3+NIRCam observations. Here also, retrieval was performed with the correlated-k method.}
    \label{fig:WN_corner}
\end{figure*}

\section{Equilibrium Chemistry Heatmaps} \label{app: heatmaps}
In this section we show heatmaps of chemical abundances
across various combinations of metallicity and C/O ratios.
Figure \ref{Change_C} shows the abundance heatmap obtained by varying
carbon (C/H) to vary the C/O ratio, while Figure \ref{Change_O} shows the
heatmap resulting from varying oxygen (O/H )to vary C/O
ratio.

\begin{figure*}[h!]
    \includegraphics[width=\textwidth]{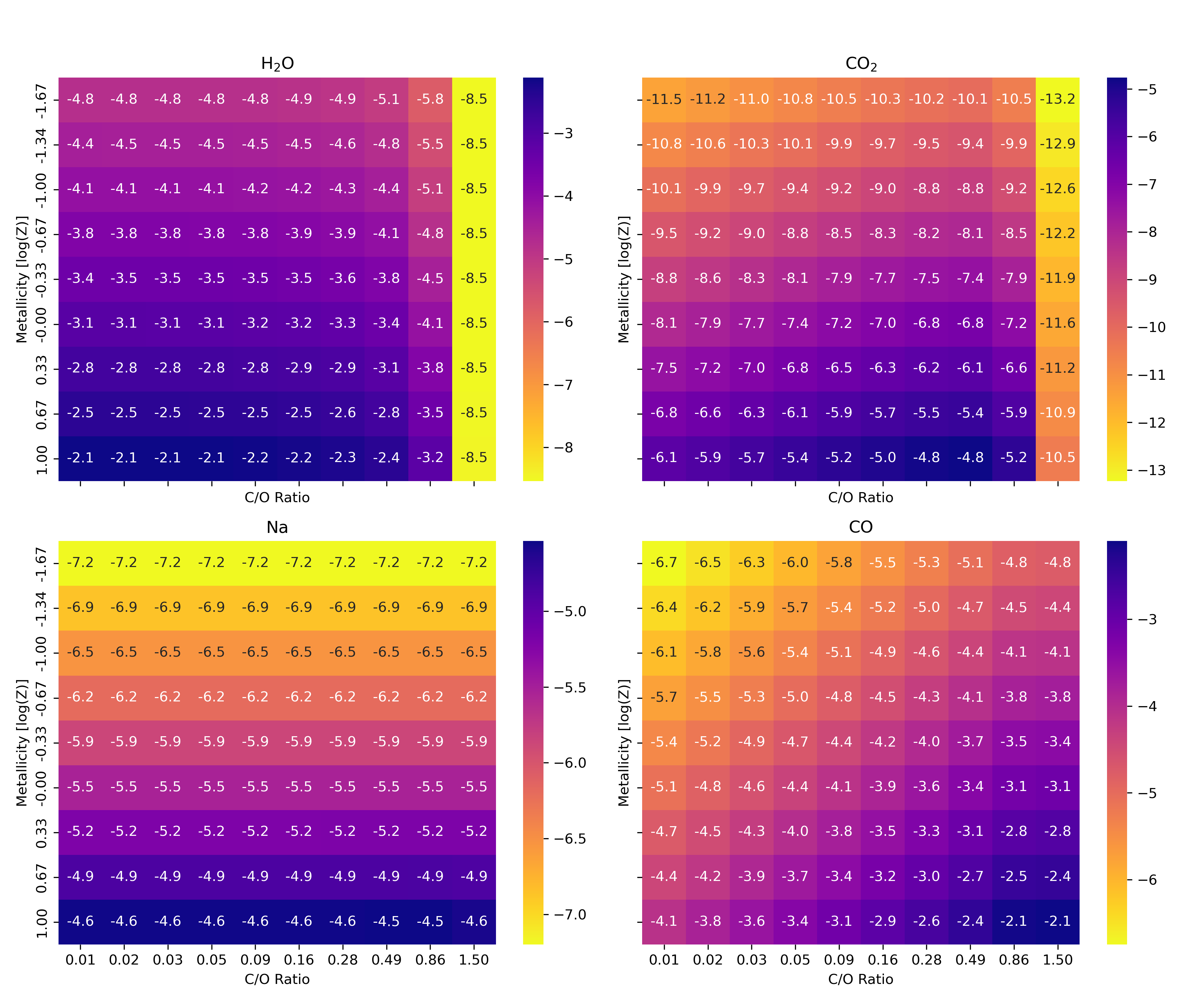}
    \caption{Heatmap showing the predicted abundances of major chemical species in HD~209458~b at 1mbar pressure level. These values are obtained using \textsc{FastChem} while varying C/H for different C/O values}
    \label{Change_C}
\end{figure*}

\begin{figure*}[h!]
    \includegraphics[width=\textwidth]{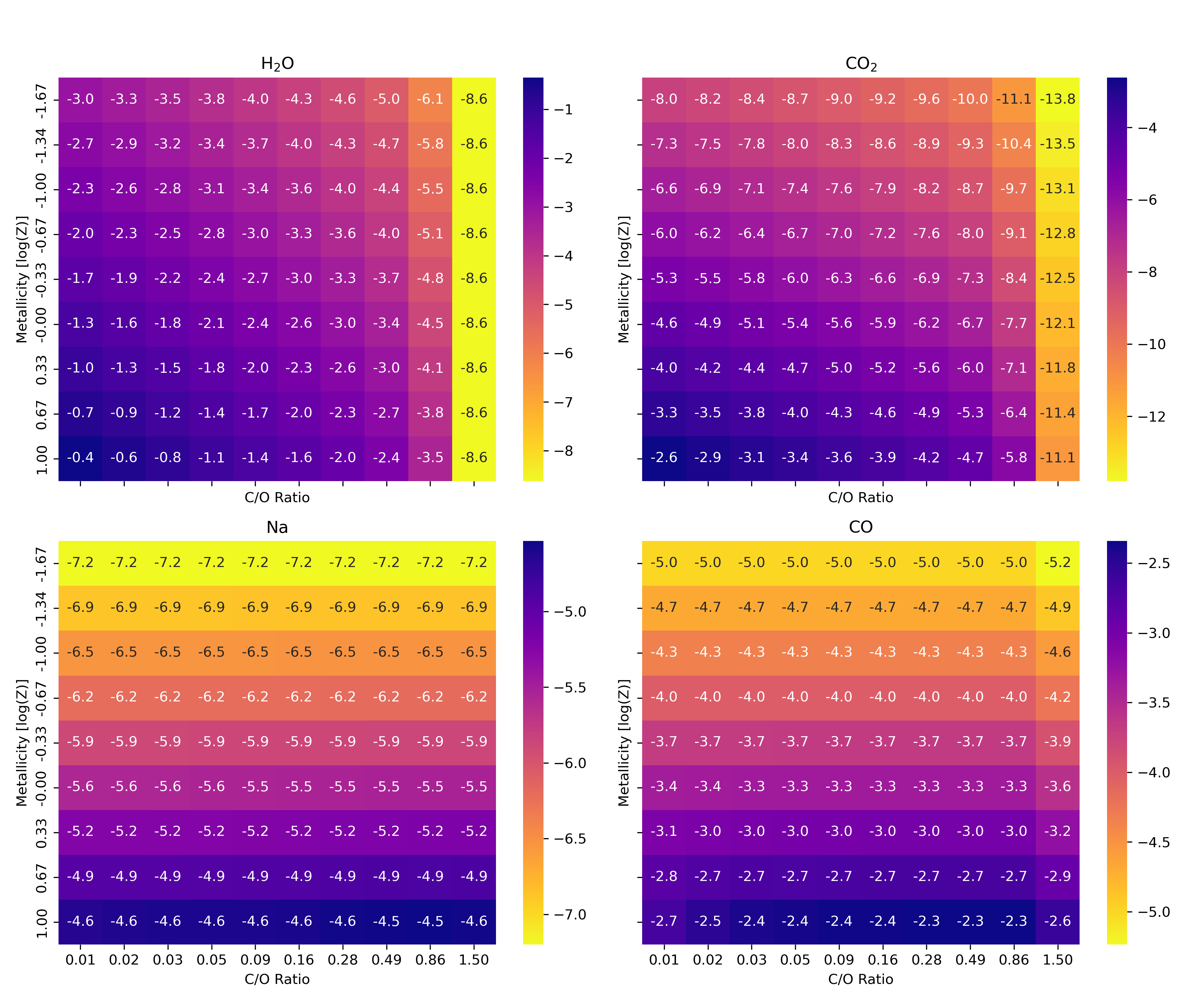}
    \caption{Heatmap showing the predicted abundances of major chemical species in HD~209458~b at 1mbar pressure level. These values are obtained using \textsc{FastChem} while varying O/H for different C/O values}
    \label{Change_O}
\end{figure*}

\section{Line List Sources} \label{app: linelist_source}

\begin{deluxetable*}{lcc}
    \label{tab:line-list sources}

    \renewcommand{\arraystretch}{1.5}
    \tabletypesize{\footnotesize}
    \tablecolumns{3} 
    \tablecaption{Line list sources of chemical species used in this work}
    \tablehead{Species & Line list & Partition function }
    \startdata
    \ce{H2O} & \citet{Barber2006} & \citet{Barber2006}\\
    \ce{CO2} & \citet{Tashkun2011} & \citet{ROTHMAN2009533}\\ 
    \ce{CO} & \citet{ROTHMAN2010} & \citet{ROTHMAN2009533} \\
    \ce{CH4} & \citet{Yurchenko2014} & \citet{Yurchenko2014} \\
    \ce{NH3} & \citet{Yurchenko2011} & \citet{Sauval1984} \\
    \ce{HCN} & \citet{Harris2006} & \citet{Harris2006} \\
    \ce{C2H2} & \citet{ROTHMAN2013} & \citet{ROTHMAN2013} \\
    \ce{H2S} & \citet{ROTHMAN2013} & \citet{ROTHMAN2013} \\
    \ce{SO2} & \citet{Underwood2016} & \citet{Underwood2016} \\
    \ce{Na} & VALD3$^{1}$ & \citet{Sauval1984} \\
    \ce{K} & VALD3$^{1}$ & \citet{Sauval1984} \\
    \ce{H2-H2} CIA & \citet{RICHARD2012} & N/A \\
    \ce{H2-He} CIA & \citet{RICHARD2012} & N/A \\
    \enddata
    \tablecomments{$^1$\citet{Heiter2008};\citet{Ryabchikova2015}. The prescription of line wings for Na and K are from \citet{Allard2003} \& \citet{Allard2007}.}
\end{deluxetable*}

\end{document}